\documentclass[superscriptaddress]{revtex4-2}

\usepackage{physics}
\usepackage{braket}
\usepackage{bm}
\usepackage{graphicx}
\usepackage{pifont}
\usepackage{amsmath}
\usepackage{amsfonts}
\usepackage{amssymb}
\usepackage{mathrsfs}
\usepackage{subfigure}
\usepackage{multirow}
\usepackage{ulem}
\usepackage{color}
\usepackage[table]{xcolor}
\usepackage[colorlinks,linkcolor={blue},citecolor={blue},urlcolor={blue}]{hyperref}

\begin{document}

\title{F\"orster valley-orbit coupling and topological lattice of hybrid moir\'e excitons}


\author{Huiyuan Zheng}
\affiliation{New Cornerstone Science Laboratory, Department of Physics, University of Hong Kong, Hong Kong, China}
\affiliation{HK Institute of Quantum Science \& Technology, University of Hong Kong, Hong Kong, China}

\author{Ci Li}
\affiliation{School of Physics and Electronics, Hunan University, Changsha
410082, China}

\author{Hongyi Yu}
\affiliation{School of Physics and Astronomy, Sun Yat-sen University (Zhuhai
Campus), Zhuhai 519082, China}

\author{Wang Yao}
\email{wangyao@hku.hk}
\affiliation{New Cornerstone Science Laboratory, Department of Physics, University of Hong Kong, Hong Kong, China}
\affiliation{HK Institute of Quantum Science \& Technology, University of Hong Kong, Hong Kong, China}

\date{\today}

\begin{abstract}
Hybrid excitons in moir\'e superlattices of two-dimensional (2D) semiconductors inherit the electric dipole, strong moir\'e trapping, and stacking optical selection rules from its interlayer part, whereas the intralayer part is intended for enhancing optical coupling strength.  
Here, we show that electron-hole Coulomb exchange, or Förster type multipole-multipole coupling, in the intralayer component qualitatively alters the properties of moir\'e excitons, enabling their coherent hopping between moir\'e traps laterally separated over 10 nm and/or across layers, where their kinetic propagation is completely suppressed.
Valley-flip hopping channels are found as significant as the valley-conserving ones, leading to rich possibilities to tailor valley-orbit-couplings and introduce non-trivial topology to the moir\'e exciton superlattice. In twisted MoTe$_2$ where hybrid moir\'e excitons feature a symmetry protection from radiative recombination, we show that Förster valley-orbit-coupling can give rise to a rich topological phase diagram.

\end{abstract}

\maketitle

\section{Introduction}
In bilayer 2D semiconductors, moir\'e pattern introduced by twist and/or lattice mismatch can endow excitons a plethora of intriguing opportunities, from applications in optoelectronics to the exploration of quantum matters of fundamental interest~\cite{yu2015anomalous, wu2017topological, yu2017moire, rivera2018interlayer, seyler2019signatures, tran2019evidence, jin2019observation, alexeev2019resonantly, huang2022excitons}.
In heterobilayers, the lowest energy excitons have an interlayer configuration carrying permanent electric dipole, which underlies electric field tunability of exciton energy and pronounced exciton-exciton interaction~\cite{rivera2016valley}.
In the moir\'e defined periodic landscape~\cite{zhang2017interlayer}, these interlayer excitons are tightly confined in ordered array of moir\'e traps, exhibiting behavior akin to quantum-dot-like single-photon emitters~\cite{yu2017moire, seyler2019signatures, brotons2020spin, baek2020highly}. 
Such quantum emitters uniquely feature optical selection rules conditioned on the spin and valley indices as well as local atomic registries~\cite{yu2017moire, yu2018brightened, jin2019identification}, implying rich optoelectronic control possibilities. 
While the layer separation of electron and hole components has led to a weak coupling to light, the resultant long radiative lifetime is on the other hand favorable for the exploration of exciton many-body physics in Bose-Hubbard lattices with the strong dipolar interaction~\cite{li2020dipolar, xiong2023correlated, park2023dipole}.

\begin{figure}
\includegraphics[width=0.48\textwidth]{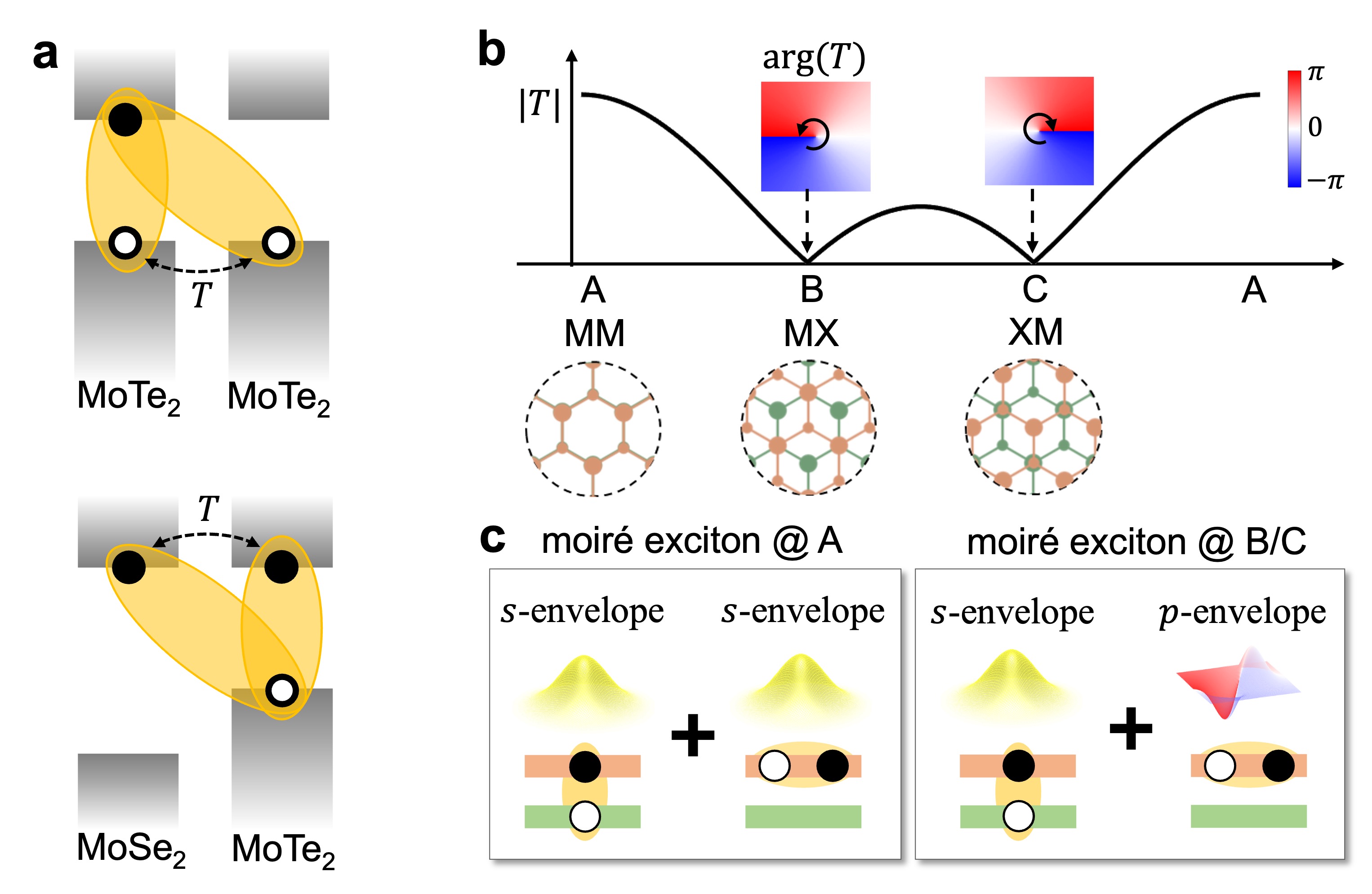}%
\caption{Exciton hybridization in TMD moir\'e structures. $\mathbf{a}$. Schematics of hybrid excitons in TMD homobilayers (upper) and heterobilayers (lower). Black (white) dots denote the electron (hole). $T$ denotes interlayer hopping.
$\mathbf{b}$. Spatially varying interlayer hopping in R-stacking bilayers. Hopping vanishes at the high symmetry points B (MX) and C (XM), and has a $p$-wave form in their vicinity. The insets below show their local stacking. The phase of $T$ is color coded to show the counterclockwise and clockwise winding patterns around the B and C points respectively. At the vicinity of A point of MM stacking, the hopping has a $s$-wave form.
$\mathbf{c}$. Schematics of the center-of-mass envelope functions of the hybrid exciton at high symmetry points. The minor intralayer component exhibits a $s$-wave and a $p$-wave envelope at A point and B/C points, respectively, whereas the major interlayer component always has an $s$-wave envelope for the lowest energy state in the confinement. Black (white) dots represent the electron (hole), indicating their intralayer/interlayer components.}
\label{fig_selectionrule}
\end{figure}

When light-coupling is favored, the interlayer moir\'e excitons can be brightened through hybridization with a nearly resonant intralayer exciton. 
In heterobilayers this is made possible under various compound choices that have nearly aligned conduction or valence bands~\cite{hsu2019tailoring, alexeev2019resonantly, zhang2020twist, zhao2024hybrid} (Fig.~\ref{fig_selectionrule}a). 
In moir\'e traps of different local stacking registries, the $C_3$ rotational symmetry 
dictates different center-of-mass envelope forms of the intralayer component hybridized to a tightly trapped interlayer exciton wavepacket (Fig.~\ref{fig_selectionrule}c), whereas an $s$-wave envelope leads to brightening. 
The intralayer wavefunction overlap between electron and hole also underlies pronounced electron-hole Coulomb exchange that can non-locally transfer exciton either in momentum space (i.e. between valleys~\cite{yu2014dirac, qiu2015nonanalyticity,liu2025direct}), or in real space~\cite{selig2019theory,baimuratov2020valley,hichri2021resonance,li2023cross}. This is essentially the F\"orster non-radiative dipole-dipole coupling underlying the fluorescence resonance energy transfer~\cite{clegg2006the}. 

Hybrid moir\'e excitons are also hosted by homobilayers of near 0$^{\circ}$ twisting, such as twisted MoTe$_2$ (t-MoTe$_2$) that has caught great attention for the fractional quantum anomalous Hall effects~\cite{cai2023signatures,park2023observation,zeng2023thermodynamic, xu2023observation}. 
Because of the stacking dependent interfacial electrical polarization~\cite{weston2022interfacial, wang2022interfacial}, interlayer excitons are trapped at the MX and XM stacking regions with opposite layer configuration (electric dipole), where trapping energy can largely compensate the binding energy difference from an intralayer exciton, leading to their hybridization~\cite{yu2021luminescence}.
The $C_3$ rotational symmetry, however, dictates that the hybridized intralayer component has a $p$-wave envelope that leads to vanishing optical transition dipole, therefore protected from radiative recombination. Such homobilayers offer a platform to explore dipolar ordering of long-lived hybrid moir\'e excitons, where spontaneous $C_3$ symmetry breaking manifests a unique optical signature~\cite{yu2021luminescence}.

Here, we show that the moir\'e exciton is qualitatively altered by its intralayer component, in which sizable F\"orster type multipole-multipole coupling enables the hybrid exciton to hop coherently and non-locally between moir\'e traps across layers and/or over 10 nm apart laterally, where kinetic propagation is completely suppressed.
Valley-flip hopping channels are found as significant as the valley-conserving ones, and we identify their non-monotonic dependence on hopping distance, with phase pattern dependent on valley indices and envelope forms of intralayer component. 
This leads to rich possibilities to tailor valley-orbit-coupling (VOC) forms, which can introduce non-trivial topology to the moir\'e exciton superlattice.
We demonstrate this possibility for the long-lived excitons in twisted MoTe$_2$ bilayers, where F\"orster type quadrupole-quadrupole couplings link the two otherwise isolated exciton sublattices of opposite layer configurations and give rise to a rich topological phase diagram.

\section{Results}
\subsection{Layer hybridization of moir\'e trapped excitons.}
In a long-wavelength moir\'e pattern, the $C_3$ rotational symmetry dictates that the potential energy minima and maxima of exciton must correspond to the three high symmetry points in a supercell where local stacking registries are $C_3$ invariant. We focus on moir\'e excitons trapped at these points with a major interlayer component of an $s$-wave envelope, and a minor intralayer component (Throughout this paper, `envelope function' refers specifically to the exciton's center-of-mass wave function.). Their hybridization is determined by the interlayer tunneling matrix element $T$ between conduction or valence band edges. At the $\pm K$ valleys, the large crystal momentum leads to a stacking selection rule that dictates $T$ must vanish at two out of the three high symmetry stacking points~\cite{tong2017topological}. In the first harmonic approximation~\cite{rafi2011moire}, $T$ as a function of space coordinate $\mathbf{R}$ takes the form of $\sum_j \exp(i C_3^j \delta \mathbf{K} \cdot \mathbf{R})$~\cite{tong2017topological, wang2017interlayer}, $\delta \mathbf{K}$ being the valley mismatch between layers. The example for the R-type bilayer is shown in Fig.~\ref{fig_selectionrule}b. In the vicinity of B (MX stacking) and C (XM) points, interlayer tunneling takes the $p$-wave forms $T \propto \delta R_x + i \delta R_y$ and $\delta R_x - i \delta R_y$ respectively, where $\delta \mathbf{R}$ denotes the displacement from the high symmetry points. In the vicinity of A (MM) point, $T$ is approximately a constant. Lattice reconstructions can quantitatively change the function $T(\mathbf{R})$, but not these symmetry dictated leading order forms near the three high symmetry points. These determine the envelope function of the minor intralayer component.
For instance, hybrid moir\'e exciton has an intralayer component in $p$-wave envelope at B and C points in R-type bilayer (Fig.~\ref{fig_selectionrule}c).

Table.~\ref{table_selection rule} summaries the symmetry dictated envelope forms of intralayer component in moir\'e hybrid excitons at the high symmetry points in R- (near $0^{\circ}$) and H-type (near $60^{\circ}$) bilayers, which can also be deduced from optical selection rules (see Supplementary Note 1). 
In a hybrid moir\'e exciton, the allowed intralayer component must have an optical dipole with the same polarization as the interlayer component. 
However, the latter has a valley optical selection rule that also depends on the stacking registries (c.f. Table.~\ref{table_selection rule})~\cite{yu2017moire}. 
At high symmetry points where the intra- and inter-layer valley selection rules differ, a $p$-type envelope is  necessitated to reshape the optical dipole polarization of the intralayer component. Such hybrid excitons have a tiny optical dipole and long radiative lifetime (see Supplementary Note 2), just like their interlayer component. Nevertheless, having an intralayer component can activate F\"orster coupling which enables a coherent non-local hopping. 

\begin{table}
\caption{Symmetry dictated properties of hybrid moir\'e excitons trapped at high symmetry points in R- and H-stacking bilayers. 
The quantities listed: direction of electric dipole; optical selection rule of interlayer exciton ($\sigma_+, \sigma_-$ circularly and out-of-plane ($\updownarrow$) linearly polarized light); and envelope form of the minor intralayer component. The hybrid moir\'e exciton considered is the lowest energy configuration involving only A excitons, and for H-stacking this means the hybridization is through the electron hopping only. The major interlayer component is assumed an $s$-wave form, and only the spin singlet exciton with spin down hole is shown. The highlighted columns correspond to moir\'e excitons in R-homobilayer. }%
\label{table_selection rule}%
\renewcommand{\arraystretch}{1.2}
\setlength{\tabcolsep}{2.5mm}{
\begin{tabular}{r|ccc|ccc|ccc|ccc}
\hline \hline 
 & \multicolumn{6}{c|}{R-stacking} & \multicolumn{6}{c}{H-stacking} \\
\hline Electric dipole & \multicolumn{3}{c|}{$\mathbf{z}$} & \multicolumn{3}{c|}{$-\mathbf{z}$} & \multicolumn{3}{c|}{$\mathbf{z}$} & \multicolumn{3}{c}{$-\mathbf{z}$} \\
\hline High symmetry point & MM & MX & \cellcolor{yellow!70}{XM} & MM & \cellcolor{yellow!70}{MX} & XM & MM & MX & XX & MM & MX & XX \\
\hline Optical selection rule & $\sigma_+$ & $\updownarrow$ & \cellcolor{yellow!70}{$\sigma_-$} & $\sigma_+$ & \cellcolor{yellow!70}{$\sigma_-$} & $\updownarrow$ & $\sigma_+$ & $\sigma_-$ & $\updownarrow$ & $\sigma_+$ & $\sigma_-$ & $\updownarrow$ \\
\hline Intralayer envelope & $s$ & $p_-$ & \cellcolor{yellow!70}{$p_+$} & $s$ & \cellcolor{yellow!70}{$p_+$} & $p_-$ & $s$ & $p_+$ & $p_-$  & $s$ & $p_+$ & $p_-$ \\
\hline \hline
\end{tabular}}
\end{table}

\subsection{F\"orster type multipole-multipole coupling of intralayer exciton wavepackets.} 
\begin{figure}
\includegraphics[width=0.48\textwidth]{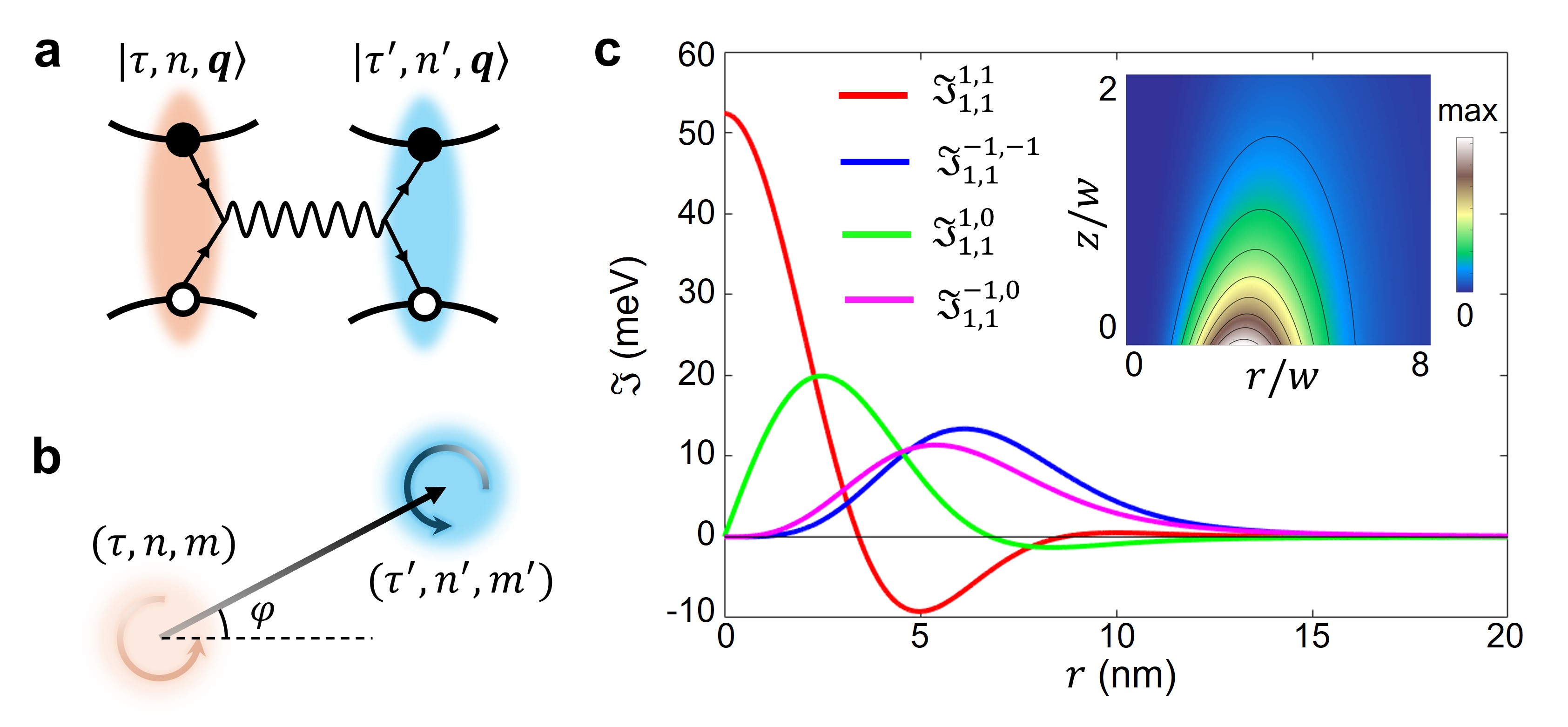}%
\caption{F\"orster coupling of moir\'e exciton wavepackets. $\mathbf{a}$. Schematic of Coulomb exchange that annihilates an electron-hole pair in valley $\tau$ at layer $n$ and creates one in valley $\tau'$ of layer $n'$. 
$\mathbf{b}$. The process in $\mathbf{a}$ effectively realizes non-local hopping - F\"orster coupling - of moir\'e exciton wavepackets, which is further determined by the envelope form (azimuthal quantum number $m$, wavepacket width $w$) of the intralayer component.
$\mathbf{c}$. F\"orster coupling strength of various channels $\mathfrak{J}^{\tau',m'}_{\tau,m}$, 
as function of in-plane distance ($r$) evaluated at out-of-plane distance $z=0$. The subscripts (superscripts) denote the valley index $\tau$ ($\tau'$) and azimuthal index $m$ ($m'$) of the initial (final) states. The inset shows the $\mathfrak{J}^{-1,-1}_{1,1}$ channel as function of the dimensionless in-plane and out-of-plane distances $r/w$, $z/w$. Wavepacket width $w = 2$ nm is used in calculation. }
\label{fig_forster}
\end{figure}

Electron-hole Coulomb exchange can annihilate an intralayer exciton at valley $\tau$ of layer $n$ and create another at $\tau'$ valley of layer $n'$ (Fig.~\ref{fig_forster}a). In the basis of exciton momentum eigenstates $\ket{\tau,n,\mathbf{q}}$, such a F\"orster coupling takes the form: $\hat{J}^{n',n}_{\tau',\tau}(\mathbf{q}) = (-1)^{\frac{\tau'-\tau}{2}}  e^{-i (\tau' \theta_{n'} - \tau \theta_{n})} e^{-i (\tau'-\tau) \varphi_q} J e^{-q z} \frac{q}{K}  $~\cite{li2023cross,selig2019theory}, where $\tau,\tau' = \pm 1$, $\mathbf{q}$ is the exciton center-of-mass momentum with $\varphi_q$ being the azimuthal angle, $z$ is the distance between layer $n$ and $n'$, and $\theta_n$ denotes the twisting angle of layer $n$.
Now we consider this coupling between localized exciton wavepacket states of $C_3$ rotational symmetry (Fig.~\ref{fig_forster}b), which are further classified by the azimuthal quantum number $m$: $\ket{\psi_{\tau,n,m}(\mathbf{r}_c)} = \sum_{\mathbf{q}} e^{-i \mathbf{q} \cdot \mathbf{r}_c} f_m(q) e^{i m \varphi_q} \ket{\tau,n,\mathbf{q}}$, $m = 0$ or $\pm 1$, corresponding to $s$- or $p$-wave envelope. $\mathbf{r}_c$ denotes the wavepacket center, and $f_m(q)$ is a real function accounting for the radial dependence. F\"orster coupling between a pair of exciton wavepacket states of in-plane displacement $\mathbf{r}= (r, \varphi_r) = \mathbf{r}'_c - \mathbf{r}_c $ is of the form (see Supplementary Note 3 for derivation details):
\begin{widetext}
    \begin{eqnarray}
        J^{\tau', n', m' }_{\tau, n, m}(\mathbf{r}) = \braket{\psi_{\tau',n',m'}(\mathbf{r}'_c)|\hat{J}|\psi_{\tau,n,m}(\mathbf{r}_c)} = e^{-i (m'-m) \frac{\pi}{2}}  e^{-i (\tau' \theta_{n'} - \tau \theta_{n})} e^{-i(\tau'-\tau + m'-m)\varphi_r}  \mathfrak{J}^{\tau',m'}_{\tau,m} (r,z,w)
        \label{eq_Forster}
    \end{eqnarray}
\end{widetext}
The angular part is in the form of a phase factor, where $(\tau'-\tau + m'-m)$ are integers between $\pm 4$. The radial part $\mathfrak{J}_{\tau,m}^{\tau',m'} (r)$ is a real function, with dependence on the layer distance $z$ and wavepacket width $w$. For calculations hereafter,  we adopt the wavepacket envelopes to be eigenfunctions of 2D harmonic traps~\cite{baimuratov2020valley}: $f_m(q) \propto w^{|m|+1} q^{|m|} e^{-w^2 q^2/2}$. Figure~\ref{fig_forster}c plots the radial dependence of several representative intra- and inter-valley F\"orster coupling channels. Notably, compared to the intravalley ones, the intervalley channels peak at a larger hopping distance $r \sim 5- 10$ nm, where the overlap between the initial and final state wavepackets is already negligible.
We also identify scaling behaviors (see Supplementary Note 3): (1) $w \mathfrak{J}^{\tau',m'}_{\tau,m}$ is a function of the dimensionless distances $r/w$ and $z/w$ only; (2) For $r > 10 w$, $ \mathfrak{J}^{\tau',m'}_{\tau,m}$ is 
asymptotic to $r^{-(3+|m|+|m'|)}$, suggesting that $p$-envelope excitons are coupled via a quadrupole moment~\cite{baer2008theory}. 
For hybrid excitons, their F\"orster coupling can be obtained by multiplying a factor $\eta \equiv \psi_i \psi_f^*$, where $\psi_i$ ($\psi_f$) is the normalized probability amplitude of intralayer component in the initial (final) state.

\subsection{F\"orster valley-orbit coupling in t-MoTe$_2$.}
\begin{figure}
\includegraphics[width=0.48\textwidth]{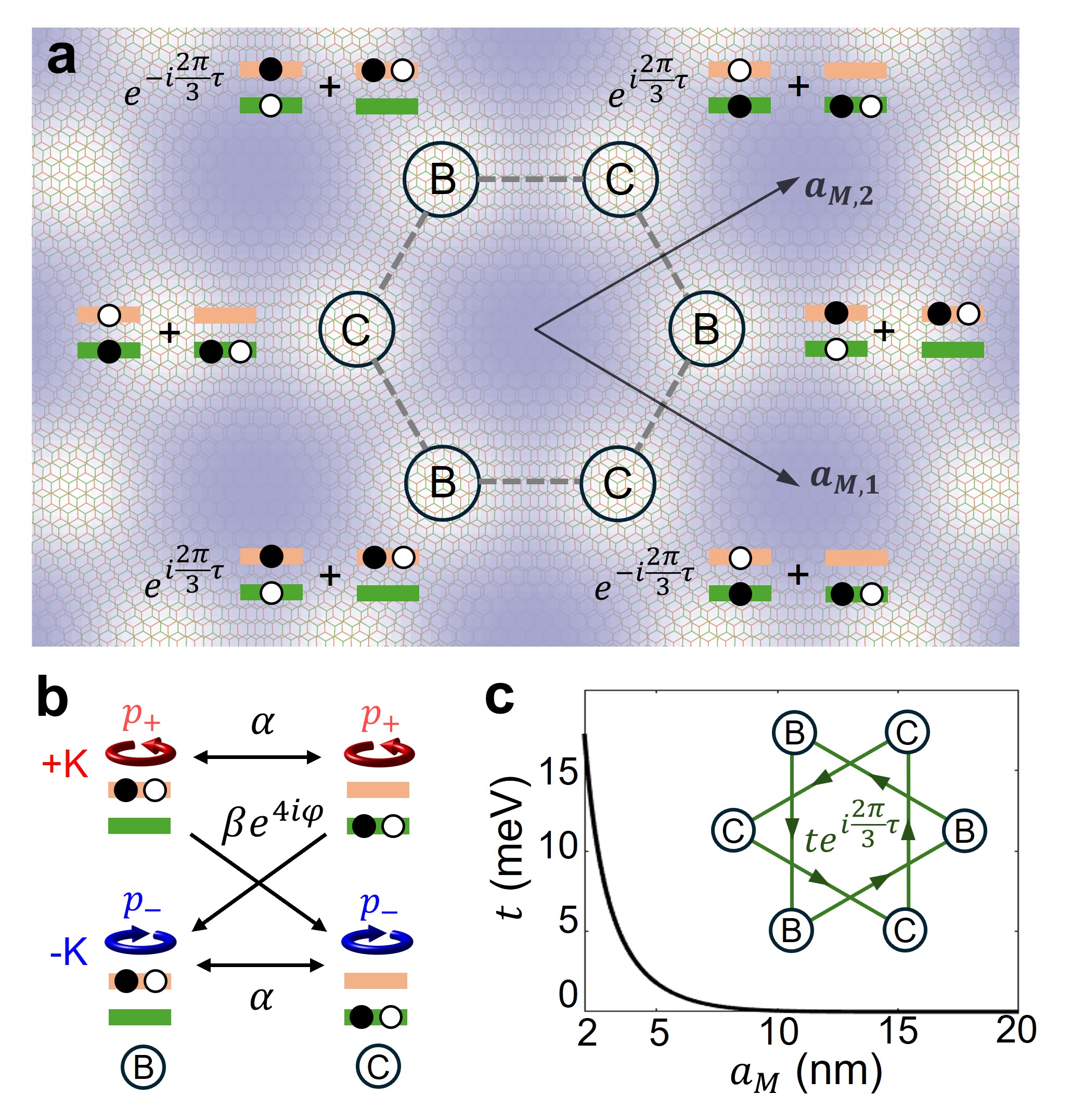}%
\caption{Hybrid moir\'e exciton lattice in the twisted MoTe$_2$ homobilayer. $\mathbf{a}$. Schematics of hybrid moir\'e excitons trapped at the B and C sites in t-MoTe$_2$. Black (white) dots denote the electron (hole), and they are distributed in top (orange) and bottom (green) layer. The symbols of inter- and intra-layer configurations correspond to wavepackets of the form $\ket{\psi^{\text{inter/intra}}_{\tau,m} (\mathbf{r}_c)} = \sum_{\mathbf{q}} e^{-i \mathbf{q} \cdot \mathbf{r}_c} f(q) e^{i m \varphi_q} \ket{X_{\tau,\mathbf{q}}^{\text{inter/intra}}}$, $\mathbf{r}_c$ denoting wavepacket center, and this gauge choice leads to the $\mathbf{r}_c$-dependent phase factor in the linear superposition (see Supplementary Note 2).
$\mathbf{b}$. Nearest-neighbor intravalley and intervalley F\"orster coupling between moir\'e exciton states at B and C site, which have $p_+$ ($p_-$) intralayer envelope at $K (-K)$ valley.
$\mathbf{c}$. Kinetic propagation amplitude as a function of moir\'e period. The inset shows the kinetic propagation as a next-nearest-neighbor complex hopping, $\tau = \pm 1$ is the valley index. 
\label{fig_TBmodel}}
\end{figure}

In the following, we focus on R-stacking t-MoTe$_2$, where moir\'e excitons are trapped at the B (MX) and C (XM) points with opposite layer configurations (Fig.~\ref{fig_TBmodel}a), such that kinetic propagation has no nearest-neighbor terms, leaving the two sublattices uncoupled. 
Pronounced hybrdization of a $p$-wave intralayer component is expected~\cite{yu2021luminescence}, where our analysis finds an azimuthal index locked to the valley $m=\tau$ (see Table~\ref{table_selection rule}). Through such intralayer component, F\"orster coupling not only links the two sublattices, but also introduces a pronounced valley-orbit coupling~(VOC). 
The relevant intra- and inter-valley channels are $\mathfrak{J}^{1,1}_{1,1}$ and $\mathfrak{J}^{-1,-1}_{1,1}$, the latter having a phase factor dependent on hopping direction $e^{4i\varphi}$ (Fig.~\ref{fig_TBmodel}b, c.f. Eq.~(\ref{eq_Forster})). The F\"orster VOC in the honeycomb superlattice can be written as (details in Supplementary Note 5):
\begin{equation}
    H^{voc}_{i' i} = b_{i'}^{\dag} \left[
    \alpha_{i' i} - \beta_{i' i} \left( \tau_x \cos 4\varphi_{i' i}  + \tau_y \sin 4  \varphi_{i' i} \right) \right] b_i,
    \label{eq_ForsterHamiltonian}
\end{equation}
where $b_i \equiv (b_{i, K}, b_{i, - K})^T$ is the exciton annihilation operator at site $i$, $\tau_{x,y,z}$ are the Pauli matrices for the valley pseudospin, and $\varphi_{i' i}$ denotes the hopping angle from site $i$ to $i'$. $\alpha_{i' i} = \eta \mathfrak{J}^{1,1}_{1,1}$ and $\beta_{i' i} = \eta \mathfrak{J}^{-1,-1}_{1,1}$, and $\mathfrak{J}$ are evaluated at the corresponding hopping distance, whose non-monotonic dependence can be used to tailor the relative strengths of $\alpha$ and $\beta$ through tuning the twisting angle. 
$\eta$ is the weight of the intralayer component in the alike initial and final states, and it depends on several key energy scales (c.f. Ref.~\cite{yu2021luminescence}): (i) the amplitude of interlayer carrier hopping; (ii) the depth of moir\'e trapping potential; (iii) the binding energy difference between inter- and intra-layer excitons. The last quantity can be tunable through the dielectric environment, which underlies the possibility to tailor $\eta$ value. In regime of small $\eta$, the effect of F\"orster coupling can be neglected as in earlier discussion of moire excitons in twisted MoTe2~\cite{yu2021luminescence}. Hereafter, we consider a regime of relatively large $\eta$ where F\"orster coupling can qualitatively change the properties of hybrid exciton lattice.

\subsection{Topological lattice of hybrid moir\'e excitons.}
With the $p$-wave intralayer component, moir\'e hybrid excitons can have a long radiative lifetime, favorable for the exploration of correlation and topological matters of these composite bosons~\cite{xie2024long-lived}. Here we show how nontrivial topology can emerge from the F\"orster VOC in the exciton superlattice, which can be tunable by twisting angle, as well as modest interlayer bias. The latter introduces, through the opposite electric dipoles on the two sublattices, a staggered onsite energy $\varepsilon_i = M$ and $-M$ respectively on B and C sites. The exciton tight-binding Hamiltonian is then,
\begin{align}
  H = & \sum_i \varepsilon_i b_i^{\dag}b_i - t \sum_{\braket{\braket{i' i}}} b_{i'}^{\dag} \exp \left ( i \frac{2 \pi}{3} \nu_{i' i} \tau_z  \right ) b_i
   \notag \\
   & + \sum_{\braket{i' i}} b_{i'}^{\dag} \left[
    \alpha - \beta ( \tau_x \cos 4\varphi_{i' i} + \tau_y \sin4  \varphi_{i' i} ) \right]  b_i \label{eq_TBreal} 
\end{align}
The second term is the valley-conserving kinetic propagation in the moir\'e potential landscape, which is absent between sublattices, and we keep its the leading order effect within each sublattice, i.e. a NNN hopping. Its strength $t$ is an exponential function of moir\'e period (Fig.~\ref{fig_TBmodel}c)~\cite{yu2017moire}.
The momentum space mismatch between the valleys of two layers leads to a constant hopping phase, where
$\nu_{i' i} = - \nu_{i i'} = \pm 1$ following the convention shown in Fig.~\ref{fig_TBmodel}c~\cite{yu2017moire} (Supplementary Note 2). The last term is the F\"orster coupling which we retain only the NN term where $\varphi_{i' i}= 0, \pm \frac{2\pi}{3}$. The $4\varphi_{i' i}$ dependence is then equivalent to $\varphi_{i' i}$, such that intervalley F\"orster coupling behaves like a Dressalhaus spin-orbit coupling.

\begin{figure}
\includegraphics[width=0.9\textwidth]{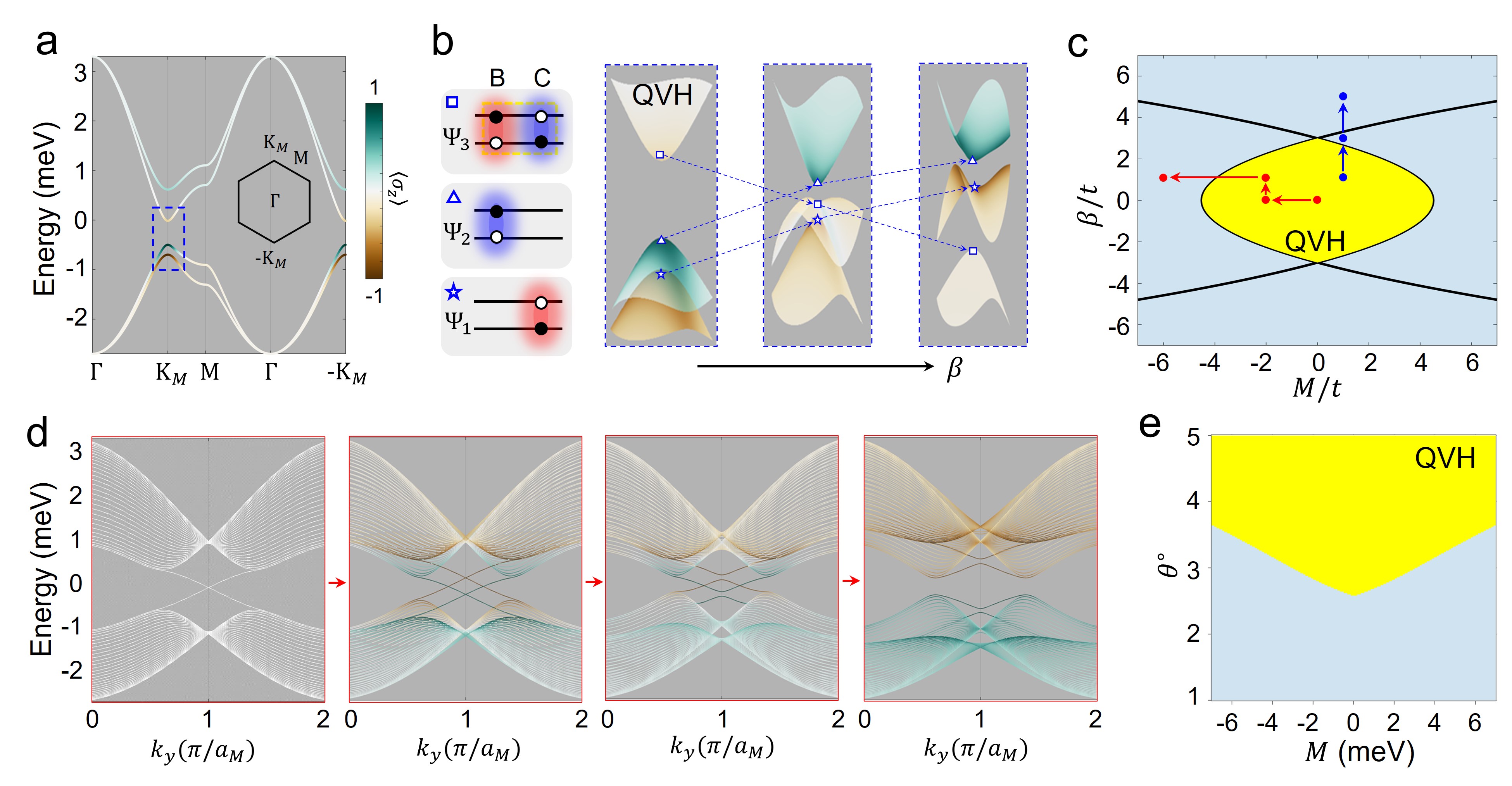}%
\caption{Dispersion and phase diagram of hybrid moir\'e exciton lattice. $\mathbf{a}$. An exemplary dispersion of hybrid moir\'e exciton in t-MoTe$_2$, color coded with the sublattice polarization $\braket{\sigma_z}$.
$t = 0.1$ meV, $\alpha = 1$ meV, $\beta = M = 0.1$ meV.
$\mathbf{b}$. Left: layer and valley configurations of three states at the mini-Brillouin zone corner $K_M$ enclosed by the dashed box in $\mathbf{a}$. Red (blue) denote $K$ ($-K$) valley excitons.
Right: inversions between the three states, upon the increase of intervalley F\"orster coupling $\beta$. 
$\mathbf{c}$. Topological phase diagram as function of $\beta$ and interlayer bias $M$. $t = 0.1$ meV and $\alpha = 1$ meV. The two parabolas mark the points of inversion of $\Psi_3$ with $\Psi_1$ and $\Psi_2$, respectively (c.f. Eq~(\ref{eq_phaseboundary})). The two consecutive band dispersions in $\mathbf{b}$ are along the path marked by blue dots. The yellow region is a quantum valley Hall (QVH) phase.
$\mathbf{d}$. Evolution of the topological edge states, along the path marked by red dots in $\mathbf{c}$. Color stands for the sublattice polarization $\braket{\sigma_z}$.
$\mathbf{e}$. Phase diagram as a function of twisting angle and interlayer bias $M$, for hybrid moir\'e excitons with a 25\% intralayer component.}
\label{fig_phasediagram}
\end{figure}

At $M=0$ and $\beta=0$, this Hamiltonian has two decoupled valley subspaces described by Haldane model of opposite Chern numbers respectively~\cite{haldane1988model}, so the system is in a quantum valley Hall phase (QVH), with a topological gap at the mini-Brillouin zone corners $\pm \text{K}_M$. 
It is known that the staggered potential $M$ tends to open a trivial gap at $\pm \text{K}_M$~\cite{haldane1988model}.
To elucidate the general situation at finite $M$ and $\beta$, we track the evolution of the eigenstates at K$_M$ point under these two control parameters (Fig.~\ref{fig_phasediagram}c). Specifically, the first group of states, $\Psi_{1}$ and $\Psi_{2}$ with eigenenergies $E_{1,2} = -6t \mp M$, are polarized in a single sublattice with valley-sublattice locking.
The other two states, $\Psi_{3}$ and $\Psi_{4}$ with energy $E_{3,4} = 3t \mp \sqrt{9\beta^2 + M^2}$, are spread over both sublattices.
The crossover between the two groups defines phase-space boundaries (c.f. Fig.~\ref{fig_phasediagram}b-c):
\begin{eqnarray}
    \frac{M}{t} = \pm \left[ \frac{9}{2} - \frac{1}{2}\left( \frac{\beta}{t} \right)^2 \right].
    \label{eq_phaseboundary}
\end{eqnarray}
Inside the area enclosed by the two parabolas, $\Psi_{3}$ and $\Psi_{4}$ have higher energy than the valley-sublattice polarized states $\Psi_{1}$ and $\Psi_{2}$, and we find the system in QVH phase with gapless edge states (Fig.~\ref{fig_phasediagram}d, see Supplementary Note 6 for details). Upon the inversion of $\Psi_3$ with $\Psi_1$ or $\Psi_2$ across the boundary defined by Eq.~(\ref{eq_phaseboundary}), the edge states become gapped.  

Lastly, we can map this topological phase diagram to the parameter space spanned by the physical control parameters of twisting angle $\theta$ and interlayer bias $M$. $\theta$ determines the moir\'e period and therefore the NNN kinetic propagation amplitude $t$ (Fig.~\ref{fig_TBmodel}c), as well as 
the intra- and inter-valley F\"orster coupling strength $\alpha$ and $\beta$ between the NN (Fig.~\ref{fig_forster}c). The details of their twisting angle dependence are given in Supplementary Note 7). The amplitudes $\alpha$ and $\beta$ are also controlled by the hybridization weight $\eta$ and envelope width $w$ of the hybrid moir\'e exciton. Taking $\eta = 0.25$ and $w=2$ nm leads to the topological phase diagram shown in Fig.~\ref{fig_phasediagram}e, where the QVH states emerge at twist angle $\theta \geq 2.6^\circ$. 
We note several simplifications that may quantitatively affect the phase diagram. First of all, the hybridization weight $\eta$ can be manipulated by the interlayer bias, and the twist angle, and therefore implies its dependence on $M$ and $\theta$. This dependence may quantitatively affect how the phase diagram in the parameter space of $M/t$ and $\beta/t$ (Fig.~\ref{fig_phasediagram}c) is mapped to the phase diagram in the $M$ and $\theta$ parameter space (Fig.~\ref{fig_phasediagram}e), due to the dependence of F\"orster coupling strength $\beta$ on $M$ and $\theta$. Secondly, our estimation of the F\"orster coupling is based on wavepackets trapped at high symmetry MX and XM sites, which can serve as a good basis since F\"orster coupling is a non-local process not relying on wavefunction overlap. In general, the spread out of the wavefunction shall also depend on the twist angle and moir\'e potential that is affected by the interlayer bias $M$, which we did not take into account. On the other hand, as long as the dependence is not drastic, the qualitative feature of the phase diagram shall remain.


\section*{Data Availability}
Data sharing not applicable to this article as no datasets were generated or analyzed during the current study.

\begin{acknowledgments}
This work is supported by the National Key R\&D Program of China (2020YFA0309600),
the Research Grant Council of Hong Kong (AoE/P-701/20, HKU SRFS2122-7S05, A-HKU705/21), and New Cornerstone Science Foundation.
\end{acknowledgments}

\section*{Author contributions}
W.Y. conceived of the presented idea. H.Z., C.L. and H.Y. performed the analytical calculations. All authors discussed the results and contributed to the final manuscript.

\section*{Competing Interests}
The authors declare no competing interests.

\section*{References}
\bibliography{hybridX_ref}

\begin{thebibliography}{42}%
\makeatletter
\providecommand \@ifxundefined [1]{%
 \@ifx{#1\undefined}
}%
\providecommand \@ifnum [1]{%
 \ifnum #1\expandafter \@firstoftwo
 \else \expandafter \@secondoftwo
 \fi
}%
\providecommand \@ifx [1]{%
 \ifx #1\expandafter \@firstoftwo
 \else \expandafter \@secondoftwo
 \fi
}%
\providecommand \natexlab [1]{#1}%
\providecommand \enquote  [1]{``#1''}%
\providecommand \bibnamefont  [1]{#1}%
\providecommand \bibfnamefont [1]{#1}%
\providecommand \citenamefont [1]{#1}%
\providecommand \href@noop [0]{\@secondoftwo}%
\providecommand \href [0]{\begingroup \@sanitize@url \@href}%
\providecommand \@href[1]{\@@startlink{#1}\@@href}%
\providecommand \@@href[1]{\endgroup#1\@@endlink}%
\providecommand \@sanitize@url [0]{\catcode `\\12\catcode `\$12\catcode `\&12\catcode `\#12\catcode `\^12\catcode `\_12\catcode `\%12\relax}%
\providecommand \@@startlink[1]{}%
\providecommand \@@endlink[0]{}%
\providecommand \url  [0]{\begingroup\@sanitize@url \@url }%
\providecommand \@url [1]{\endgroup\@href {#1}{\urlprefix }}%
\providecommand \urlprefix  [0]{URL }%
\providecommand \Eprint [0]{\href }%
\providecommand \doibase [0]{https://doi.org/}%
\providecommand \selectlanguage [0]{\@gobble}%
\providecommand \bibinfo  [0]{\@secondoftwo}%
\providecommand \bibfield  [0]{\@secondoftwo}%
\providecommand \translation [1]{[#1]}%
\providecommand \BibitemOpen [0]{}%
\providecommand \bibitemStop [0]{}%
\providecommand \bibitemNoStop [0]{.\EOS\space}%
\providecommand \EOS [0]{\spacefactor3000\relax}%
\providecommand \BibitemShut  [1]{\csname bibitem#1\endcsname}%
\let\auto@bib@innerbib\@empty
\bibitem [{\citenamefont {Yu}\ \emph {et~al.}(2015)\citenamefont {Yu}, \citenamefont {Wang}, \citenamefont {Tong}, \citenamefont {Xu},\ and\ \citenamefont {Yao}}]{yu2015anomalous}%
  \BibitemOpen
  \bibfield  {author} {\bibinfo {author} {\bibfnamefont {H.}~\bibnamefont {Yu}}, \bibinfo {author} {\bibfnamefont {Y.}~\bibnamefont {Wang}}, \bibinfo {author} {\bibfnamefont {Q.}~\bibnamefont {Tong}}, \bibinfo {author} {\bibfnamefont {X.}~\bibnamefont {Xu}},\ and\ \bibinfo {author} {\bibfnamefont {W.}~\bibnamefont {Yao}},\ }\bibfield  {title} {\bibinfo {title} {Anomalous light cones and valley optical selection rules of interlayer excitons in twisted heterobilayers},\ }\href {https://doi.org/10.1103/PhysRevLett.115.187002} {\bibfield  {journal} {\bibinfo  {journal} {Phys. Rev. Lett.}\ }\textbf {\bibinfo {volume} {115}},\ \bibinfo {pages} {187002} (\bibinfo {year} {2015})}\BibitemShut {NoStop}%
\bibitem [{\citenamefont {Wu}\ \emph {et~al.}(2017)\citenamefont {Wu}, \citenamefont {Lovorn},\ and\ \citenamefont {MacDonald}}]{wu2017topological}%
  \BibitemOpen
  \bibfield  {author} {\bibinfo {author} {\bibfnamefont {F.}~\bibnamefont {Wu}}, \bibinfo {author} {\bibfnamefont {T.}~\bibnamefont {Lovorn}},\ and\ \bibinfo {author} {\bibfnamefont {A.~H.}\ \bibnamefont {MacDonald}},\ }\bibfield  {title} {\bibinfo {title} {Topological exciton bands in moir\'e heterojunctions},\ }\href {https://doi.org/10.1103/PhysRevLett.118.147401} {\bibfield  {journal} {\bibinfo  {journal} {Phys. Rev. Lett.}\ }\textbf {\bibinfo {volume} {118}},\ \bibinfo {pages} {147401} (\bibinfo {year} {2017})}\BibitemShut {NoStop}%
\bibitem [{\citenamefont {Yu}\ \emph {et~al.}(2017)\citenamefont {Yu}, \citenamefont {Liu}, \citenamefont {Tang}, \citenamefont {Xu},\ and\ \citenamefont {Yao}}]{yu2017moire}%
  \BibitemOpen
  \bibfield  {author} {\bibinfo {author} {\bibfnamefont {H.}~\bibnamefont {Yu}}, \bibinfo {author} {\bibfnamefont {G.-B.}\ \bibnamefont {Liu}}, \bibinfo {author} {\bibfnamefont {J.}~\bibnamefont {Tang}}, \bibinfo {author} {\bibfnamefont {X.}~\bibnamefont {Xu}},\ and\ \bibinfo {author} {\bibfnamefont {W.}~\bibnamefont {Yao}},\ }\bibfield  {title} {\bibinfo {title} {Moir\'e excitons: From programmable quantum emitter arrays to spin-orbit–coupled artificial lattices},\ }\href {https://doi.org/10.1126/sciadv.1701696} {\bibfield  {journal} {\bibinfo  {journal} {Sci. Adv.}\ }\textbf {\bibinfo {volume} {3}},\ \bibinfo {pages} {e1701696} (\bibinfo {year} {2017})}\BibitemShut {NoStop}%
\bibitem [{\citenamefont {Rivera}\ \emph {et~al.}(2018)\citenamefont {Rivera}, \citenamefont {Yu}, \citenamefont {Seyler}, \citenamefont {Wilson}, \citenamefont {Yao},\ and\ \citenamefont {Xu}}]{rivera2018interlayer}%
  \BibitemOpen
  \bibfield  {author} {\bibinfo {author} {\bibfnamefont {P.}~\bibnamefont {Rivera}}, \bibinfo {author} {\bibfnamefont {H.}~\bibnamefont {Yu}}, \bibinfo {author} {\bibfnamefont {K.~L.}\ \bibnamefont {Seyler}}, \bibinfo {author} {\bibfnamefont {N.~P.}\ \bibnamefont {Wilson}}, \bibinfo {author} {\bibfnamefont {W.}~\bibnamefont {Yao}},\ and\ \bibinfo {author} {\bibfnamefont {X.}~\bibnamefont {Xu}},\ }\bibfield  {title} {\bibinfo {title} {Interlayer valley excitons in heterobilayers of transition metal dichalcogenides},\ }\href {https://doi.org/10.1038/s41565-018-0193-0} {\bibfield  {journal} {\bibinfo  {journal} {Nat. Nanotechnol.}\ }\textbf {\bibinfo {volume} {13}},\ \bibinfo {pages} {1004} (\bibinfo {year} {2018})}\BibitemShut {NoStop}%
\bibitem [{\citenamefont {Seyler}\ \emph {et~al.}(2019)\citenamefont {Seyler}, \citenamefont {Rivera}, \citenamefont {Yu}, \citenamefont {Wilson}, \citenamefont {Ray}, \citenamefont {Mandrus}, \citenamefont {Yan}, \citenamefont {Yao},\ and\ \citenamefont {Xu}}]{seyler2019signatures}%
  \BibitemOpen
  \bibfield  {author} {\bibinfo {author} {\bibfnamefont {K.~L.}\ \bibnamefont {Seyler}}, \bibinfo {author} {\bibfnamefont {P.}~\bibnamefont {Rivera}}, \bibinfo {author} {\bibfnamefont {H.}~\bibnamefont {Yu}}, \bibinfo {author} {\bibfnamefont {N.~P.}\ \bibnamefont {Wilson}}, \bibinfo {author} {\bibfnamefont {E.~L.}\ \bibnamefont {Ray}}, \bibinfo {author} {\bibfnamefont {D.~G.}\ \bibnamefont {Mandrus}}, \bibinfo {author} {\bibfnamefont {J.}~\bibnamefont {Yan}}, \bibinfo {author} {\bibfnamefont {W.}~\bibnamefont {Yao}},\ and\ \bibinfo {author} {\bibfnamefont {X.}~\bibnamefont {Xu}},\ }\bibfield  {title} {\bibinfo {title} {Signatures of moir{\'e}-trapped valley excitons in $\text{MoSe}_2/\text{WSe}_2$ heterobilayers},\ }\href {https://doi.org/10.1038/s41586-019-0957-1} {\bibfield  {journal} {\bibinfo  {journal} {Nature}\ }\textbf {\bibinfo {volume} {567}},\ \bibinfo {pages} {66} (\bibinfo {year} {2019})}\BibitemShut {NoStop}%
\bibitem [{\citenamefont {Tran}\ \emph {et~al.}(2019)\citenamefont {Tran}, \citenamefont {Moody}, \citenamefont {Wu}, \citenamefont {Lu}, \citenamefont {Choi}, \citenamefont {Kim}, \citenamefont {Rai}, \citenamefont {Sanchez}, \citenamefont {Quan}, \citenamefont {Singh}, \citenamefont {Embley}, \citenamefont {Zepeda}, \citenamefont {Campbell}, \citenamefont {Autry}, \citenamefont {Taniguchi}, \citenamefont {Watanabe}, \citenamefont {Lu}, \citenamefont {Banerjee}, \citenamefont {Silverman}, \citenamefont {Kim}, \citenamefont {Tutuc}, \citenamefont {Yang}, \citenamefont {MacDonald},\ and\ \citenamefont {Li}}]{tran2019evidence}%
  \BibitemOpen
  \bibfield  {author} {\bibinfo {author} {\bibfnamefont {K.}~\bibnamefont {Tran}}, \bibinfo {author} {\bibfnamefont {G.}~\bibnamefont {Moody}}, \bibinfo {author} {\bibfnamefont {F.}~\bibnamefont {Wu}}, \bibinfo {author} {\bibfnamefont {X.}~\bibnamefont {Lu}}, \bibinfo {author} {\bibfnamefont {J.}~\bibnamefont {Choi}}, \bibinfo {author} {\bibfnamefont {K.}~\bibnamefont {Kim}}, \bibinfo {author} {\bibfnamefont {A.}~\bibnamefont {Rai}}, \bibinfo {author} {\bibfnamefont {D.~A.}\ \bibnamefont {Sanchez}}, \bibinfo {author} {\bibfnamefont {J.}~\bibnamefont {Quan}}, \bibinfo {author} {\bibfnamefont {A.}~\bibnamefont {Singh}}, \bibinfo {author} {\bibfnamefont {J.}~\bibnamefont {Embley}}, \bibinfo {author} {\bibfnamefont {A.}~\bibnamefont {Zepeda}}, \bibinfo {author} {\bibfnamefont {M.}~\bibnamefont {Campbell}}, \bibinfo {author} {\bibfnamefont {T.}~\bibnamefont {Autry}}, \bibinfo {author} {\bibfnamefont {T.}~\bibnamefont {Taniguchi}}, \bibinfo {author} {\bibfnamefont {K.}~\bibnamefont {Watanabe}}, \bibinfo {author}
  {\bibfnamefont {N.}~\bibnamefont {Lu}}, \bibinfo {author} {\bibfnamefont {S.~K.}\ \bibnamefont {Banerjee}}, \bibinfo {author} {\bibfnamefont {K.~L.}\ \bibnamefont {Silverman}}, \bibinfo {author} {\bibfnamefont {S.}~\bibnamefont {Kim}}, \bibinfo {author} {\bibfnamefont {E.}~\bibnamefont {Tutuc}}, \bibinfo {author} {\bibfnamefont {L.}~\bibnamefont {Yang}}, \bibinfo {author} {\bibfnamefont {A.~H.}\ \bibnamefont {MacDonald}},\ and\ \bibinfo {author} {\bibfnamefont {X.}~\bibnamefont {Li}},\ }\bibfield  {title} {\bibinfo {title} {Evidence for moir{\'e} excitons in van der waals heterostructures},\ }\href {https://doi.org/10.1038/s41586-019-0975-z} {\bibfield  {journal} {\bibinfo  {journal} {Nature}\ }\textbf {\bibinfo {volume} {567}},\ \bibinfo {pages} {71} (\bibinfo {year} {2019})}\BibitemShut {NoStop}%
\bibitem [{\citenamefont {Jin}\ \emph {et~al.}(2019{\natexlab{a}})\citenamefont {Jin}, \citenamefont {Regan}, \citenamefont {Yan}, \citenamefont {Iqbal Bakti~Utama}, \citenamefont {Wang}, \citenamefont {Zhao}, \citenamefont {Qin}, \citenamefont {Yang}, \citenamefont {Zheng}, \citenamefont {Shi}, \citenamefont {Watanabe}, \citenamefont {Taniguchi}, \citenamefont {Tongay}, \citenamefont {Zettl},\ and\ \citenamefont {Wang}}]{jin2019observation}%
  \BibitemOpen
  \bibfield  {author} {\bibinfo {author} {\bibfnamefont {C.}~\bibnamefont {Jin}}, \bibinfo {author} {\bibfnamefont {E.~C.}\ \bibnamefont {Regan}}, \bibinfo {author} {\bibfnamefont {A.}~\bibnamefont {Yan}}, \bibinfo {author} {\bibfnamefont {M.}~\bibnamefont {Iqbal Bakti~Utama}}, \bibinfo {author} {\bibfnamefont {D.}~\bibnamefont {Wang}}, \bibinfo {author} {\bibfnamefont {S.}~\bibnamefont {Zhao}}, \bibinfo {author} {\bibfnamefont {Y.}~\bibnamefont {Qin}}, \bibinfo {author} {\bibfnamefont {S.}~\bibnamefont {Yang}}, \bibinfo {author} {\bibfnamefont {Z.}~\bibnamefont {Zheng}}, \bibinfo {author} {\bibfnamefont {S.}~\bibnamefont {Shi}}, \bibinfo {author} {\bibfnamefont {K.}~\bibnamefont {Watanabe}}, \bibinfo {author} {\bibfnamefont {T.}~\bibnamefont {Taniguchi}}, \bibinfo {author} {\bibfnamefont {S.}~\bibnamefont {Tongay}}, \bibinfo {author} {\bibfnamefont {A.}~\bibnamefont {Zettl}},\ and\ \bibinfo {author} {\bibfnamefont {F.}~\bibnamefont {Wang}},\ }\bibfield  {title} {\bibinfo {title} {Observation of moir{\'e}
  excitons in $\text{WSe}_2/\text{WS}_2$ heterostructure superlattices},\ }\href {https://doi.org/10.1038/s41586-019-0976-y} {\bibfield  {journal} {\bibinfo  {journal} {Nature}\ }\textbf {\bibinfo {volume} {567}},\ \bibinfo {pages} {76} (\bibinfo {year} {2019}{\natexlab{a}})}\BibitemShut {NoStop}%
\bibitem [{\citenamefont {Alexeev}\ \emph {et~al.}(2019)\citenamefont {Alexeev}, \citenamefont {Ruiz-Tijerina}, \citenamefont {Danovich}, \citenamefont {Hamer}, \citenamefont {Terry}, \citenamefont {Nayak}, \citenamefont {Ahn}, \citenamefont {Pak}, \citenamefont {Lee}, \citenamefont {Sohn}, \citenamefont {Molas}, \citenamefont {Koperski}, \citenamefont {Watanabe}, \citenamefont {Taniguchi}, \citenamefont {Novoselov}, \citenamefont {Gorbachev}, \citenamefont {Shin}, \citenamefont {Fal'ko},\ and\ \citenamefont {Tartakovskii}}]{alexeev2019resonantly}%
  \BibitemOpen
  \bibfield  {author} {\bibinfo {author} {\bibfnamefont {E.~M.}\ \bibnamefont {Alexeev}}, \bibinfo {author} {\bibfnamefont {D.~A.}\ \bibnamefont {Ruiz-Tijerina}}, \bibinfo {author} {\bibfnamefont {M.}~\bibnamefont {Danovich}}, \bibinfo {author} {\bibfnamefont {M.~J.}\ \bibnamefont {Hamer}}, \bibinfo {author} {\bibfnamefont {D.~J.}\ \bibnamefont {Terry}}, \bibinfo {author} {\bibfnamefont {P.~K.}\ \bibnamefont {Nayak}}, \bibinfo {author} {\bibfnamefont {S.}~\bibnamefont {Ahn}}, \bibinfo {author} {\bibfnamefont {S.}~\bibnamefont {Pak}}, \bibinfo {author} {\bibfnamefont {J.}~\bibnamefont {Lee}}, \bibinfo {author} {\bibfnamefont {J.~I.}\ \bibnamefont {Sohn}}, \bibinfo {author} {\bibfnamefont {M.~R.}\ \bibnamefont {Molas}}, \bibinfo {author} {\bibfnamefont {M.}~\bibnamefont {Koperski}}, \bibinfo {author} {\bibfnamefont {K.}~\bibnamefont {Watanabe}}, \bibinfo {author} {\bibfnamefont {T.}~\bibnamefont {Taniguchi}}, \bibinfo {author} {\bibfnamefont {K.~S.}\ \bibnamefont {Novoselov}}, \bibinfo {author} {\bibfnamefont
  {R.~V.}\ \bibnamefont {Gorbachev}}, \bibinfo {author} {\bibfnamefont {H.~S.}\ \bibnamefont {Shin}}, \bibinfo {author} {\bibfnamefont {V.~I.}\ \bibnamefont {Fal'ko}},\ and\ \bibinfo {author} {\bibfnamefont {A.~I.}\ \bibnamefont {Tartakovskii}},\ }\bibfield  {title} {\bibinfo {title} {Resonantly hybridized excitons in moir{\'e} superlattices in van der waals heterostructures},\ }\href {https://doi.org/10.1038/s41586-019-0986-9} {\bibfield  {journal} {\bibinfo  {journal} {Nature}\ }\textbf {\bibinfo {volume} {567}},\ \bibinfo {pages} {81} (\bibinfo {year} {2019})}\BibitemShut {NoStop}%
\bibitem [{\citenamefont {Huang}\ \emph {et~al.}(2022)\citenamefont {Huang}, \citenamefont {Choi}, \citenamefont {Shih},\ and\ \citenamefont {Li}}]{huang2022excitons}%
  \BibitemOpen
  \bibfield  {author} {\bibinfo {author} {\bibfnamefont {D.}~\bibnamefont {Huang}}, \bibinfo {author} {\bibfnamefont {J.}~\bibnamefont {Choi}}, \bibinfo {author} {\bibfnamefont {C.-K.}\ \bibnamefont {Shih}},\ and\ \bibinfo {author} {\bibfnamefont {X.}~\bibnamefont {Li}},\ }\bibfield  {title} {\bibinfo {title} {Excitons in semiconductor moir{\'e} superlattices},\ }\href {https://doi.org/10.1038/s41565-021-01068-y} {\bibfield  {journal} {\bibinfo  {journal} {Nat. Nanotechnol.}\ }\textbf {\bibinfo {volume} {17}},\ \bibinfo {pages} {227} (\bibinfo {year} {2022})}\BibitemShut {NoStop}%
\bibitem [{\citenamefont {Rivera}\ \emph {et~al.}(2016)\citenamefont {Rivera}, \citenamefont {Seyler}, \citenamefont {Yu}, \citenamefont {Schaibley}, \citenamefont {Yan}, \citenamefont {Mandrus}, \citenamefont {Yao},\ and\ \citenamefont {Xu}}]{rivera2016valley}%
  \BibitemOpen
  \bibfield  {author} {\bibinfo {author} {\bibfnamefont {P.}~\bibnamefont {Rivera}}, \bibinfo {author} {\bibfnamefont {K.~L.}\ \bibnamefont {Seyler}}, \bibinfo {author} {\bibfnamefont {H.}~\bibnamefont {Yu}}, \bibinfo {author} {\bibfnamefont {J.~R.}\ \bibnamefont {Schaibley}}, \bibinfo {author} {\bibfnamefont {J.}~\bibnamefont {Yan}}, \bibinfo {author} {\bibfnamefont {D.~G.}\ \bibnamefont {Mandrus}}, \bibinfo {author} {\bibfnamefont {W.}~\bibnamefont {Yao}},\ and\ \bibinfo {author} {\bibfnamefont {X.}~\bibnamefont {Xu}},\ }\bibfield  {title} {\bibinfo {title} {Valley-polarized exciton dynamics in a $\text{2D}$ semiconductor heterostructure},\ }\href {https://doi.org/10.1126/science.aac7820} {\bibfield  {journal} {\bibinfo  {journal} {Science}\ }\textbf {\bibinfo {volume} {351}},\ \bibinfo {pages} {688} (\bibinfo {year} {2016})}\BibitemShut {NoStop}%
\bibitem [{\citenamefont {Zhang}\ \emph {et~al.}(2017)\citenamefont {Zhang}, \citenamefont {Chuu}, \citenamefont {Ren}, \citenamefont {Li}, \citenamefont {Li}, \citenamefont {Jin}, \citenamefont {Chou},\ and\ \citenamefont {Shih}}]{zhang2017interlayer}%
  \BibitemOpen
  \bibfield  {author} {\bibinfo {author} {\bibfnamefont {C.}~\bibnamefont {Zhang}}, \bibinfo {author} {\bibfnamefont {C.-P.}\ \bibnamefont {Chuu}}, \bibinfo {author} {\bibfnamefont {X.}~\bibnamefont {Ren}}, \bibinfo {author} {\bibfnamefont {M.-Y.}\ \bibnamefont {Li}}, \bibinfo {author} {\bibfnamefont {L.-J.}\ \bibnamefont {Li}}, \bibinfo {author} {\bibfnamefont {C.}~\bibnamefont {Jin}}, \bibinfo {author} {\bibfnamefont {M.-Y.}\ \bibnamefont {Chou}},\ and\ \bibinfo {author} {\bibfnamefont {C.-K.}\ \bibnamefont {Shih}},\ }\bibfield  {title} {\bibinfo {title} {Interlayer couplings, moir\'e patterns, and $\text{2D}$ electronic superlattices in $\text{MoS}_2/\text{WSe}_2$ hetero-bilayers},\ }\href {https://doi.org/10.1126/sciadv.1601459} {\bibfield  {journal} {\bibinfo  {journal} {Sci. Adv.}\ }\textbf {\bibinfo {volume} {3}},\ \bibinfo {pages} {e1601459} (\bibinfo {year} {2017})}\BibitemShut {NoStop}%
\bibitem [{\citenamefont {Brotons-Gisbert}\ \emph {et~al.}(2020)\citenamefont {Brotons-Gisbert}, \citenamefont {Baek}, \citenamefont {Molina-S{\'a}nchez}, \citenamefont {Campbell}, \citenamefont {Scerri}, \citenamefont {White}, \citenamefont {Watanabe}, \citenamefont {Taniguchi}, \citenamefont {Bonato},\ and\ \citenamefont {Gerardot}}]{brotons2020spin}%
  \BibitemOpen
  \bibfield  {author} {\bibinfo {author} {\bibfnamefont {M.}~\bibnamefont {Brotons-Gisbert}}, \bibinfo {author} {\bibfnamefont {H.}~\bibnamefont {Baek}}, \bibinfo {author} {\bibfnamefont {A.}~\bibnamefont {Molina-S{\'a}nchez}}, \bibinfo {author} {\bibfnamefont {A.}~\bibnamefont {Campbell}}, \bibinfo {author} {\bibfnamefont {E.}~\bibnamefont {Scerri}}, \bibinfo {author} {\bibfnamefont {D.}~\bibnamefont {White}}, \bibinfo {author} {\bibfnamefont {K.}~\bibnamefont {Watanabe}}, \bibinfo {author} {\bibfnamefont {T.}~\bibnamefont {Taniguchi}}, \bibinfo {author} {\bibfnamefont {C.}~\bibnamefont {Bonato}},\ and\ \bibinfo {author} {\bibfnamefont {B.~D.}\ \bibnamefont {Gerardot}},\ }\bibfield  {title} {\bibinfo {title} {Spin--layer locking of interlayer excitons trapped in moir{\'e} potentials},\ }\href {https://doi.org/10.1038/s41563-020-0687-7} {\bibfield  {journal} {\bibinfo  {journal} {Nat. Mater.}\ }\textbf {\bibinfo {volume} {19}},\ \bibinfo {pages} {630} (\bibinfo {year} {2020})}\BibitemShut {NoStop}%
\bibitem [{\citenamefont {Baek}\ \emph {et~al.}(2020)\citenamefont {Baek}, \citenamefont {Brotons-Gisbert}, \citenamefont {Koong}, \citenamefont {Campbell}, \citenamefont {Rambach}, \citenamefont {Watanabe}, \citenamefont {Taniguchi},\ and\ \citenamefont {Gerardot}}]{baek2020highly}%
  \BibitemOpen
  \bibfield  {author} {\bibinfo {author} {\bibfnamefont {H.}~\bibnamefont {Baek}}, \bibinfo {author} {\bibfnamefont {M.}~\bibnamefont {Brotons-Gisbert}}, \bibinfo {author} {\bibfnamefont {Z.~X.}\ \bibnamefont {Koong}}, \bibinfo {author} {\bibfnamefont {A.}~\bibnamefont {Campbell}}, \bibinfo {author} {\bibfnamefont {M.}~\bibnamefont {Rambach}}, \bibinfo {author} {\bibfnamefont {K.}~\bibnamefont {Watanabe}}, \bibinfo {author} {\bibfnamefont {T.}~\bibnamefont {Taniguchi}},\ and\ \bibinfo {author} {\bibfnamefont {B.~D.}\ \bibnamefont {Gerardot}},\ }\bibfield  {title} {\bibinfo {title} {Highly energy-tunable quantum light from moiré-trapped excitons},\ }\href {https://doi.org/10.1126/sciadv.aba8526} {\bibfield  {journal} {\bibinfo  {journal} {Sci. Adv.}\ }\textbf {\bibinfo {volume} {6}},\ \bibinfo {pages} {eaba8526} (\bibinfo {year} {2020})}\BibitemShut {NoStop}%
\bibitem [{\citenamefont {Yu}\ \emph {et~al.}(2018)\citenamefont {Yu}, \citenamefont {Liu},\ and\ \citenamefont {Yao}}]{yu2018brightened}%
  \BibitemOpen
  \bibfield  {author} {\bibinfo {author} {\bibfnamefont {H.}~\bibnamefont {Yu}}, \bibinfo {author} {\bibfnamefont {G.-B.}\ \bibnamefont {Liu}},\ and\ \bibinfo {author} {\bibfnamefont {W.}~\bibnamefont {Yao}},\ }\bibfield  {title} {\bibinfo {title} {Brightened spin-triplet interlayer excitons and optical selection rules in van der waals heterobilayers},\ }\href {https://doi.org/10.1088/2053-1583/aac065} {\bibfield  {journal} {\bibinfo  {journal} {2D Mater.}\ }\textbf {\bibinfo {volume} {5}},\ \bibinfo {pages} {035021} (\bibinfo {year} {2018})}\BibitemShut {NoStop}%
\bibitem [{\citenamefont {Jin}\ \emph {et~al.}(2019{\natexlab{b}})\citenamefont {Jin}, \citenamefont {Regan}, \citenamefont {Wang}, \citenamefont {Iqbal Bakti~Utama}, \citenamefont {Yang}, \citenamefont {Cain}, \citenamefont {Qin}, \citenamefont {Shen}, \citenamefont {Zheng}, \citenamefont {Watanabe}, \citenamefont {Taniguchi}, \citenamefont {Tongay}, \citenamefont {Zettl},\ and\ \citenamefont {Wang}}]{jin2019identification}%
  \BibitemOpen
  \bibfield  {author} {\bibinfo {author} {\bibfnamefont {C.}~\bibnamefont {Jin}}, \bibinfo {author} {\bibfnamefont {E.~C.}\ \bibnamefont {Regan}}, \bibinfo {author} {\bibfnamefont {D.}~\bibnamefont {Wang}}, \bibinfo {author} {\bibfnamefont {M.}~\bibnamefont {Iqbal Bakti~Utama}}, \bibinfo {author} {\bibfnamefont {C.-S.}\ \bibnamefont {Yang}}, \bibinfo {author} {\bibfnamefont {J.}~\bibnamefont {Cain}}, \bibinfo {author} {\bibfnamefont {Y.}~\bibnamefont {Qin}}, \bibinfo {author} {\bibfnamefont {Y.}~\bibnamefont {Shen}}, \bibinfo {author} {\bibfnamefont {Z.}~\bibnamefont {Zheng}}, \bibinfo {author} {\bibfnamefont {K.}~\bibnamefont {Watanabe}}, \bibinfo {author} {\bibfnamefont {T.}~\bibnamefont {Taniguchi}}, \bibinfo {author} {\bibfnamefont {S.}~\bibnamefont {Tongay}}, \bibinfo {author} {\bibfnamefont {A.}~\bibnamefont {Zettl}},\ and\ \bibinfo {author} {\bibfnamefont {F.}~\bibnamefont {Wang}},\ }\bibfield  {title} {\bibinfo {title} {Identification of spin, valley and moir{\'e} quasi-angular momentum of interlayer
  excitons},\ }\href {https://doi.org/10.1038/s41567-019-0631-4} {\bibfield  {journal} {\bibinfo  {journal} {Nat. Phys.}\ }\textbf {\bibinfo {volume} {15}},\ \bibinfo {pages} {1140} (\bibinfo {year} {2019}{\natexlab{b}})}\BibitemShut {NoStop}%
\bibitem [{\citenamefont {Li}\ \emph {et~al.}(2020)\citenamefont {Li}, \citenamefont {Lu}, \citenamefont {Dubey}, \citenamefont {Devenica},\ and\ \citenamefont {Srivastava}}]{li2020dipolar}%
  \BibitemOpen
  \bibfield  {author} {\bibinfo {author} {\bibfnamefont {W.}~\bibnamefont {Li}}, \bibinfo {author} {\bibfnamefont {X.}~\bibnamefont {Lu}}, \bibinfo {author} {\bibfnamefont {S.}~\bibnamefont {Dubey}}, \bibinfo {author} {\bibfnamefont {L.}~\bibnamefont {Devenica}},\ and\ \bibinfo {author} {\bibfnamefont {A.}~\bibnamefont {Srivastava}},\ }\bibfield  {title} {\bibinfo {title} {Dipolar interactions between localized interlayer excitons in van der waals heterostructures},\ }\href {https://doi.org/10.1038/s41563-020-0661-4} {\bibfield  {journal} {\bibinfo  {journal} {Nat. Mater.}\ }\textbf {\bibinfo {volume} {19}},\ \bibinfo {pages} {624} (\bibinfo {year} {2020})}\BibitemShut {NoStop}%
\bibitem [{\citenamefont {Xiong}\ \emph {et~al.}(2023)\citenamefont {Xiong}, \citenamefont {Nie}, \citenamefont {Brantly}, \citenamefont {Hays}, \citenamefont {Sailus}, \citenamefont {Watanabe}, \citenamefont {Taniguchi}, \citenamefont {Tongay},\ and\ \citenamefont {Jin}}]{xiong2023correlated}%
  \BibitemOpen
  \bibfield  {author} {\bibinfo {author} {\bibfnamefont {R.}~\bibnamefont {Xiong}}, \bibinfo {author} {\bibfnamefont {J.~H.}\ \bibnamefont {Nie}}, \bibinfo {author} {\bibfnamefont {S.~L.}\ \bibnamefont {Brantly}}, \bibinfo {author} {\bibfnamefont {P.}~\bibnamefont {Hays}}, \bibinfo {author} {\bibfnamefont {R.}~\bibnamefont {Sailus}}, \bibinfo {author} {\bibfnamefont {K.}~\bibnamefont {Watanabe}}, \bibinfo {author} {\bibfnamefont {T.}~\bibnamefont {Taniguchi}}, \bibinfo {author} {\bibfnamefont {S.}~\bibnamefont {Tongay}},\ and\ \bibinfo {author} {\bibfnamefont {C.}~\bibnamefont {Jin}},\ }\bibfield  {title} {\bibinfo {title} {Correlated insulator of excitons in $\text{WSe}_2/\text{WS}_2$ moir\'e superlattices},\ }\href {https://doi.org/10.1126/science.add5574} {\bibfield  {journal} {\bibinfo  {journal} {Science}\ }\textbf {\bibinfo {volume} {380}},\ \bibinfo {pages} {860} (\bibinfo {year} {2023})}\BibitemShut {NoStop}%
\bibitem [{\citenamefont {Park}\ \emph {et~al.}(2023{\natexlab{a}})\citenamefont {Park}, \citenamefont {Zhu}, \citenamefont {Wang}, \citenamefont {Wang}, \citenamefont {Holtzmann}, \citenamefont {Taniguchi}, \citenamefont {Watanabe}, \citenamefont {Yan}, \citenamefont {Fu}, \citenamefont {Cao}, \citenamefont {Xiao}, \citenamefont {Gamelin}, \citenamefont {Yu}, \citenamefont {Yao},\ and\ \citenamefont {Xu}}]{park2023dipole}%
  \BibitemOpen
  \bibfield  {author} {\bibinfo {author} {\bibfnamefont {H.}~\bibnamefont {Park}}, \bibinfo {author} {\bibfnamefont {J.}~\bibnamefont {Zhu}}, \bibinfo {author} {\bibfnamefont {X.}~\bibnamefont {Wang}}, \bibinfo {author} {\bibfnamefont {Y.}~\bibnamefont {Wang}}, \bibinfo {author} {\bibfnamefont {W.}~\bibnamefont {Holtzmann}}, \bibinfo {author} {\bibfnamefont {T.}~\bibnamefont {Taniguchi}}, \bibinfo {author} {\bibfnamefont {K.}~\bibnamefont {Watanabe}}, \bibinfo {author} {\bibfnamefont {J.}~\bibnamefont {Yan}}, \bibinfo {author} {\bibfnamefont {L.}~\bibnamefont {Fu}}, \bibinfo {author} {\bibfnamefont {T.}~\bibnamefont {Cao}}, \bibinfo {author} {\bibfnamefont {D.}~\bibnamefont {Xiao}}, \bibinfo {author} {\bibfnamefont {D.~R.}\ \bibnamefont {Gamelin}}, \bibinfo {author} {\bibfnamefont {H.}~\bibnamefont {Yu}}, \bibinfo {author} {\bibfnamefont {W.}~\bibnamefont {Yao}},\ and\ \bibinfo {author} {\bibfnamefont {X.}~\bibnamefont {Xu}},\ }\bibfield  {title} {\bibinfo {title} {Dipole ladders with large hubbard
  interaction in a moir{\'e} exciton lattice},\ }\href {https://doi.org/10.1038/s41567-023-02077-5} {\bibfield  {journal} {\bibinfo  {journal} {Nat. Phys.}\ }\textbf {\bibinfo {volume} {19}},\ \bibinfo {pages} {1286} (\bibinfo {year} {2023}{\natexlab{a}})}\BibitemShut {NoStop}%
\bibitem [{\citenamefont {Hsu}\ \emph {et~al.}(2019)\citenamefont {Hsu}, \citenamefont {Lin}, \citenamefont {Lu}, \citenamefont {Lee}, \citenamefont {Chu}, \citenamefont {Li}, \citenamefont {Yao}, \citenamefont {Chang},\ and\ \citenamefont {Shih}}]{hsu2019tailoring}%
  \BibitemOpen
  \bibfield  {author} {\bibinfo {author} {\bibfnamefont {W.-T.}\ \bibnamefont {Hsu}}, \bibinfo {author} {\bibfnamefont {B.-H.}\ \bibnamefont {Lin}}, \bibinfo {author} {\bibfnamefont {L.-S.}\ \bibnamefont {Lu}}, \bibinfo {author} {\bibfnamefont {M.-H.}\ \bibnamefont {Lee}}, \bibinfo {author} {\bibfnamefont {M.-W.}\ \bibnamefont {Chu}}, \bibinfo {author} {\bibfnamefont {L.-J.}\ \bibnamefont {Li}}, \bibinfo {author} {\bibfnamefont {W.}~\bibnamefont {Yao}}, \bibinfo {author} {\bibfnamefont {W.-H.}\ \bibnamefont {Chang}},\ and\ \bibinfo {author} {\bibfnamefont {C.-K.}\ \bibnamefont {Shih}},\ }\bibfield  {title} {\bibinfo {title} {Tailoring excitonic states of van der waals bilayers through stacking configuration, band alignment, and valley spin},\ }\href {https://doi.org/10.1126/sciadv.aax7407} {\bibfield  {journal} {\bibinfo  {journal} {Sci. Adv.}\ }\textbf {\bibinfo {volume} {5}},\ \bibinfo {pages} {eaax7407} (\bibinfo {year} {2019})}\BibitemShut {NoStop}%
\bibitem [{\citenamefont {Zhang}\ \emph {et~al.}(2020)\citenamefont {Zhang}, \citenamefont {Zhang}, \citenamefont {Wu}, \citenamefont {Wang}, \citenamefont {Gogna}, \citenamefont {Hou}, \citenamefont {Watanabe}, \citenamefont {Taniguchi}, \citenamefont {Kulkarni}, \citenamefont {Kuo}, \citenamefont {Forrest},\ and\ \citenamefont {Deng}}]{zhang2020twist}%
  \BibitemOpen
  \bibfield  {author} {\bibinfo {author} {\bibfnamefont {L.}~\bibnamefont {Zhang}}, \bibinfo {author} {\bibfnamefont {Z.}~\bibnamefont {Zhang}}, \bibinfo {author} {\bibfnamefont {F.}~\bibnamefont {Wu}}, \bibinfo {author} {\bibfnamefont {D.}~\bibnamefont {Wang}}, \bibinfo {author} {\bibfnamefont {R.}~\bibnamefont {Gogna}}, \bibinfo {author} {\bibfnamefont {S.}~\bibnamefont {Hou}}, \bibinfo {author} {\bibfnamefont {K.}~\bibnamefont {Watanabe}}, \bibinfo {author} {\bibfnamefont {T.}~\bibnamefont {Taniguchi}}, \bibinfo {author} {\bibfnamefont {K.}~\bibnamefont {Kulkarni}}, \bibinfo {author} {\bibfnamefont {T.}~\bibnamefont {Kuo}}, \bibinfo {author} {\bibfnamefont {S.~R.}\ \bibnamefont {Forrest}},\ and\ \bibinfo {author} {\bibfnamefont {H.}~\bibnamefont {Deng}},\ }\bibfield  {title} {\bibinfo {title} {Twist-angle dependence of moir{\'e} excitons in $\text{WS}_2/\text{MoSe}_2$ heterobilayers},\ }\href {https://doi.org/10.1038/s41467-020-19466-6} {\bibfield  {journal} {\bibinfo  {journal} {Nat. Commun.}\ }\textbf
  {\bibinfo {volume} {11}},\ \bibinfo {pages} {5888} (\bibinfo {year} {2020})}\BibitemShut {NoStop}%
\bibitem [{\citenamefont {Zhao}\ \emph {et~al.}(2024)\citenamefont {Zhao}, \citenamefont {Huang}, \citenamefont {Gillen}, \citenamefont {Li}, \citenamefont {Liu}, \citenamefont {Watanabe}, \citenamefont {Taniguchi}, \citenamefont {Maultzsch}, \citenamefont {Hone}, \citenamefont {H{\"o}gele},\ and\ \citenamefont {Baimuratov}}]{zhao2024hybrid}%
  \BibitemOpen
  \bibfield  {author} {\bibinfo {author} {\bibfnamefont {S.}~\bibnamefont {Zhao}}, \bibinfo {author} {\bibfnamefont {X.}~\bibnamefont {Huang}}, \bibinfo {author} {\bibfnamefont {R.}~\bibnamefont {Gillen}}, \bibinfo {author} {\bibfnamefont {Z.}~\bibnamefont {Li}}, \bibinfo {author} {\bibfnamefont {S.}~\bibnamefont {Liu}}, \bibinfo {author} {\bibfnamefont {K.}~\bibnamefont {Watanabe}}, \bibinfo {author} {\bibfnamefont {T.}~\bibnamefont {Taniguchi}}, \bibinfo {author} {\bibfnamefont {J.}~\bibnamefont {Maultzsch}}, \bibinfo {author} {\bibfnamefont {J.}~\bibnamefont {Hone}}, \bibinfo {author} {\bibfnamefont {A.}~\bibnamefont {H{\"o}gele}},\ and\ \bibinfo {author} {\bibfnamefont {A.~S.}\ \bibnamefont {Baimuratov}},\ }\bibfield  {title} {\bibinfo {title} {Hybrid moir{\'e} excitons and trions in twisted $\text{MoTe}_2$--$\text{MoSe}_2$ heterobilayers},\ }\href {https://doi.org/10.1021/acs.nanolett.4c00541} {\bibfield  {journal} {\bibinfo  {journal} {Nano Lett.}\ }\textbf {\bibinfo {volume} {24}},\ \bibinfo {pages}
  {4917} (\bibinfo {year} {2024})}\BibitemShut {NoStop}%
\bibitem [{\citenamefont {Yu}\ \emph {et~al.}(2014)\citenamefont {Yu}, \citenamefont {Liu}, \citenamefont {Gong}, \citenamefont {Xu},\ and\ \citenamefont {Yao}}]{yu2014dirac}%
  \BibitemOpen
  \bibfield  {author} {\bibinfo {author} {\bibfnamefont {H.}~\bibnamefont {Yu}}, \bibinfo {author} {\bibfnamefont {G.-B.}\ \bibnamefont {Liu}}, \bibinfo {author} {\bibfnamefont {P.}~\bibnamefont {Gong}}, \bibinfo {author} {\bibfnamefont {X.}~\bibnamefont {Xu}},\ and\ \bibinfo {author} {\bibfnamefont {W.}~\bibnamefont {Yao}},\ }\bibfield  {title} {\bibinfo {title} {Dirac cones and dirac saddle points of bright excitons in monolayer transition metal dichalcogenides},\ }\href {https://doi.org/10.1038/ncomms4876} {\bibfield  {journal} {\bibinfo  {journal} {Nat. Commun.}\ }\textbf {\bibinfo {volume} {5}},\ \bibinfo {pages} {3876} (\bibinfo {year} {2014})}\BibitemShut {NoStop}%
\bibitem [{\citenamefont {Qiu}\ \emph {et~al.}(2015)\citenamefont {Qiu}, \citenamefont {Cao},\ and\ \citenamefont {Louie}}]{qiu2015nonanalyticity}%
  \BibitemOpen
  \bibfield  {author} {\bibinfo {author} {\bibfnamefont {D.~Y.}\ \bibnamefont {Qiu}}, \bibinfo {author} {\bibfnamefont {T.}~\bibnamefont {Cao}},\ and\ \bibinfo {author} {\bibfnamefont {S.~G.}\ \bibnamefont {Louie}},\ }\bibfield  {title} {\bibinfo {title} {Nonanalyticity, valley quantum phases, and lightlike exciton dispersion in monolayer transition metal dichalcogenides: Theory and first-principles calculations},\ }\href {https://doi.org/10.1103/PhysRevLett.115.176801} {\bibfield  {journal} {\bibinfo  {journal} {Phys. Rev. Lett.}\ }\textbf {\bibinfo {volume} {115}},\ \bibinfo {pages} {176801} (\bibinfo {year} {2015})}\BibitemShut {NoStop}%
\bibitem [{\citenamefont {Liu}\ \emph {et~al.}()\citenamefont {Liu}, \citenamefont {Woo}, \citenamefont {Wu}, \citenamefont {Hou}, \citenamefont {Su},\ and\ \citenamefont {Qiu}}]{liu2025direct}%
  \BibitemOpen
  \bibfield  {author} {\bibinfo {author} {\bibfnamefont {L.~Y.}\ \bibnamefont {Liu}}, \bibinfo {author} {\bibfnamefont {S.~Y.}\ \bibnamefont {Woo}}, \bibinfo {author} {\bibfnamefont {J.}~\bibnamefont {Wu}}, \bibinfo {author} {\bibfnamefont {B.}~\bibnamefont {Hou}}, \bibinfo {author} {\bibfnamefont {C.}~\bibnamefont {Su}},\ and\ \bibinfo {author} {\bibfnamefont {D.~Y.}\ \bibnamefont {Qiu}},\ }\href@noop {} {\bibinfo {title} {Direct observation of massless excitons and linear exciton dispersion}},\ \Eprint {https://arxiv.org/abs/2502.20454} {arXiv:2502.20454} \BibitemShut {NoStop}%
\bibitem [{\citenamefont {Selig}\ \emph {et~al.}(2019)\citenamefont {Selig}, \citenamefont {Malic}, \citenamefont {Ahn}, \citenamefont {Koch},\ and\ \citenamefont {Knorr}}]{selig2019theory}%
  \BibitemOpen
  \bibfield  {author} {\bibinfo {author} {\bibfnamefont {M.}~\bibnamefont {Selig}}, \bibinfo {author} {\bibfnamefont {E.}~\bibnamefont {Malic}}, \bibinfo {author} {\bibfnamefont {K.~J.}\ \bibnamefont {Ahn}}, \bibinfo {author} {\bibfnamefont {N.}~\bibnamefont {Koch}},\ and\ \bibinfo {author} {\bibfnamefont {A.}~\bibnamefont {Knorr}},\ }\bibfield  {title} {\bibinfo {title} {Theory of optically induced $\text{F\"orster}$ coupling in van der waals coupled heterostructures},\ }\href {https://doi.org/10.1103/PhysRevB.99.035420} {\bibfield  {journal} {\bibinfo  {journal} {Phys. Rev. B}\ }\textbf {\bibinfo {volume} {99}},\ \bibinfo {pages} {035420} (\bibinfo {year} {2019})}\BibitemShut {NoStop}%
\bibitem [{\citenamefont {Baimuratov}\ and\ \citenamefont {H{\"o}gele}(2020)}]{baimuratov2020valley}%
  \BibitemOpen
  \bibfield  {author} {\bibinfo {author} {\bibfnamefont {A.~S.}\ \bibnamefont {Baimuratov}}\ and\ \bibinfo {author} {\bibfnamefont {A.}~\bibnamefont {H{\"o}gele}},\ }\bibfield  {title} {\bibinfo {title} {Valley-selective energy transfer between quantum dots in atomically thin semiconductors},\ }\href {https://doi.org/10.1038/s41598-020-73688-8} {\bibfield  {journal} {\bibinfo  {journal} {Sci. Rep.}\ }\textbf {\bibinfo {volume} {10}},\ \bibinfo {pages} {16971} (\bibinfo {year} {2020})}\BibitemShut {NoStop}%
\bibitem [{\citenamefont {Hichri}\ \emph {et~al.}(2021)\citenamefont {Hichri}, \citenamefont {Amand},\ and\ \citenamefont {Jaziri}}]{hichri2021resonance}%
  \BibitemOpen
  \bibfield  {author} {\bibinfo {author} {\bibfnamefont {A.}~\bibnamefont {Hichri}}, \bibinfo {author} {\bibfnamefont {T.}~\bibnamefont {Amand}},\ and\ \bibinfo {author} {\bibfnamefont {S.}~\bibnamefont {Jaziri}},\ }\bibfield  {title} {\bibinfo {title} {Resonance energy transfer from moir\'e-trapped excitons in $\text{MoSe}_{2}/\text{WSe}_{2}$ heterobilayers to graphene: Dielectric environment effect},\ }\href {https://doi.org/10.1103/PhysRevMaterials.5.114002} {\bibfield  {journal} {\bibinfo  {journal} {Phys. Rev. Mater.}\ }\textbf {\bibinfo {volume} {5}},\ \bibinfo {pages} {114002} (\bibinfo {year} {2021})}\BibitemShut {NoStop}%
\bibitem [{\citenamefont {Li}\ and\ \citenamefont {Yao}(2023)}]{li2023cross}%
  \BibitemOpen
  \bibfield  {author} {\bibinfo {author} {\bibfnamefont {C.}~\bibnamefont {Li}}\ and\ \bibinfo {author} {\bibfnamefont {W.}~\bibnamefont {Yao}},\ }\bibfield  {title} {\bibinfo {title} {Cross-dimensional valley excitons from $\text{F\"orster}$ coupling in arbitrarily twisted stacks of monolayer semiconductors},\ }\href {https://doi.org/10.1088/2053-1583/ad0403} {\bibfield  {journal} {\bibinfo  {journal} {2D Mater.}\ }\textbf {\bibinfo {volume} {11}},\ \bibinfo {pages} {015006} (\bibinfo {year} {2023})}\BibitemShut {NoStop}%
\bibitem [{\citenamefont {Clegg}(2006)}]{clegg2006the}%
  \BibitemOpen
  \bibfield  {author} {\bibinfo {author} {\bibfnamefont {R.~M.}\ \bibnamefont {Clegg}},\ }\bibinfo {title} {The history of fret},\ in\ \href {https://doi.org/10.1007/0-387-33016-X_1} {\emph {\bibinfo {booktitle} {Reviews in Fluorescence 2006}}},\ \bibinfo {editor} {edited by\ \bibinfo {editor} {\bibfnamefont {C.~D.}\ \bibnamefont {Geddes}}\ and\ \bibinfo {editor} {\bibfnamefont {J.~R.}\ \bibnamefont {Lakowicz}}}\ (\bibinfo  {publisher} {Springer US},\ \bibinfo {address} {Boston, MA},\ \bibinfo {year} {2006})\ pp.\ \bibinfo {pages} {1--45}\BibitemShut {NoStop}%
\bibitem [{\citenamefont {Cai}\ \emph {et~al.}(2023)\citenamefont {Cai}, \citenamefont {Anderson}, \citenamefont {Wang}, \citenamefont {Zhang}, \citenamefont {Liu}, \citenamefont {Holtzmann}, \citenamefont {Zhang}, \citenamefont {Fan}, \citenamefont {Taniguchi}, \citenamefont {Watanabe}, \citenamefont {Ran}, \citenamefont {Cao}, \citenamefont {Fu}, \citenamefont {Xiao}, \citenamefont {Yao},\ and\ \citenamefont {Xu}}]{cai2023signatures}%
  \BibitemOpen
  \bibfield  {author} {\bibinfo {author} {\bibfnamefont {J.}~\bibnamefont {Cai}}, \bibinfo {author} {\bibfnamefont {E.}~\bibnamefont {Anderson}}, \bibinfo {author} {\bibfnamefont {C.}~\bibnamefont {Wang}}, \bibinfo {author} {\bibfnamefont {X.}~\bibnamefont {Zhang}}, \bibinfo {author} {\bibfnamefont {X.}~\bibnamefont {Liu}}, \bibinfo {author} {\bibfnamefont {W.}~\bibnamefont {Holtzmann}}, \bibinfo {author} {\bibfnamefont {Y.}~\bibnamefont {Zhang}}, \bibinfo {author} {\bibfnamefont {F.}~\bibnamefont {Fan}}, \bibinfo {author} {\bibfnamefont {T.}~\bibnamefont {Taniguchi}}, \bibinfo {author} {\bibfnamefont {K.}~\bibnamefont {Watanabe}}, \bibinfo {author} {\bibfnamefont {Y.}~\bibnamefont {Ran}}, \bibinfo {author} {\bibfnamefont {T.}~\bibnamefont {Cao}}, \bibinfo {author} {\bibfnamefont {L.}~\bibnamefont {Fu}}, \bibinfo {author} {\bibfnamefont {D.}~\bibnamefont {Xiao}}, \bibinfo {author} {\bibfnamefont {W.}~\bibnamefont {Yao}},\ and\ \bibinfo {author} {\bibfnamefont {X.}~\bibnamefont {Xu}},\ }\bibfield  {title}
  {\bibinfo {title} {Signatures of fractional quantum anomalous hall states in twisted $\text{MoTe}_2$},\ }\href {https://doi.org/10.1038/s41586-023-06289-w} {\bibfield  {journal} {\bibinfo  {journal} {Nature}\ }\textbf {\bibinfo {volume} {622}},\ \bibinfo {pages} {63} (\bibinfo {year} {2023})}\BibitemShut {NoStop}%
\bibitem [{\citenamefont {Park}\ \emph {et~al.}(2023{\natexlab{b}})\citenamefont {Park}, \citenamefont {Cai}, \citenamefont {Anderson}, \citenamefont {Zhang}, \citenamefont {Zhu}, \citenamefont {Liu}, \citenamefont {Wang}, \citenamefont {Holtzmann}, \citenamefont {Hu}, \citenamefont {Liu}, \citenamefont {Taniguchi}, \citenamefont {Watanabe}, \citenamefont {Chu}, \citenamefont {Cao}, \citenamefont {Fu}, \citenamefont {Yao}, \citenamefont {Chang}, \citenamefont {Cobden}, \citenamefont {Xiao},\ and\ \citenamefont {Xu}}]{park2023observation}%
  \BibitemOpen
  \bibfield  {author} {\bibinfo {author} {\bibfnamefont {H.}~\bibnamefont {Park}}, \bibinfo {author} {\bibfnamefont {J.}~\bibnamefont {Cai}}, \bibinfo {author} {\bibfnamefont {E.}~\bibnamefont {Anderson}}, \bibinfo {author} {\bibfnamefont {Y.}~\bibnamefont {Zhang}}, \bibinfo {author} {\bibfnamefont {J.}~\bibnamefont {Zhu}}, \bibinfo {author} {\bibfnamefont {X.}~\bibnamefont {Liu}}, \bibinfo {author} {\bibfnamefont {C.}~\bibnamefont {Wang}}, \bibinfo {author} {\bibfnamefont {W.}~\bibnamefont {Holtzmann}}, \bibinfo {author} {\bibfnamefont {C.}~\bibnamefont {Hu}}, \bibinfo {author} {\bibfnamefont {Z.}~\bibnamefont {Liu}}, \bibinfo {author} {\bibfnamefont {T.}~\bibnamefont {Taniguchi}}, \bibinfo {author} {\bibfnamefont {K.}~\bibnamefont {Watanabe}}, \bibinfo {author} {\bibfnamefont {J.-H.}\ \bibnamefont {Chu}}, \bibinfo {author} {\bibfnamefont {T.}~\bibnamefont {Cao}}, \bibinfo {author} {\bibfnamefont {L.}~\bibnamefont {Fu}}, \bibinfo {author} {\bibfnamefont {W.}~\bibnamefont {Yao}}, \bibinfo {author}
  {\bibfnamefont {C.-Z.}\ \bibnamefont {Chang}}, \bibinfo {author} {\bibfnamefont {D.}~\bibnamefont {Cobden}}, \bibinfo {author} {\bibfnamefont {D.}~\bibnamefont {Xiao}},\ and\ \bibinfo {author} {\bibfnamefont {X.}~\bibnamefont {Xu}},\ }\bibfield  {title} {\bibinfo {title} {Observation of fractionally quantized anomalous hall effect},\ }\href {https://doi.org/10.1038/s41586-023-06536-0} {\bibfield  {journal} {\bibinfo  {journal} {Nature}\ }\textbf {\bibinfo {volume} {622}},\ \bibinfo {pages} {74} (\bibinfo {year} {2023}{\natexlab{b}})}\BibitemShut {NoStop}%
\bibitem [{\citenamefont {Zeng}\ \emph {et~al.}(2023)\citenamefont {Zeng}, \citenamefont {Xia}, \citenamefont {Kang}, \citenamefont {Zhu}, \citenamefont {Kn{\"u}ppel}, \citenamefont {Vaswani}, \citenamefont {Watanabe}, \citenamefont {Taniguchi}, \citenamefont {Mak},\ and\ \citenamefont {Shan}}]{zeng2023thermodynamic}%
  \BibitemOpen
  \bibfield  {author} {\bibinfo {author} {\bibfnamefont {Y.}~\bibnamefont {Zeng}}, \bibinfo {author} {\bibfnamefont {Z.}~\bibnamefont {Xia}}, \bibinfo {author} {\bibfnamefont {K.}~\bibnamefont {Kang}}, \bibinfo {author} {\bibfnamefont {J.}~\bibnamefont {Zhu}}, \bibinfo {author} {\bibfnamefont {P.}~\bibnamefont {Kn{\"u}ppel}}, \bibinfo {author} {\bibfnamefont {C.}~\bibnamefont {Vaswani}}, \bibinfo {author} {\bibfnamefont {K.}~\bibnamefont {Watanabe}}, \bibinfo {author} {\bibfnamefont {T.}~\bibnamefont {Taniguchi}}, \bibinfo {author} {\bibfnamefont {K.~F.}\ \bibnamefont {Mak}},\ and\ \bibinfo {author} {\bibfnamefont {J.}~\bibnamefont {Shan}},\ }\bibfield  {title} {\bibinfo {title} {Thermodynamic evidence of fractional chern insulator in moir{\'e} $\text{MoTe}_2$},\ }\href {https://doi.org/10.1038/s41586-023-06452-3} {\bibfield  {journal} {\bibinfo  {journal} {Nature}\ }\textbf {\bibinfo {volume} {622}},\ \bibinfo {pages} {69} (\bibinfo {year} {2023})}\BibitemShut {NoStop}%
\bibitem [{\citenamefont {Xu}\ \emph {et~al.}(2023)\citenamefont {Xu}, \citenamefont {Sun}, \citenamefont {Jia}, \citenamefont {Liu}, \citenamefont {Xu}, \citenamefont {Li}, \citenamefont {Gu}, \citenamefont {Watanabe}, \citenamefont {Taniguchi}, \citenamefont {Tong}, \citenamefont {Jia}, \citenamefont {Shi}, \citenamefont {Jiang}, \citenamefont {Zhang}, \citenamefont {Liu},\ and\ \citenamefont {Li}}]{xu2023observation}%
  \BibitemOpen
  \bibfield  {author} {\bibinfo {author} {\bibfnamefont {F.}~\bibnamefont {Xu}}, \bibinfo {author} {\bibfnamefont {Z.}~\bibnamefont {Sun}}, \bibinfo {author} {\bibfnamefont {T.}~\bibnamefont {Jia}}, \bibinfo {author} {\bibfnamefont {C.}~\bibnamefont {Liu}}, \bibinfo {author} {\bibfnamefont {C.}~\bibnamefont {Xu}}, \bibinfo {author} {\bibfnamefont {C.}~\bibnamefont {Li}}, \bibinfo {author} {\bibfnamefont {Y.}~\bibnamefont {Gu}}, \bibinfo {author} {\bibfnamefont {K.}~\bibnamefont {Watanabe}}, \bibinfo {author} {\bibfnamefont {T.}~\bibnamefont {Taniguchi}}, \bibinfo {author} {\bibfnamefont {B.}~\bibnamefont {Tong}}, \bibinfo {author} {\bibfnamefont {J.}~\bibnamefont {Jia}}, \bibinfo {author} {\bibfnamefont {Z.}~\bibnamefont {Shi}}, \bibinfo {author} {\bibfnamefont {S.}~\bibnamefont {Jiang}}, \bibinfo {author} {\bibfnamefont {Y.}~\bibnamefont {Zhang}}, \bibinfo {author} {\bibfnamefont {X.}~\bibnamefont {Liu}},\ and\ \bibinfo {author} {\bibfnamefont {T.}~\bibnamefont {Li}},\ }\bibfield  {title} {\bibinfo {title}
  {Observation of integer and fractional quantum anomalous hall effects in twisted bilayer $\text{MoTe}_2$},\ }\href {https://doi.org/10.1103/PhysRevX.13.031037} {\bibfield  {journal} {\bibinfo  {journal} {Phys. Rev. X}\ }\textbf {\bibinfo {volume} {13}},\ \bibinfo {pages} {031037} (\bibinfo {year} {2023})}\BibitemShut {NoStop}%
\bibitem [{\citenamefont {Weston}\ \emph {et~al.}(2022)\citenamefont {Weston}, \citenamefont {Castanon}, \citenamefont {Enaldiev}, \citenamefont {Ferreira}, \citenamefont {Bhattacharjee}, \citenamefont {Xu}, \citenamefont {Corte-Le{\'o}n}, \citenamefont {Wu}, \citenamefont {Clark}, \citenamefont {Summerfield}, \citenamefont {Hashimoto}, \citenamefont {Gao}, \citenamefont {Wang}, \citenamefont {Hamer}, \citenamefont {Read}, \citenamefont {Fumagalli}, \citenamefont {Kretinin}, \citenamefont {Haigh}, \citenamefont {Kazakova}, \citenamefont {Geim}, \citenamefont {Fal'ko},\ and\ \citenamefont {Gorbachev}}]{weston2022interfacial}%
  \BibitemOpen
  \bibfield  {author} {\bibinfo {author} {\bibfnamefont {A.}~\bibnamefont {Weston}}, \bibinfo {author} {\bibfnamefont {E.~G.}\ \bibnamefont {Castanon}}, \bibinfo {author} {\bibfnamefont {V.}~\bibnamefont {Enaldiev}}, \bibinfo {author} {\bibfnamefont {F.}~\bibnamefont {Ferreira}}, \bibinfo {author} {\bibfnamefont {S.}~\bibnamefont {Bhattacharjee}}, \bibinfo {author} {\bibfnamefont {S.}~\bibnamefont {Xu}}, \bibinfo {author} {\bibfnamefont {H.}~\bibnamefont {Corte-Le{\'o}n}}, \bibinfo {author} {\bibfnamefont {Z.}~\bibnamefont {Wu}}, \bibinfo {author} {\bibfnamefont {N.}~\bibnamefont {Clark}}, \bibinfo {author} {\bibfnamefont {A.}~\bibnamefont {Summerfield}}, \bibinfo {author} {\bibfnamefont {T.}~\bibnamefont {Hashimoto}}, \bibinfo {author} {\bibfnamefont {Y.}~\bibnamefont {Gao}}, \bibinfo {author} {\bibfnamefont {W.}~\bibnamefont {Wang}}, \bibinfo {author} {\bibfnamefont {M.}~\bibnamefont {Hamer}}, \bibinfo {author} {\bibfnamefont {H.}~\bibnamefont {Read}}, \bibinfo {author} {\bibfnamefont {L.}~\bibnamefont
  {Fumagalli}}, \bibinfo {author} {\bibfnamefont {A.~V.}\ \bibnamefont {Kretinin}}, \bibinfo {author} {\bibfnamefont {S.~J.}\ \bibnamefont {Haigh}}, \bibinfo {author} {\bibfnamefont {O.}~\bibnamefont {Kazakova}}, \bibinfo {author} {\bibfnamefont {A.~K.}\ \bibnamefont {Geim}}, \bibinfo {author} {\bibfnamefont {V.~I.}\ \bibnamefont {Fal'ko}},\ and\ \bibinfo {author} {\bibfnamefont {R.}~\bibnamefont {Gorbachev}},\ }\bibfield  {title} {\bibinfo {title} {Interfacial ferroelectricity in marginally twisted $\text{2D}$ semiconductors},\ }\href {https://doi.org/10.1038/s41565-022-01072-w} {\bibfield  {journal} {\bibinfo  {journal} {Nat. Nanotechnol.}\ }\textbf {\bibinfo {volume} {17}},\ \bibinfo {pages} {390} (\bibinfo {year} {2022})}\BibitemShut {NoStop}%
\bibitem [{\citenamefont {Wang}\ \emph {et~al.}(2022)\citenamefont {Wang}, \citenamefont {Yasuda}, \citenamefont {Zhang}, \citenamefont {Liu}, \citenamefont {Watanabe}, \citenamefont {Taniguchi}, \citenamefont {Hone}, \citenamefont {Fu},\ and\ \citenamefont {Jarillo-Herrero}}]{wang2022interfacial}%
  \BibitemOpen
  \bibfield  {author} {\bibinfo {author} {\bibfnamefont {X.}~\bibnamefont {Wang}}, \bibinfo {author} {\bibfnamefont {K.}~\bibnamefont {Yasuda}}, \bibinfo {author} {\bibfnamefont {Y.}~\bibnamefont {Zhang}}, \bibinfo {author} {\bibfnamefont {S.}~\bibnamefont {Liu}}, \bibinfo {author} {\bibfnamefont {K.}~\bibnamefont {Watanabe}}, \bibinfo {author} {\bibfnamefont {T.}~\bibnamefont {Taniguchi}}, \bibinfo {author} {\bibfnamefont {J.}~\bibnamefont {Hone}}, \bibinfo {author} {\bibfnamefont {L.}~\bibnamefont {Fu}},\ and\ \bibinfo {author} {\bibfnamefont {P.}~\bibnamefont {Jarillo-Herrero}},\ }\bibfield  {title} {\bibinfo {title} {Interfacial ferroelectricity in rhombohedral-stacked bilayer transition metal dichalcogenides},\ }\href {https://doi.org/10.1038/s41565-021-01059-z} {\bibfield  {journal} {\bibinfo  {journal} {Nat. Nanotechnol.}\ }\textbf {\bibinfo {volume} {17}},\ \bibinfo {pages} {367} (\bibinfo {year} {2022})}\BibitemShut {NoStop}%
\bibitem [{\citenamefont {Yu}\ and\ \citenamefont {Yao}(2021)}]{yu2021luminescence}%
  \BibitemOpen
  \bibfield  {author} {\bibinfo {author} {\bibfnamefont {H.}~\bibnamefont {Yu}}\ and\ \bibinfo {author} {\bibfnamefont {W.}~\bibnamefont {Yao}},\ }\bibfield  {title} {\bibinfo {title} {Luminescence anomaly of dipolar valley excitons in homobilayer semiconductor moir\'e superlattices},\ }\href {https://doi.org/10.1103/PhysRevX.11.021042} {\bibfield  {journal} {\bibinfo  {journal} {Phys. Rev. X}\ }\textbf {\bibinfo {volume} {11}},\ \bibinfo {pages} {021042} (\bibinfo {year} {2021})}\BibitemShut {NoStop}%
\bibitem [{\citenamefont {Tong}\ \emph {et~al.}(2017)\citenamefont {Tong}, \citenamefont {Yu}, \citenamefont {Zhu}, \citenamefont {Wang}, \citenamefont {Xu},\ and\ \citenamefont {Yao}}]{tong2017topological}%
  \BibitemOpen
  \bibfield  {author} {\bibinfo {author} {\bibfnamefont {Q.}~\bibnamefont {Tong}}, \bibinfo {author} {\bibfnamefont {H.}~\bibnamefont {Yu}}, \bibinfo {author} {\bibfnamefont {Q.}~\bibnamefont {Zhu}}, \bibinfo {author} {\bibfnamefont {Y.}~\bibnamefont {Wang}}, \bibinfo {author} {\bibfnamefont {X.}~\bibnamefont {Xu}},\ and\ \bibinfo {author} {\bibfnamefont {W.}~\bibnamefont {Yao}},\ }\bibfield  {title} {\bibinfo {title} {Topological mosaics in moir{\'e} superlattices of van der waals heterobilayers},\ }\href {https://doi.org/10.1038/nphys3968} {\bibfield  {journal} {\bibinfo  {journal} {Nat. Phys.}\ }\textbf {\bibinfo {volume} {13}},\ \bibinfo {pages} {356} (\bibinfo {year} {2017})}\BibitemShut {NoStop}%
\bibitem [{\citenamefont {Bistritzer}\ and\ \citenamefont {MacDonald}(2011)}]{rafi2011moire}%
  \BibitemOpen
  \bibfield  {author} {\bibinfo {author} {\bibfnamefont {R.}~\bibnamefont {Bistritzer}}\ and\ \bibinfo {author} {\bibfnamefont {A.~H.}\ \bibnamefont {MacDonald}},\ }\bibfield  {title} {\bibinfo {title} {Moiré bands in twisted double-layer graphene},\ }\href {https://doi.org/10.1073/pnas.1108174108} {\bibfield  {journal} {\bibinfo  {journal} {Proc. Natl. Acad. Sci. USA}\ }\textbf {\bibinfo {volume} {108}},\ \bibinfo {pages} {12233} (\bibinfo {year} {2011})}\BibitemShut {NoStop}%
\bibitem [{\citenamefont {Wang}\ \emph {et~al.}(2017)\citenamefont {Wang}, \citenamefont {Wang}, \citenamefont {Yao}, \citenamefont {Liu},\ and\ \citenamefont {Yu}}]{wang2017interlayer}%
  \BibitemOpen
  \bibfield  {author} {\bibinfo {author} {\bibfnamefont {Y.}~\bibnamefont {Wang}}, \bibinfo {author} {\bibfnamefont {Z.}~\bibnamefont {Wang}}, \bibinfo {author} {\bibfnamefont {W.}~\bibnamefont {Yao}}, \bibinfo {author} {\bibfnamefont {G.-B.}\ \bibnamefont {Liu}},\ and\ \bibinfo {author} {\bibfnamefont {H.}~\bibnamefont {Yu}},\ }\bibfield  {title} {\bibinfo {title} {Interlayer coupling in commensurate and incommensurate bilayer structures of transition-metal dichalcogenides},\ }\href {https://doi.org/10.1103/PhysRevB.95.115429} {\bibfield  {journal} {\bibinfo  {journal} {Phys. Rev. B}\ }\textbf {\bibinfo {volume} {95}},\ \bibinfo {pages} {115429} (\bibinfo {year} {2017})}\BibitemShut {NoStop}%
\bibitem [{\citenamefont {Baer}\ and\ \citenamefont {Rabani}(2008)}]{baer2008theory}%
  \BibitemOpen
  \bibfield  {author} {\bibinfo {author} {\bibfnamefont {R.}~\bibnamefont {Baer}}\ and\ \bibinfo {author} {\bibfnamefont {E.}~\bibnamefont {Rabani}},\ }\bibfield  {title} {\bibinfo {title} {{Theory of resonance energy transfer involving nanocrystals: The role of high multipoles}},\ }\href {https://doi.org/10.1063/1.2913247} {\bibfield  {journal} {\bibinfo  {journal} {J. Chem. Phys}\ }\textbf {\bibinfo {volume} {128}},\ \bibinfo {pages} {184710} (\bibinfo {year} {2008})}\BibitemShut {NoStop}%
\bibitem [{\citenamefont {Xie}\ \emph {et~al.}(2024)\citenamefont {Xie}, \citenamefont {Hafezi},\ and\ \citenamefont {Das~Sarma}}]{xie2024long-lived}%
  \BibitemOpen
  \bibfield  {author} {\bibinfo {author} {\bibfnamefont {M.}~\bibnamefont {Xie}}, \bibinfo {author} {\bibfnamefont {M.}~\bibnamefont {Hafezi}},\ and\ \bibinfo {author} {\bibfnamefont {S.}~\bibnamefont {Das~Sarma}},\ }\bibfield  {title} {\bibinfo {title} {Long-lived topological flatband excitons in semiconductor moir\'e heterostructures: A bosonic kane-mele model platform},\ }\href {https://doi.org/10.1103/PhysRevLett.133.136403} {\bibfield  {journal} {\bibinfo  {journal} {Phys. Rev. Lett.}\ }\textbf {\bibinfo {volume} {133}},\ \bibinfo {pages} {136403} (\bibinfo {year} {2024})}\BibitemShut {NoStop}%
\bibitem [{\citenamefont {Haldane}(1988)}]{haldane1988model}%
  \BibitemOpen
  \bibfield  {author} {\bibinfo {author} {\bibfnamefont {F.~D.~M.}\ \bibnamefont {Haldane}},\ }\bibfield  {title} {\bibinfo {title} {Model for a quantum hall effect without landau levels: Condensed-matter realization of the ``parity anomaly"},\ }\href {https://doi.org/10.1103/PhysRevLett.61.2015} {\bibfield  {journal} {\bibinfo  {journal} {Phys. Rev. Lett.}\ }\textbf {\bibinfo {volume} {61}},\ \bibinfo {pages} {2015} (\bibinfo {year} {1988})}\BibitemShut {NoStop}%
\end{thebibliography}%


\begin{thebibliography}{11}%
\makeatletter
\providecommand \@ifxundefined [1]{%
 \@ifx{#1\undefined}
}%
\providecommand \@ifnum [1]{%
 \ifnum #1\expandafter \@firstoftwo
 \else \expandafter \@secondoftwo
 \fi
}%
\providecommand \@ifx [1]{%
 \ifx #1\expandafter \@firstoftwo
 \else \expandafter \@secondoftwo
 \fi
}%
\providecommand \natexlab [1]{#1}%
\providecommand \enquote  [1]{``#1''}%
\providecommand \bibnamefont  [1]{#1}%
\providecommand \bibfnamefont [1]{#1}%
\providecommand \citenamefont [1]{#1}%
\providecommand \href@noop [0]{\@secondoftwo}%
\providecommand \href [0]{\begingroup \@sanitize@url \@href}%
\providecommand \@href[1]{\@@startlink{#1}\@@href}%
\providecommand \@@href[1]{\endgroup#1\@@endlink}%
\providecommand \@sanitize@url [0]{\catcode `\\12\catcode `\$12\catcode `\&12\catcode `\#12\catcode `\^12\catcode `\_12\catcode `\%12\relax}%
\providecommand \@@startlink[1]{}%
\providecommand \@@endlink[0]{}%
\providecommand \url  [0]{\begingroup\@sanitize@url \@url }%
\providecommand \@url [1]{\endgroup\@href {#1}{\urlprefix }}%
\providecommand \urlprefix  [0]{URL }%
\providecommand \Eprint [0]{\href }%
\providecommand \doibase [0]{https://doi.org/}%
\providecommand \selectlanguage [0]{\@gobble}%
\providecommand \bibinfo  [0]{\@secondoftwo}%
\providecommand \bibfield  [0]{\@secondoftwo}%
\providecommand \translation [1]{[#1]}%
\providecommand \BibitemOpen [0]{}%
\providecommand \bibitemStop [0]{}%
\providecommand \bibitemNoStop [0]{.\EOS\space}%
\providecommand \EOS [0]{\spacefactor3000\relax}%
\providecommand \BibitemShut  [1]{\csname bibitem#1\endcsname}%
\let\auto@bib@innerbib\@empty
\bibitem [{\citenamefont {Xiao}\ \emph {et~al.}(2012)\citenamefont {Xiao}, \citenamefont {Liu}, \citenamefont {Feng}, \citenamefont {Xu},\ and\ \citenamefont {Yao}}]{xiao2012coupled}%
  \BibitemOpen
  \bibfield  {author} {\bibinfo {author} {\bibfnamefont {D.}~\bibnamefont {Xiao}}, \bibinfo {author} {\bibfnamefont {G.-B.}\ \bibnamefont {Liu}}, \bibinfo {author} {\bibfnamefont {W.}~\bibnamefont {Feng}}, \bibinfo {author} {\bibfnamefont {X.}~\bibnamefont {Xu}},\ and\ \bibinfo {author} {\bibfnamefont {W.}~\bibnamefont {Yao}},\ }\bibfield  {title} {\bibinfo {title} {Coupled spin and valley physics in monolayers of ${\text{mos}}_{2}$ and other group-vi dichalcogenides},\ }\href {https://doi.org/10.1103/PhysRevLett.108.196802} {\bibfield  {journal} {\bibinfo  {journal} {Phys. Rev. Lett.}\ }\textbf {\bibinfo {volume} {108}},\ \bibinfo {pages} {196802} (\bibinfo {year} {2012})}\BibitemShut {NoStop}%
\bibitem [{\citenamefont {Yu}\ \emph {et~al.}(2017)\citenamefont {Yu}, \citenamefont {Liu}, \citenamefont {Tang}, \citenamefont {Xu},\ and\ \citenamefont {Yao}}]{yu2017moire}%
  \BibitemOpen
  \bibfield  {author} {\bibinfo {author} {\bibfnamefont {H.}~\bibnamefont {Yu}}, \bibinfo {author} {\bibfnamefont {G.-B.}\ \bibnamefont {Liu}}, \bibinfo {author} {\bibfnamefont {J.}~\bibnamefont {Tang}}, \bibinfo {author} {\bibfnamefont {X.}~\bibnamefont {Xu}},\ and\ \bibinfo {author} {\bibfnamefont {W.}~\bibnamefont {Yao}},\ }\bibfield  {title} {\bibinfo {title} {Moir\'e excitons: From programmable quantum emitter arrays to spin-orbit–coupled artificial lattices},\ }\href {https://doi.org/10.1126/sciadv.1701696} {\bibfield  {journal} {\bibinfo  {journal} {Sci. Adv.}\ }\textbf {\bibinfo {volume} {3}},\ \bibinfo {pages} {e1701696} (\bibinfo {year} {2017})}\BibitemShut {NoStop}%
\bibitem [{\citenamefont {Bistritzer}\ and\ \citenamefont {MacDonald}(2011)}]{rafi2011moire}%
  \BibitemOpen
  \bibfield  {author} {\bibinfo {author} {\bibfnamefont {R.}~\bibnamefont {Bistritzer}}\ and\ \bibinfo {author} {\bibfnamefont {A.~H.}\ \bibnamefont {MacDonald}},\ }\bibfield  {title} {\bibinfo {title} {Moiré bands in twisted double-layer graphene},\ }\href {https://doi.org/10.1073/pnas.1108174108} {\bibfield  {journal} {\bibinfo  {journal} {Proc. Natl. Acad. Sci. USA}\ }\textbf {\bibinfo {volume} {108}},\ \bibinfo {pages} {12233} (\bibinfo {year} {2011})}\BibitemShut {NoStop}%
\bibitem [{\citenamefont {Wang}\ \emph {et~al.}(2017)\citenamefont {Wang}, \citenamefont {Wang}, \citenamefont {Yao}, \citenamefont {Liu},\ and\ \citenamefont {Yu}}]{wang2017interlayer}%
  \BibitemOpen
  \bibfield  {author} {\bibinfo {author} {\bibfnamefont {Y.}~\bibnamefont {Wang}}, \bibinfo {author} {\bibfnamefont {Z.}~\bibnamefont {Wang}}, \bibinfo {author} {\bibfnamefont {W.}~\bibnamefont {Yao}}, \bibinfo {author} {\bibfnamefont {G.-B.}\ \bibnamefont {Liu}},\ and\ \bibinfo {author} {\bibfnamefont {H.}~\bibnamefont {Yu}},\ }\bibfield  {title} {\bibinfo {title} {Interlayer coupling in commensurate and incommensurate bilayer structures of transition-metal dichalcogenides},\ }\href {https://doi.org/10.1103/PhysRevB.95.115429} {\bibfield  {journal} {\bibinfo  {journal} {Phys. Rev. B}\ }\textbf {\bibinfo {volume} {95}},\ \bibinfo {pages} {115429} (\bibinfo {year} {2017})}\BibitemShut {NoStop}%
\bibitem [{\citenamefont {Li}\ and\ \citenamefont {Yao}(2023)}]{li2023cross}%
  \BibitemOpen
  \bibfield  {author} {\bibinfo {author} {\bibfnamefont {C.}~\bibnamefont {Li}}\ and\ \bibinfo {author} {\bibfnamefont {W.}~\bibnamefont {Yao}},\ }\bibfield  {title} {\bibinfo {title} {Cross-dimensional valley excitons from $\text{F\"orster}$ coupling in arbitrarily twisted stacks of monolayer semiconductors},\ }\href {https://doi.org/10.1088/2053-1583/ad0403} {\bibfield  {journal} {\bibinfo  {journal} {2D Mater.}\ }\textbf {\bibinfo {volume} {11}},\ \bibinfo {pages} {015006} (\bibinfo {year} {2023})}\BibitemShut {NoStop}%
\bibitem [{\citenamefont {Yu}\ \emph {et~al.}(2014)\citenamefont {Yu}, \citenamefont {Liu}, \citenamefont {Gong}, \citenamefont {Xu},\ and\ \citenamefont {Yao}}]{yu2014dirac}%
  \BibitemOpen
  \bibfield  {author} {\bibinfo {author} {\bibfnamefont {H.}~\bibnamefont {Yu}}, \bibinfo {author} {\bibfnamefont {G.-B.}\ \bibnamefont {Liu}}, \bibinfo {author} {\bibfnamefont {P.}~\bibnamefont {Gong}}, \bibinfo {author} {\bibfnamefont {X.}~\bibnamefont {Xu}},\ and\ \bibinfo {author} {\bibfnamefont {W.}~\bibnamefont {Yao}},\ }\bibfield  {title} {\bibinfo {title} {Dirac cones and dirac saddle points of bright excitons in monolayer transition metal dichalcogenides},\ }\href {https://doi.org/10.1038/ncomms4876} {\bibfield  {journal} {\bibinfo  {journal} {Nat. Commun.}\ }\textbf {\bibinfo {volume} {5}},\ \bibinfo {pages} {3876} (\bibinfo {year} {2014})}\BibitemShut {NoStop}%
\bibitem [{\citenamefont {Liu}\ \emph {et~al.}()\citenamefont {Liu}, \citenamefont {Woo}, \citenamefont {Wu}, \citenamefont {Hou}, \citenamefont {Su},\ and\ \citenamefont {Qiu}}]{liu2025direct}%
  \BibitemOpen
  \bibfield  {author} {\bibinfo {author} {\bibfnamefont {L.~Y.}\ \bibnamefont {Liu}}, \bibinfo {author} {\bibfnamefont {S.~Y.}\ \bibnamefont {Woo}}, \bibinfo {author} {\bibfnamefont {J.}~\bibnamefont {Wu}}, \bibinfo {author} {\bibfnamefont {B.}~\bibnamefont {Hou}}, \bibinfo {author} {\bibfnamefont {C.}~\bibnamefont {Su}},\ and\ \bibinfo {author} {\bibfnamefont {D.~Y.}\ \bibnamefont {Qiu}},\ }\href@noop {} {\bibinfo {title} {Direct observation of massless excitons and linear exciton dispersion}},\ \Eprint {https://arxiv.org/abs/2502.20454} {arXiv:2502.20454} \BibitemShut {NoStop}%
\bibitem [{\citenamefont {Baer}\ and\ \citenamefont {Rabani}(2008)}]{baer2008theory}%
  \BibitemOpen
  \bibfield  {author} {\bibinfo {author} {\bibfnamefont {R.}~\bibnamefont {Baer}}\ and\ \bibinfo {author} {\bibfnamefont {E.}~\bibnamefont {Rabani}},\ }\bibfield  {title} {\bibinfo {title} {{Theory of resonance energy transfer involving nanocrystals: The role of high multipoles}},\ }\href {https://doi.org/10.1063/1.2913247} {\bibfield  {journal} {\bibinfo  {journal} {J. Chem. Phys}\ }\textbf {\bibinfo {volume} {128}},\ \bibinfo {pages} {184710} (\bibinfo {year} {2008})}\BibitemShut {NoStop}%
\bibitem [{\citenamefont {Kruchinin}\ \emph {et~al.}(2008)\citenamefont {Kruchinin}, \citenamefont {Fedorov}, \citenamefont {Baranov}, \citenamefont {Perova},\ and\ \citenamefont {Berwick}}]{kruchinin2008resonant}%
  \BibitemOpen
  \bibfield  {author} {\bibinfo {author} {\bibfnamefont {S.~Y.}\ \bibnamefont {Kruchinin}}, \bibinfo {author} {\bibfnamefont {A.~V.}\ \bibnamefont {Fedorov}}, \bibinfo {author} {\bibfnamefont {A.~V.}\ \bibnamefont {Baranov}}, \bibinfo {author} {\bibfnamefont {T.~S.}\ \bibnamefont {Perova}},\ and\ \bibinfo {author} {\bibfnamefont {K.}~\bibnamefont {Berwick}},\ }\bibfield  {title} {\bibinfo {title} {Resonant energy transfer in quantum dots: Frequency-domain luminescent spectroscopy},\ }\href {https://doi.org/10.1103/PhysRevB.78.125311} {\bibfield  {journal} {\bibinfo  {journal} {Phys. Rev. B}\ }\textbf {\bibinfo {volume} {78}},\ \bibinfo {pages} {125311} (\bibinfo {year} {2008})}\BibitemShut {NoStop}%
\bibitem [{\citenamefont {Baimuratov}\ and\ \citenamefont {H{\"o}gele}(2020)}]{baimuratov2020valley}%
  \BibitemOpen
  \bibfield  {author} {\bibinfo {author} {\bibfnamefont {A.~S.}\ \bibnamefont {Baimuratov}}\ and\ \bibinfo {author} {\bibfnamefont {A.}~\bibnamefont {H{\"o}gele}},\ }\bibfield  {title} {\bibinfo {title} {Valley-selective energy transfer between quantum dots in atomically thin semiconductors},\ }\href {https://doi.org/10.1038/s41598-020-73688-8} {\bibfield  {journal} {\bibinfo  {journal} {Sci. Rep.}\ }\textbf {\bibinfo {volume} {10}},\ \bibinfo {pages} {16971} (\bibinfo {year} {2020})}\BibitemShut {NoStop}%
\bibitem [{\citenamefont {Haug}\ and\ \citenamefont {Koch}(2009)}]{haug2009quantum}%
  \BibitemOpen
  \bibfield  {author} {\bibinfo {author} {\bibfnamefont {H.}~\bibnamefont {Haug}}\ and\ \bibinfo {author} {\bibfnamefont {S.~W.}\ \bibnamefont {Koch}},\ }\href {https://doi.org/10.1142/7184} {\emph {\bibinfo {title} {Quantum Theory of the Optical and Electronic Properties of Semiconductors}}},\ \bibinfo {edition} {5th}\ ed.\ (\bibinfo  {publisher} {WORLD SCIENTIFIC},\ \bibinfo {year} {2009})\BibitemShut {NoStop}%
\end{thebibliography}%

\end{document}


\title{Supplementary Information for ``F\"orster valley-orbit coupling and topological lattice of hybrid moir\'e excitons"}

\author{Huiyuan Zheng}
\affiliation{New Cornerstone Science Laboratory, Department of Physics, University of Hong Kong, Hong Kong, China}
\affiliation{HK Institute of Quantum Science \& Technology, University of Hong Kong, Hong Kong, China}

\author{Ci Li}
\affiliation{School of Physics and Electronics, Hunan University, Changsha
410082, China}

\author{Hongyi Yu}
\affiliation{School of Physics and Astronomy, Sun Yat-sen University (Zhuhai
Campus), Zhuhai 519082, China}

\author{Wang Yao}
\email{wangyao@hku.hk}
\affiliation{New Cornerstone Science Laboratory, Department of Physics, University of Hong Kong, Hong Kong, China}
\affiliation{HK Institute of Quantum Science \& Technology, University of Hong Kong, Hong Kong, China}

\maketitle
\tableofcontents

\section{Supplementary Note 1: Analysis of allowed envelope forms of hybridized intralayer excitons}

In this section, we use symmetry analysis to determine the center-of-mass
(COM) envelope form of intralayer excitons which hybridize with an interlayer
exciton in $s$-type envelope at different high symmetry points. For an
intralayer exciton, its optical selection rule is straightforward: $K (- K)$
valley components only emit $\sigma_+ (\sigma_-)$ circularly polarized light~\cite{xiao2012coupled}. While for an interlayer exciton, the optical selection rule
additionally depends on the local atomic registry, since the electron and hole
are separated in different layers. Due to angular momentum conservation, a
permitted hybridization channel requires the consistant optical selection rule
for an initial and a final state. The interlayer exciton can couple to
distinct light polarization at different high symmetry points, the final
intralayer state should carry different envelope forms to reshape the optical
dipole polarization.

We use $\psi^{\mu \nu}_{m, \tau}$ to denote an exciton from $\tau K$ valley
with its electron and hole constituents labeled by $\mu = c, c'$ and $\nu = v,
v'$, respectively. The prime notation represents that the band comes from the bottom layer. Its
envelope carries an azimuthal quantum number $m$. There are three parts
in the exciton wave packets depending on the COM coordinates: the COM envelope
function $W_m (\mathbf{R})$, the electron, and the hole Bloch constituents~\cite{yu2017moire}. At high symmetry points, the envelope function are the
eigenfunction of the threefold rotation operator $\hat{C}_3$ with
$m$-dependent eigenvalues:
\begin{eqnarray}
  \hat{C}_3 W_m (\mathbf{R}) & = & W_m (\hat{C}_3^{- 1} \mathbf{R}) = e^{-
  i \frac{2 \pi}{3} m} W_m (\mathbf{R}).
  \label{eqS_C3envelope}
\end{eqnarray}
Thus, the envelope function has $s$-type ($m = 0$), $p_+$-type ($m = + 1$) and
$p_-$-type ($m = - 1$). 

The electron and hole constituents inherit the
conduction and valence Bloch bands at $\tau K$ valley, respectively. 
In R-stacking bilayers, the interlayer tunneling matrix element between conduction bands and between valence bands can be approximated as~\cite{rafi2011moire,wang2017interlayer}
\begin{eqnarray}
    T_{v v' (c c'),\tau}(\mathbf{R}) = t_{v(c)} (e^{i \tau \delta \mathbf{K} \cdot \mathbf{R}} + e^{i \tau \hat{C}_3 \delta \mathbf{K} \cdot \mathbf{R}} + e^{i \tau \hat{C}_3^2 \delta \mathbf{K} \cdot \mathbf{R}}),
\end{eqnarray}
where $\delta \mathbf{K}$ denotes the valley mismatch between layers. We use R-stacking TMD homobilayer as an example. There are a pair of high
symmetry points, i.e. B (MX) and C (XM), in one supercell, and they are related by the out-of-plane mirror symmetry. At the vicinity $\mathbf{R} = \mathbf{R}_\text{B(C)} + \mathbf{r}$, the eigenvalue of the interlayer matrix element under $C_3$ takes the form (c.f. Fig. 1b in the main text):
\begin{eqnarray}
    \hat{C}_3 T_{v v'(c c'),\tau}(\mathbf{r}; \mathbf{R}_\text{B}) \hat{C}_3^\dag &=& T_{v v'(c c'),\tau}(\hat{C}_3^{-1} \mathbf{r}; \mathbf{R}_\text{B}) = e^{-i\frac{2\pi}{3}\tau} T_{v v'(c c'),\tau}(\mathbf{r}; \mathbf{R}_\text{B}), \\
    \hat{C}_3 T_{v v'(c c'),\tau}(\mathbf{r}; \mathbf{R}_\text{C}) \hat{C}_3^\dag &=& T_{v v'(c c'),\tau}(\hat{C}_3^{-1} \mathbf{r}; \mathbf{R}_\text{C}) = e^{+i\frac{2\pi}{3}\tau} T_{v v'(c c'),\tau}(\mathbf{r}; \mathbf{R}_\text{C}).
\end{eqnarray}
Considering the dominant valence band hopping~\cite{wang2017interlayer}, the hybridization matrix element should satisfy the rotational symmetry:
\begin{eqnarray}
\braket{\psi^{cv}_{m,\tau}|\hat{T}|\psi^{cv'}_{0,\tau}} &=& 
\braket{W_m|T_{vv',\tau}(\mathbf{R}_\text{B})|W_0} = \braket{\hat{C}_3 W_m| \hat{C}_3 T_{vv',\tau}(\mathbf{R}_\text{B}) \hat{C}_3^\dag| \hat{C}_3 W_0} = e^{i\frac{2\pi}{3}(m-\tau)}\braket{\psi^{cv}_{m,\tau}|\hat{T}|\psi^{cv'}_{0,\tau}}, \\
\braket{\psi^{c'v'}_{m,\tau}|\hat{T}|\psi^{c'v}_{0,\tau}} &=& 
\braket{W_m|T_{vv',\tau}^{*}(\mathbf{R}_\text{C})|W_0} = \braket{\hat{C}_3 W_m| \hat{C}_3 T_{vv',\tau}^{*}(\mathbf{R}_\text{C}) \hat{C}_3^\dag| \hat{C}_3 W_0} = e^{i\frac{2\pi}{3}(m-\tau)}\braket{\psi^{c'v'}_{m,\tau}|\hat{T}|\psi^{c'v}_{0,\tau}}.
\end{eqnarray}
Therefore, to keep the hybridization finite, a $p_+$- ($p_-$-) type envelope is
required for $K (- K)$ valley, regardless of B or C sites. 

In short, to couple with $\sigma_-$- and $z$-polarized light, an intralayer
exciton from $K$ valley, which intrinsically emit $\sigma_+$ light, should
carries $p_+$- and $p_-$-type envelope to compensate the angular momentum.
Therefore, the required envelope forms of hybridized intralayer components in
Table. I in the main text can be found in a similar way.

\begin{figure}
\includegraphics[width=0.8\textwidth]{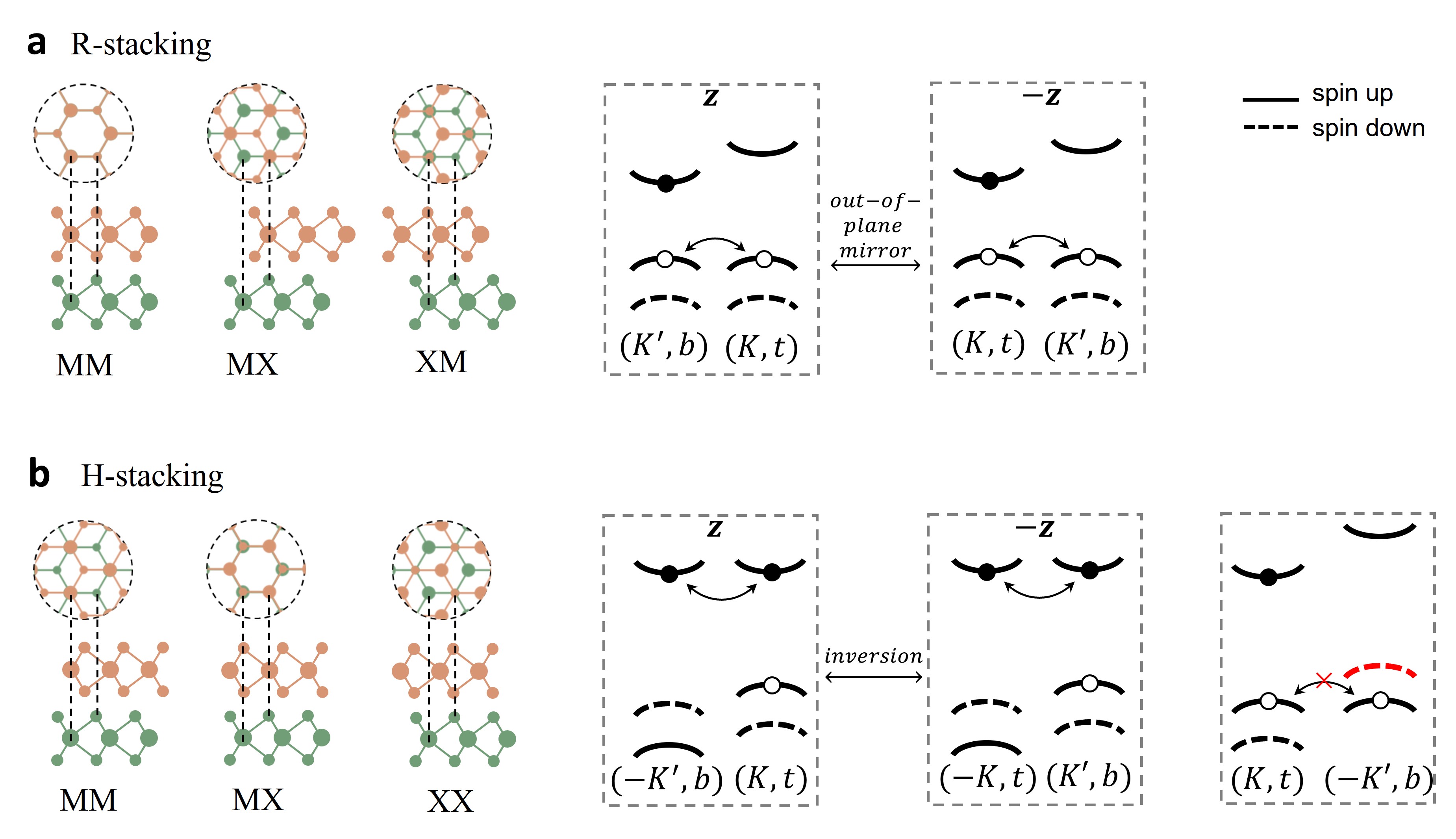}%
\caption{$\mathbf{a}$. For R-stacking bilayers, the top view and side view of local atomic registries at high symmetry points (left panel), and configuration in momentum space of hybrid excitons (right panel). Black (white) dots denote the electron (hole). For each band, valley and layer indices are denoted by ($K/-K, t/b$), and the prime notation represents that the valley comes from bottom layer. 
$\mathbf{b}$. Similar information for H-stacking bilayers. The right-most panel shows the configuration of hole hopping hybridization, which results in a B-type exciton.}
\label{figS_configuration}
\end{figure}

To facilitate understanding, Fig.~\ref{figS_configuration}a,b show the configurations in momentum
space of hybrid exciton with two opposite electric dipoles in R- and H-
stacking. The pair of hybrid excitons with opposite electric dipoles are linked by out-of-plane mirror and inversion symmetry in R- and H-stacking, respectively. Their selection rules thus will not change. The spin shown here is for electronic
bands, and the spin for holes should be reversed. Notably, in R-stacking
bilayers, high symmetry points MX and XM will exchange under out-of-plane mirror
symmetry, which results in the difference of the optical selection rule for
hybrid excitons with two electric dipoles. While for H-stacking, all high
symmetry points remain unchanged under inversion symmetry. Lastly, the
hole side hopping in H-stacking bilayers is not discussed here, since it
will end up with a B-type exciton (hole is not on the band edge).

\section{Supplementary Note 2: Hybridization of exciton wavepackets in twisted bilayer MoTe$_2$}

Here we consider the hybridization between an interlayer exciton in $s$-type
wavepacket and an intralayer exciton wavepacket centered at the same location
$\mathbf{r}_\text{c}$:
\begin{eqnarray}
  \ket{\psi_{m, \tau}^{\text{intra}}(\mathbf{r}_\text{c})}  & = & \sum_{\mathbf{q}}
  e^{- i \mathbf{q} \cdot \mathbf{r}_\text{c}} W_m (\mathbf{q}) |
  X^{\text{intra}}_{\tau, \mathbf{q}} \rangle, \\
  \ket{\psi_{0, \tau}^{\text{inter}}(\mathbf{r}_\text{c})}  & = & \sum_{\mathbf{q}}
  e^{- i (\mathbf{q} - \tau\delta \mathbf{K}) \cdot \mathbf{r}_\text{c}} W_0
  (\mathbf{q}) | X^{\text{inter}}_{\tau, \mathbf{q}} \rangle, 
\end{eqnarray}
where the phase factor $e^{- i \tau \delta \mathbf{K} \cdot \mathbf{r}_\text{c}}$ of
the interlayer exciton originates from the valley mismatch $\tau \delta
\mathbf{K}$ between the electron and hole separated in different layers. 
Under this gauge choice, the interlayer exciton wavepackets will carry a $\mathbf{r}_\text{c}$-dependent complex phase (c.f. Fig.3a,c in the main text). $|
X_{\tau, \mathbf{q}} \rangle$ is the momentum eigenstate of the exciton. The envelope form can be approximated by the eigenfunctions of 2D harmonic trap:
$W_m (\mathbf{q}) = f_m (q) e^{i m \varphi_q} = \sqrt{\frac{4 \pi}{\Omega}}
w^{| m | + 1} q^{| m |} e^{- w^2 q^2 / 2} e^{i m \varphi_q}$, where $\Omega$
denotes the normalization area and $w$ describes the wavepacket width. Thus, the
exciton wavepackets are normalized:
\begin{eqnarray}
  \int_{\Omega} \braket{ \psi_{m, \tau}^{\text{intra}} | \psi_{m, \tau}^{\text{intra}} } d \mathbf{R} & = &
  \sum_{\mathbf{q}, \mathbf{q}'} e^{- i (\mathbf{q} - \mathbf{q}')
  \cdot \mathbf{r}_\text{c}} W^{\ast}_m (\mathbf{q}) W_m (\mathbf{q}') \int_{\Omega}
  (X_{\tau, \mathbf{q}}^{\text{intra}})^{\ast} X_{\tau,
  \mathbf{q}'}^{\text{intra}} d \mathbf{R} \nonumber\\
  & = & \sum_{\mathbf{q}, \mathbf{q}'} e^{- i (\mathbf{q} -
  \mathbf{q}') \cdot \mathbf{r}_\text{c}} W^{\ast}_m (\mathbf{q}) W_m
  (\mathbf{q}') \delta_{\mathbf{q}, \mathbf{q}'} = \sum_{\mathbf{q}} |
  W_m (\mathbf{q}) |^2 = \frac{\Omega}{4 \pi^2} \int d \mathbf{q} | W_m
  (\mathbf{q}) |^2 = 1 .
\end{eqnarray}
The hybridization of inter- and intra-layer exciton at $\mathbf{r}_\text{c}$ is
\begin{eqnarray}
  \langle \psi^{\text{intra}}_{m, \tau} (\mathbf{r}_\text{c}) | \hat{T} | \psi_{0,
  \tau}^{\text{inter}} (\mathbf{r}_\text{c}) \rangle & = & \sum_{ \mathbf{q},
  \mathbf{q'}} e^{i (\mathbf{q} - \mathbf{q'} + \delta \mathbf{K})
  \cdot \mathbf{r}_\text{c}} W \mathcal{}_m^{\ast} (\mathbf{q}) W_0
  (\mathbf{q'}) \langle X^{\text{intra}}_{\tau,\mathbf{q}} | \hat{T} |
  X_{\tau,\mathbf{q'}}^{\text{inter}} \rangle .
\end{eqnarray}
We consider the dominant valence band hopping in the hopping matrix element
between momentum eigenstates and keep the leading order $\langle
X^{\text{intra}}_{\tau,\mathbf{q}} | \hat{T} | X_{\tau,\mathbf{q'}}^{\text{inter}}
\rangle = - t_v (\delta_{\mathbf{q'}, \tau \delta \mathbf{K} + \mathbf{q}} +
\delta_{\mathbf{q'}, \hat{C}_3 \tau \delta \mathbf{K} + \mathbf{q}} +
\delta_{\mathbf{q'}, \hat{C}_3^2 \tau \delta \mathbf{K} + \mathbf{q}})$~\cite{wang2017interlayer}.
\begin{eqnarray}
  \langle \psi^{\text{intra}}_{m, \tau} (\mathbf{r}_\text{c}) | \hat{T} | \psi_{0,
  \tau}^{\text{inter}} (\mathbf{r}_\text{c}) \rangle & = & \sum_{ \mathbf{q},
  \mathbf{q'}} e^{i (\mathbf{q} - \mathbf{q'} + \tau \delta \mathbf{K})
  \cdot \mathbf{r}_\text{c}} W \mathcal{}_m^{\ast} (\mathbf{q}) W_0
  (\mathbf{q'}) \langle X^{\text{intra}}_{\tau,\mathbf{q}} | \hat{T} |
  X_{\tau,\mathbf{q'}}^{\text{inter}} \rangle \nonumber\\
  & = & - t_v \sum_{\mathbf{q}} W_m^{\ast} (\mathbf{q}) W_0 (\mathbf{q}
  + \delta \mathbf{K}) + e^{- i \mathbf{g}_3 \cdot \mathbf{r}_\text{c}}
  W_m^{\ast} (\mathbf{q}) W_0 (\mathbf{q} + \hat{C}_3 \delta \mathbf{K})  \nonumber\\
  & + & e^{i \mathbf{g}_2 \cdot \mathbf{r}_\text{c}} W_m^{\ast} (\mathbf{q})
  W_0 (\mathbf{q} + \hat{C}_3^2 \delta \mathbf{K}) .
\end{eqnarray}
Reorder the summation of $\mathbf{q}$ and use Eq.(\ref{eqS_C3envelope}), we get
\begin{eqnarray}
  \langle \psi^{\text{intra}}_{m, \tau} (\mathbf{r}_\text{c}) | \hat{T} | \psi_{0,
  \tau}^{\text{inter}} (\mathbf{r}_\text{c}) \rangle & = & - \Gamma_{m,
  \mathbf{r}_\text{c}} t_v \sum_{\mathbf{q}} W_m^{\ast} (\mathbf{q}) W_0
  (\mathbf{q} + \delta \mathbf{K}),
\end{eqnarray}
$\Gamma_{m, \mathbf{r}_\text{c}} = 0$ or $3$ relies on whether the hybridization is
permitted or not.

Therefore, in t-MoTe$_2$, a pair of hybrid excitons are located at B and C sites
which comprise of an interlayer exciton with a perturbatively hybridized
intralayer exciton.
\begin{eqnarray}
  \ket{\Psi^{\text{hy}}_{\text{B}, \tau}} & = & \ket{\psi^{c v'}_{0, \tau} (\mathbf{R}_\text{B})} + \mathcal{W}_m
  \ket{\psi^{c v}_{m, \tau}(\mathbf{R}_\text{B})}, \nonumber \\
  \ket{\Psi^{\text{hy}}_{\text{C}, \tau}} & = & \ket{\psi^{c' v}_{0, \tau}(\mathbf{R}_\text{C})} + \mathcal{W}_m
  \ket{\psi_{m, \tau}^{c' v'}(\mathbf{R}_\text{C})}, 
\end{eqnarray}
with the azimuthal index locked with the valley index $m = \tau = \pm 1$ at
$K$ and $- K$ valley, respectively. The hybridization weight reads
\begin{eqnarray}
  \mathcal{W}_{\pm 1} & = & \frac{- 3 t_v}{\delta E} \sum_{\mathbf{q}}
  W_{\pm 1}^{\ast} (\mathbf{q}) W_0 (\mathbf{q} + \delta \mathbf{K}), 
\end{eqnarray}
where $\delta E$ represents the energy offset between inter- and intralayer components.

The hybridization coefficients $\mc{W}$ can be evaluated. The widths of wavepackets of inter- and intralayer
excitons usually differs, which are related to the moire trapping strength.
Here, we approximate $w_{\text{intra}} = \sqrt{2} w_{\text{inter}}$. Employing the eigenfucntion of 2D harmonic trap, the hybridization weight can be written as
\begin{eqnarray}
  \mathcal{W}_{\pm 1} & = & \mp i \frac{8 \sqrt{2} \pi}{9} \frac{w}{a_M} \exp
  \left( - \frac{8 \pi^2}{27} \frac{w^2}{a_M^2} \right) \times
  \frac{t_v}{\delta E} = \mp i \mathcal{C} \times \frac{t_v}{\delta E} .
\end{eqnarray}
The amplitude of F\"orster coupling between hybrid states can be obtained by
\begin{eqnarray}
  \langle \Psi^{\text{hy}}_{\text{B}, \pm K} | \hat{J} | \Psi^{\text{hy}}_{\text{C},
  \pm K} \rangle & = & | \mathcal{W}_{\pm 1} |^2 \langle \psi^{c v}_{\pm K} |
  \hat{J} | \psi^{c' v'}_{\pm K} \rangle = \frac{t_v^2}{\delta E^2}
  \mathcal{C}^2 \langle \psi^{c v}_{\pm K} | \hat{J} | \psi^{c' v'}_{\pm K}
  \rangle, \\
  \langle \Psi^{\text{hy}}_{\text{B}, \pm K} | \hat{J} | \Psi^{\text{hy}}_{\text{C},
  \mp K} \rangle & = & \mathcal{W}_{\pm 1}^{\ast} \mathcal{W}_{\mp 1} \langle
  \psi^{c v}_{\pm K} | \hat{J} | \psi^{c' v'}_{\mp K} \rangle = -
  \frac{t_v^2}{\delta E^2} \mathcal{C}^2 \langle \psi^{c v}_{\pm K} | \hat{J}
  | \psi^{c' v'}_{\mp K} \rangle, 
\end{eqnarray}
The negative sign from the intervalley term is contained in Eq. (2) in the
main text.

\section{Supplementary Note 3: F\"orster coupling between intralayer exciton wavepackets}
The F\"orster coupling of exciton wave packets can be represented by $J^{\tau', n', m' }_{ \tau, n, m} = \braket{\tau', n', m' |\hat{J}| \tau, n, m}$, where an exciton is characterized by its valley $\tau$, resided layer $n$, and angular momentum of the COM wave function $m$. If the excitons behave as a plane wave with a conserved COM momentum but not a wave packet, the form of F\"orster coupling has been derived~\cite{li2023cross}. When they behave like 0D-type wave packets, the wave functions are reshaped by $\ket{\psi_{\tau,n,m}(\mathbf{r}_\text{c})} = \sum_{\mathbf{q}} e^{-i \mathbf{q} \cdot \mathbf{r}_\text{c}} f_m(q) e^{i m \varphi_q} \ket{\tau,n,\mathbf{q}}$. $\mathbf{r}_\text{c}$ denotes the wavepacket center, and $f_m(q)$ is a real function accounting for the radial dependence. By assuming the excitonic wave packets are eigenfunctions of 2D harmonic traps. Confined by threefold rotational symmetry, the angular momentum of wave packets only takes $m = 0, \pm 1$, and we consider the lowest three states and neglect higher states. The F\"orster coupling between two momentum eigenstates takes the form
\begin{eqnarray}
    \hat{J}^{n',n}_{\tau',\tau}(\mathbf{q}) = (-1)^{\frac{\Delta \tau}{2}} J\frac{q}{K} e^{-i (\tau' \theta_{n'} - \tau \theta_{n})} e^{-(z_{n'} - z_n)q} e^{-i \Delta \tau \varphi}. 
    \label{eqS_Coulombexchange}
\end{eqnarray}
where $J= 1$eV~\cite{yu2014dirac,liu2025direct} denotes the electron-hole exchange strength, $K=4\pi/3a$ represents the length to Brillouin zone corner of TMD, with $a$ being the lattice constant. Thus, the F\"orster coupling between two wave packets can be calculated as

\begin{widetext}
\begin{eqnarray}
    J^{\tau', n', m' }_{\tau, n, m}
    &=& \sum_{\mathbf{q}} e^{-i \mathbf{q} \cdot \mathbf{r}} W^*_{m'}(\mathbf{q}) W_{m}(\mathbf{q}') \delta_{\mathbf{q},\mathbf{q}'} \left[ (-1)^{\frac{\Delta \tau}{2}} J\frac{q}{K} e^{-i (\tau' \theta_{n'} - \tau \theta_{n})} e^{-q z} e^{-i \Delta \tau \varphi} \right] \notag \\
    &=& (-1)^{\frac{\Delta \tau}{2}} e^{-i (\tau' \theta_{n'} - \tau \theta_{n})} \frac{J}{\pi K} w^{|m|+|m'|+2} \int d\mathbf{q} \, q^{|m|+|m'|+1} e^{-w^2 q^2} e^{-q z} e^{-i \mathbf{q} \cdot \mathbf{r}} e^{-i(\Delta \tau + \Delta m)\varphi} \notag \\
    &=& (-1)^{\frac{\Delta \tau}{2}} e^{-i (\tau' \theta_{n'} - \tau \theta_{n})} \frac{J}{\pi K} w^{|m|+|m'|+2} \nonumber \\
    & & \qquad \qquad *\int_0^{+\infty} dq \, q^{|m|+|m'|+2} e^{-w^2 q^2} e^{-q z} \int_0^{2\pi} d\varphi \, e^{-i q r \, \cos(\varphi-\varphi_r)} e^{-i(\Delta \tau + \Delta m)\varphi} \notag \\
    &=& (-1)^{\frac{\Delta \tau}{2}} e^{-i (\tau' \theta_{n'} - \tau \theta_{n})} \frac{J}{\pi K} w^{|m|+|m'|+2} e^{-i(\Delta \tau + \Delta m)\varphi_r} \nonumber\\
    & & \qquad \qquad * \int_0^{+\infty} dq \, q^{|m|+|m'|+2} e^{-w^2 q^2} e^{-q z} \int_0^{2\pi} d\gamma \, e^{-i q r \, \cos \gamma} e^{-i(\Delta \tau + \Delta m)\gamma} \notag \\
    &=& (-1)^{\frac{\Delta \tau}{2}} e^{-i (\tau' \theta_{n'} - \tau \theta_{n})} \frac{2 J}{w K} e^{-i(\Delta \tau + \Delta m)\varphi_r} \nonumber\\
    & & \qquad \qquad * \int_0^{+\infty} du \, u^{|m|+|m'|+2} e^{-u^2} e^{-\frac{z}{w}u} \left[\frac{1}{2\pi} \int_0^{2\pi} d\gamma \, e^{-i \frac{r}{w} u \, \cos \gamma} e^{-i(\Delta \tau + \Delta m)\gamma} \right].
\end{eqnarray}
\end{widetext}
The azimuthal integral in the last row has two equivalent forms:

\begin{eqnarray}
     \frac{1}{2\pi} \int_0^{2\pi} d\gamma \, e^{-i \frac{r}{w} u \, \cos \gamma} e^{-i(\Delta \tau + \Delta m)\gamma}
    = e^{-i |\Delta \tau + \Delta m| \frac{\pi}{2}} \mc{J}_{ |\Delta \tau + \Delta m|} \left( \frac{r}{w}u \right) 
    = e^{-i (\Delta \tau + \Delta m) \frac{\pi}{2}} \mc{J}_{\Delta \tau + \Delta m} \left( \frac{r}{w}u \right),
\end{eqnarray}
where $\mc{J}$ is the Bessel function of the first kind. Thus, the F\"orster coupling has the form:
\begin{eqnarray}
    J^{\tau', n', m' }_{\tau, n, m} &=& e^{-i \Delta m \frac{\pi}{2}} e^{-i(\Delta \tau + \Delta m)\varphi_r} e^{-i (\tau' \theta_{n'} - \tau \theta_{n})}  \mathfrak{J}^{\tau',m'}_{\tau,m} \left(\frac{r}{w},\frac{z}{w} \right). \label{eqS_ForsterCoupling}
\end{eqnarray}
The strength of the F\"orster coupling $\mathfrak{J}$ relies on $\Delta \tau + \Delta m$ and individual $m,m'$, which is contributed by the radial function integral.
\begin{eqnarray}
    \mathfrak{J}_{\tau,m}^{\tau',m'}\left( \frac{r}{w}, \frac{z}{w} \right) = \frac{2J}{wK} \int_0^{+\infty} du \, u^{|m|+|m'|+2} e^{-u^2} e^{-\frac{z}{w}u} \mc{J}_{\Delta \tau + \Delta m} \left( \frac{r}{w}u \right).
    \label{eqS_Radialintegral}
\end{eqnarray}

The hermiticity of the result can be guaranteed. Consider an initial state $\ket{\psi_{\tau,n,m}(\mathbf{r}_\text{c})}$ and a final state $\ket{\psi_{\tau',n',m'}(\mathbf{r}'_\text{c})}$, the F\"orster coupling between them can be obtained by Eq.(\ref{eqS_ForsterCoupling}). We now exchange the initial and final state, $\tau \leftrightarrow \tau'$, $m \leftrightarrow m'$,  $\mathbf{r} \rightarrow -\mathbf{r}$, the F\"orster coupling becomes

\begin{eqnarray}
    J^{\tau, n, m }_{\tau', n', m'} &=& \sum_{\mathbf{q}} e^{+i \mathbf{q} \cdot \mathbf{r}} W^*_{m}(\mathbf{q}) W_{m'}(\mathbf{q}') \delta_{\mathbf{q},\mathbf{q}'} \left[ (-1)^{\frac{-\Delta \tau}{2}} J\frac{q}{K} e^{-i (\tau \theta_{n} - \tau' \theta_{n'})} e^{-q z} e^{-i (-\Delta \tau) \varphi} \right] \notag \\
    &=& (-1)^{\frac{-\Delta \tau}{2}} e^{-i (\tau \theta_{n} - \tau' \theta_{n'})} \frac{2 J}{w K} e^{+i(\Delta \tau + \Delta m)\varphi_r} \nonumber \\
    & & \qquad \qquad *\int_0^{+\infty} du \, u^{|m|+|m'|+2} e^{-u^2} e^{-\frac{z}{w}u} \left[\frac{1}{2\pi} \int_0^{2\pi} d\gamma \, e^{+i \frac{r}{w} u \, \cos \gamma} e^{+i(\Delta \tau + \Delta m)\gamma} \right] \notag \\
    &=& (-1)^{\frac{-\Delta \tau}{2}} e^{-i (\tau \theta_{n} - \tau' \theta_{n'})} \frac{2 J}{w K} e^{+i(\Delta \tau + \Delta m)\varphi_r} \nonumber \\
    & & \qquad \qquad *\int_0^{+\infty} du \, u^{|m|+|m'|+2} e^{-u^2} e^{-\frac{z}{w}u} \left[ e^{i(\Delta \tau + \Delta m)\frac{\pi}{2}} \mc{J}_{\Delta \tau + \Delta m} \left( \frac{r}{w}u \right) \right] \notag \\
    &=& e^{i \Delta m \frac{\pi}{2}} e^{i(\Delta \tau + \Delta m)\varphi_r} e^{-i (\tau \theta_{n} - \tau' \theta_{n'})}  \mathfrak{J}^{\tau',m'}_{\tau,m} \left(\frac{r}{w},\frac{z}{w} \right),   
\end{eqnarray}
The result is a complex conjugate of $J^{\tau', n', m' }_{\tau, n, m}$. In short, in Eq.(\ref{eqS_ForsterCoupling}) (also Eq. 1 in the main text), when $\Delta \tau + \Delta m$ takes an odd value, the exchange of initial and final states will result in a sign change $e^{-i(\Delta \tau + \Delta m) \varphi_r} \rightarrow -e^{-i(\Delta \tau + \Delta m) \varphi_r}$ and $\mathfrak{J}_{\tau',m'}^{\tau,m} = -\mathfrak{J}_{\tau,m}^{\tau',m'}$ at the same time.

\subsection{Nonradiative multipole-multipole interaction as an asymptotic form}
\begin{figure}
\includegraphics[width=0.7\textwidth]{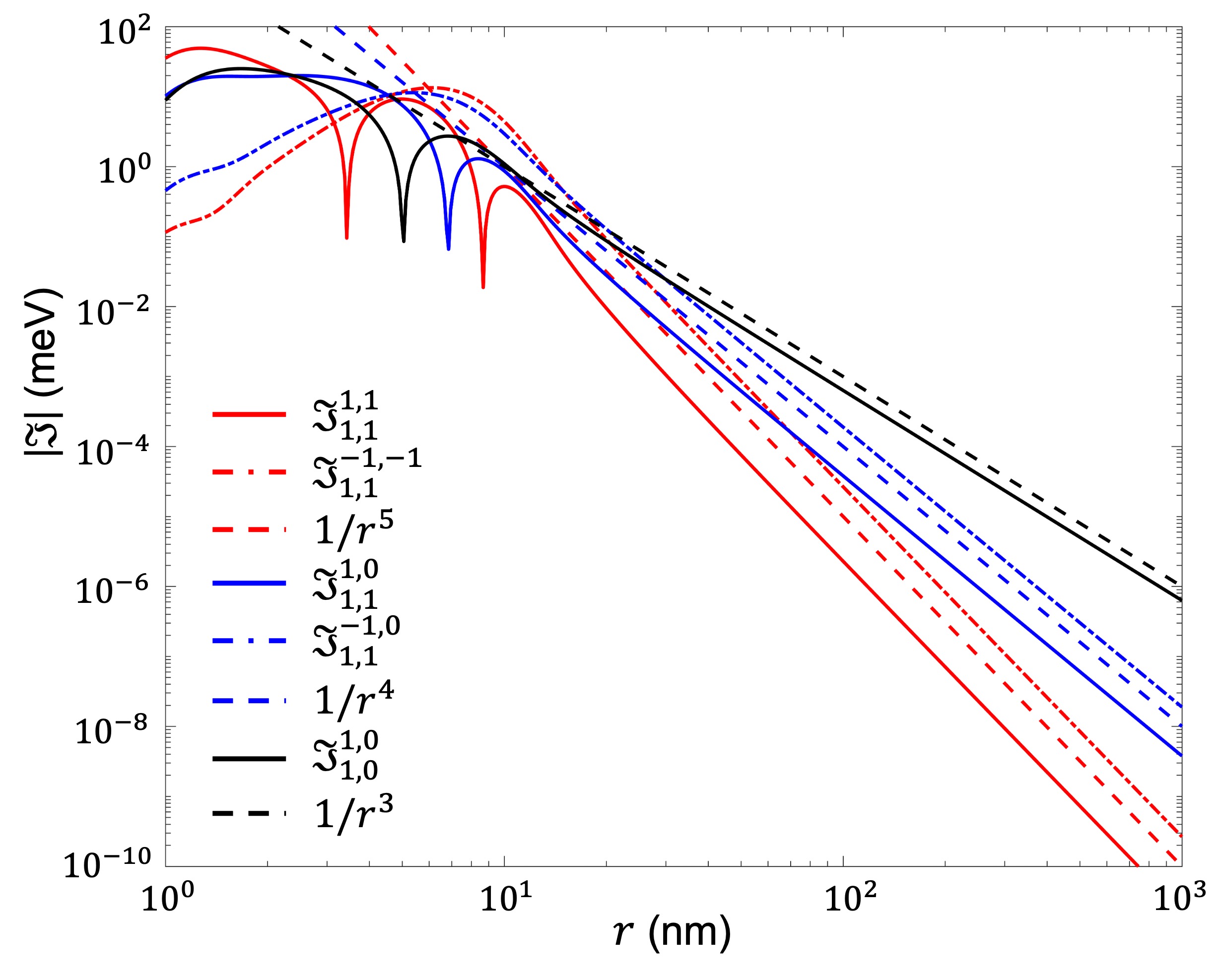}%
\caption{Asymptotic form of F\"orster coupling amplitude.
$|\mathfrak{J}_{\tau,m}^{\tau',m'}|$ with representative $\tau$ and $m$ are shown, indicating F\"orster coupling is asymptotic to nonradiative multipole-multipole interaction. Dashed lines in power law decay are shown for auxiliary. $z=0$ and $w=2$nm are used for calculation.}
\label{figS_Asymptotic}
\end{figure}

As the distance $r$ grows, the radial integral (Eq.~\ref{eqS_Radialintegral}) can be evaluated by substituting
the variable $\lambda = u r$, and it shows an asymptotic form
\begin{eqnarray}
  \mathfrak{J}_{\tau, m}^{\tau', m'} \left( \frac{r}{w}, \frac{z}{w} \right) &
  = & \frac{2 J}{w K} \int^{+ \infty}_0 d u \, u^{| m | + | m' | + 2} e^{- u^2}
  e^{- \frac{z}{w} u} \mathcal{J}_{\Delta \tau + \Delta m} \left( \frac{r}{w}
  u \right) \notag\\
  & = & \frac{2 J}{w K} \left( \frac{w}{r} \right)^{| m | + | m' | + 3} \int_0^{+ \infty} d
  \lambda \, \lambda^{| m | + | m' | + 2} e^{- \frac{w^2}{r^2} \lambda^2} e^{-
  \frac{z}{r} \lambda} \mathcal{J}_{\Delta \tau + \Delta m} \left(
  \lambda \right).
\end{eqnarray}
Since the exponential terms in the last integral decay rapidly when $\lambda \rightarrow + \infty$,
the value of the integral are dominantly contributed by integrating $\lambda$ to $O(\frac{r}{w})$,
the integral will converge to a finite value, and become independent to $r$ as $r > 10 w$. Therefore,
the amplitude of F\"orster coupling will decay in $r^{- (3 + | m | + | m' |)}$,
indicating its asymptotic form as a nonradiative multipole-multipole
interaction~\cite{baer2008theory, kruchinin2008resonant, baimuratov2020valley}. We show several F\"orster coupling amplitudes as a function of $r$ in their asymptotic forms in Fig.~\ref{figS_Asymptotic}.

Specifically, the Coulomb exchange of two exciton momentum eigenstates (Eq.~\ref{eqS_Coulombexchange})
can be recast in the form of a pair of optical dipole
\begin{eqnarray}
  \hat{J}_{\tau' \tau}^{n' n} (\mathbf{q}) & = & J \frac{q}{K} e^{- q (z_{n'} -
  z_n)} (e^{i \tau' \theta_{n'}}  \mathbf{d}_{\text{op}, \tau'}^{\dagger}
  \cdot \hat{\mathbf{q}})^{\ast} (e^{i \tau \theta_n} 
  \mathbf{d}_{\text{op}, \tau}^{\dagger} \cdot \hat{\mathbf{q}}),
\end{eqnarray}
where $\mathbf{d}_{\text{op}, \tau} = \left( \begin{array}{c}
  i \tau\\
  1
\end{array} \right)$ describes the orientation of optical dipole of exciton,
written as a column vector, its amplitude is contained in $J$.

The envelope function $W_m (\mathbf{q}) = f_m (q) e^{i m \varphi_q}$
contains a phase factor, indicating that $W_m (\mathbf{q})$ also carries a
dipole moment for $m = \pm 1$:
\begin{eqnarray}
  W_m (\mathbf{q}) & = & f_m (q) e^{i m \varphi} = f_m (q)
  (\mathbf{d}_{\text{env}, m}^T \cdot \hat{\mathbf{q}}),
\end{eqnarray}
where $\mathbf{d}_{\text{env}, m} = \left( \begin{array}{c}
  1\\
  - i m
\end{array} \right)$, describes the orientation of the envelope dipole moment.
When $m = 0$, the $s$-type wave function indicates the envelope carries a
monopole, and it will carry a quadrupole when $m = \pm 2$. Thus, as $r > 10 w$ , the F\"orster coupling (Eq.) can be discussed in three cases
with the constraint of $C_3$:

i) $m' = m = 0$
\begin{eqnarray}
  \mathcal{J}_{\tau, n, 0}^{\tau', n', 0} (r) & = & \sum_{\mathbf{q}} e^{- i
  \mathbf{q} \cdot \mathbf{r}} | f_0 (q) |^2 J_{\tau' \tau}^{n' n}(\mathbf{q}) \notag\\
  & = & \sum_{\mathbf{q}} e^{- i \mathbf{q} \cdot \mathbf{r}} | f_0 (q)
  |^2 J \frac{q}{K} e^{- q (z_{n'} - z_n)} (e^{i \tau' \theta_{n'}} 
  \mathbf{d}_{\text{op}, \tau'}^{\dagger} \cdot
  \hat{\mathbf{q}})^{\ast} (e^{i \tau \theta_n} 
  \mathbf{d}_{\text{op}, \tau}^{\dagger} \cdot \hat{\mathbf{q}}) \notag\\
  & = & e^{- i (\tau' - \tau) \varphi_r} e^{- i (\tau' \theta_{n'} - \tau
  \theta_n)} \mathfrak{J}_{\tau, 0}^{\tau', 0} \left( \frac{r}{w}, \frac{z}{w}
  \right) \notag\\
  & = & ( \mathbf{d}_{\tau', n', 0}^{\dagger} \cdot
  \hat{\mathbf{r}})^{\ast} ( \mathbf{d}_{\tau, n, 0}^{\dagger} \cdot
  \hat{\mathbf{r}}) \mathfrak{J}_{\tau, 0}^{\tau', 0} \left(
  \frac{r}{w}, \frac{z}{w} \right) \notag\\
  & \sim & \frac{1}{r^3} ( \mathbf{d}_{\tau', n', 0}^{\dagger} \cdot
  \hat{\mathbf{r}})^{\ast} ( \mathbf{d}_{\tau, n, 0}^{\dagger} \cdot
  \hat{\mathbf{r}}),
\end{eqnarray}
where $\mathbf{d}_{\tau, n, 0} = e^{- i \tau \theta_n} 
\mathbf{d}_{\text{op}, \tau}^{}$, and $\hat{\mathbf{r}}$ is the unit
vector co-directional with $\mathbf{r}$.

ii) $m' = \pm 1, m = 0$
\begin{eqnarray}
  \mathcal{J}_{\tau, n, 0}^{\tau', n', \pm 1} (r) & = & \sum_{\mathbf{q}}
  e^{- i \mathbf{q} \cdot \mathbf{r}} f_{\pm 1}^{\ast} (q) f_0 (q)
  (\mathbf{d}_{\text{env}, \pm 1}^T \cdot \hat{\mathbf{q}})^{\ast}
  J_{\tau' \tau}^{n' n}(\mathbf{q}) \notag\\
  & = & (\hat{\mathbf{r}}^T \cdot e^{i \tau' \theta_{n'}}
  \mathbf{d}_{\text{env}, \pm 1}  \mathbf{d}_{\text{op}, \tau'}^{\dagger}
  \cdot \hat{\mathbf{r}})^{\ast} (e^{i \tau \theta_n} 
  \mathbf{d}_{\text{op}, \tau}^{\dagger} \cdot \hat{\mathbf{r}})
  \mathfrak{J}_{\tau, 0}^{\tau', \pm 1} \left( \frac{r}{w}, \frac{z}{w}
  \right) \notag\\
  & \sim & \frac{1}{r^4} (\hat{\mathbf{r}}^T \cdot
  \mathbf{\Theta}_{\tau', n', \pm 1} \cdot \hat{\mathbf{r}})^{\ast} (
  \mathbf{d}_{\tau, n, 0}^{\dagger} \cdot \hat{\mathbf{r}}),
\end{eqnarray}
where $\mathbf{\Theta}_{\tau', n', \pm 1} = e^{i \tau' \theta_{n'}}
\mathbf{d}_{\text{env}, \pm 1}  \mathbf{d}_{\text{op}, \tau'}^{\dagger}$
is a tensor, featuring the quadrupole of exciton wavepacket.

iii) $m' = \pm 1, m = \pm 1$
\begin{eqnarray}
  \mathcal{J}_{\tau, n, \pm 1}^{\tau', n', \pm 1} (r) & = &
  \sum_{\mathbf{q}} e^{- i \mathbf{q} \cdot \mathbf{r}} f_{\pm 1}^{\ast}
  (q) f_{\pm 1} (q) (\mathbf{d}_{\text{env}, \pm 1}^T \cdot
  \hat{\mathbf{q}})^{\ast} (\mathbf{d}_{\text{env}, \pm 1}^T \cdot
  \hat{\mathbf{q}}) J_{\tau' \tau}^{n' n}(\mathbf{q}) \notag\\
  & = & (\hat{\mathbf{r}}^T \cdot e^{i \tau' \theta_{n'}}
  \mathbf{d}_{\text{env}, \pm 1}  \mathbf{d}_{\text{op}, \tau'}^{\dagger}
  \cdot \hat{\mathbf{r}})^{\ast} (\hat{\mathbf{r}}^T \cdot e^{i
  \tau \theta_n} \mathbf{d}_{\text{env}, \pm 1}  \mathbf{d}_{\text{op},
  \tau}^{\dagger} \cdot \hat{\mathbf{r}}) \mathfrak{J}_{\tau, \pm
  1}^{\tau', \pm 1} \left( \frac{r}{w}, \frac{z}{w} \right) \notag\\
  & \sim & \frac{1}{r^5} (\hat{\mathbf{r}}^T \cdot
  \mathbf{\Theta}_{\tau', n', \pm 1} \cdot \hat{\mathbf{r}})^{\ast}
  (\hat{\mathbf{r}}^T \cdot \mathbf{\Theta}_{\tau, n, \pm 1} \cdot
  \hat{\mathbf{r}}).
\end{eqnarray}
Therefore, as the distance between the annihilated and created exciton
wavepackets grows, the F\"orster coupling between two $s$-type wavepackets is
asymptotic to a nonradiative dipole-dipole interaction, with the amplitude
decaying in $r^{- 3}$; between a $s$-type and a $p$-type wavepackets, it will
be asymptotic to a dipole-quadrupole interaction, with the amplitude decaying
in $r^{- 4}$; between two $p$-type wavepackets, it will be asymptotic to a
quadrupole-quadrupole interaction, with the amplitude decaying in $r^{- 5}$.

\section{Supplementary Note 4: Oscillator strength of the intralayer excitonic component}
The oscillator strength of two studied hybrid
exciton wave packets is dominantly contributed by interlayer component, since
the intralayer component has a small optical dipole. To justify this, one can
estimate the ratio of oscillator strength $O_m^{\text{intra}}$ between an
intralayer exciton wave packet with $p$-type COM envelope function and that
with $s$-type COM envelope function by:
\begin{eqnarray}
  \frac{O^{\text{intra}}_{\pm 1}}{O^{\text{intra}}_0} & = & \frac{\int d
  \mathbf{q} e^{- i \mathbf{q} \cdot \mathbf{r}_\text{c}} | W_{\pm 1}
  (\mathbf{q}) \langle \mathbf{q} | \hat{H}_{\text{LM}} |
  X^{\text{intra}}_{\tau, \mathbf{q}} \rangle |^2}{\int d \mathbf{q} e^{-
  i \mathbf{q} \cdot \mathbf{r}_\text{c}} | W_0 (\mathbf{q}) \langle
  \mathbf{q} | \hat{H}_{\text{LM}} | X^{\text{intra}}_{\tau, \mathbf{q}}
  \rangle |^2} \approx \frac{\int_{\text{lc}} d \mathbf{q} | W_{\pm 1} (q)
  |^2}{\int_{\text{lc}} d \mathbf{q} | W_0 (q) |^2}
\end{eqnarray}
where $\int_{\text{lc}}$ represents that the integration is inside light cone,
whose radius is marked as $k_{\text{lc}}$. One can choose $\mathbf{r}_\text{c} = 0$
without losing generality. Inside the integral, the transition between
photonic state $| \mathbf{q} \rangle$ and the exciton through light-matter
interaction $\hat{H}_{\text{LM}}$ will cancel the angular momentum carried by
the envelope function, which is guaranteed by angular momentum conservation. The transition amplitude of the momentum eigenstate $|
X^{\text{intra}}_{\tau, \mathbf{q}} \rangle$ can be approximated uniform
inside the light cone~\cite{haug2009quantum}. Using the eigenstate of 2D harmonic oscillator $| W_m
(q) | \propto w^{| m | + 1} q^{| m |} e^{- w^2 q^2 / 2}$, the integral of $|
W_m (q) |^2$ inside light cone can be evaluated in the following form:
\begin{eqnarray}
  \int_{\text{lc}} d \mathbf{q} | W_m (\mathbf{q}) |^2 & = & 2 \pi \int_0^{w
  k_{\text{lc}}} d u u^{2 | m | + 1} e^{- u^2}
\end{eqnarray}
Typically, the range of light cone is $\sim 10^{- 3}$ of monolayer Brillouin
zone length, so $w k_{\text{lc}} = 10^{- 3} w / a \sim 3 \times 10^{- 3}$ to $10^{-2}$, where $a$ is
the monolayer lattice constant. The ratio will result in
\begin{eqnarray}
  \frac{O^{\text{intra}}_{\pm 1}}{O^{\text{intra}}_0} & \sim & 10^{- 5} \,\, \text{to} \,\, 10^{-4}
\end{eqnarray}
The intralayer exciton wave packet with $p$-type COM envelope function thus shows an oscillator strength nearly 4 to 5 orders smaller (depending on the wavepacket width), as compared to the s-type COM wavepacket.

\section{Supplementary Note 5: Tight binding models of hybrid moire excitons in t-MoTe$_2$}

\subsection{Tight binding Hamiltonian}

\begin{figure}
\includegraphics[width=0.8\textwidth]{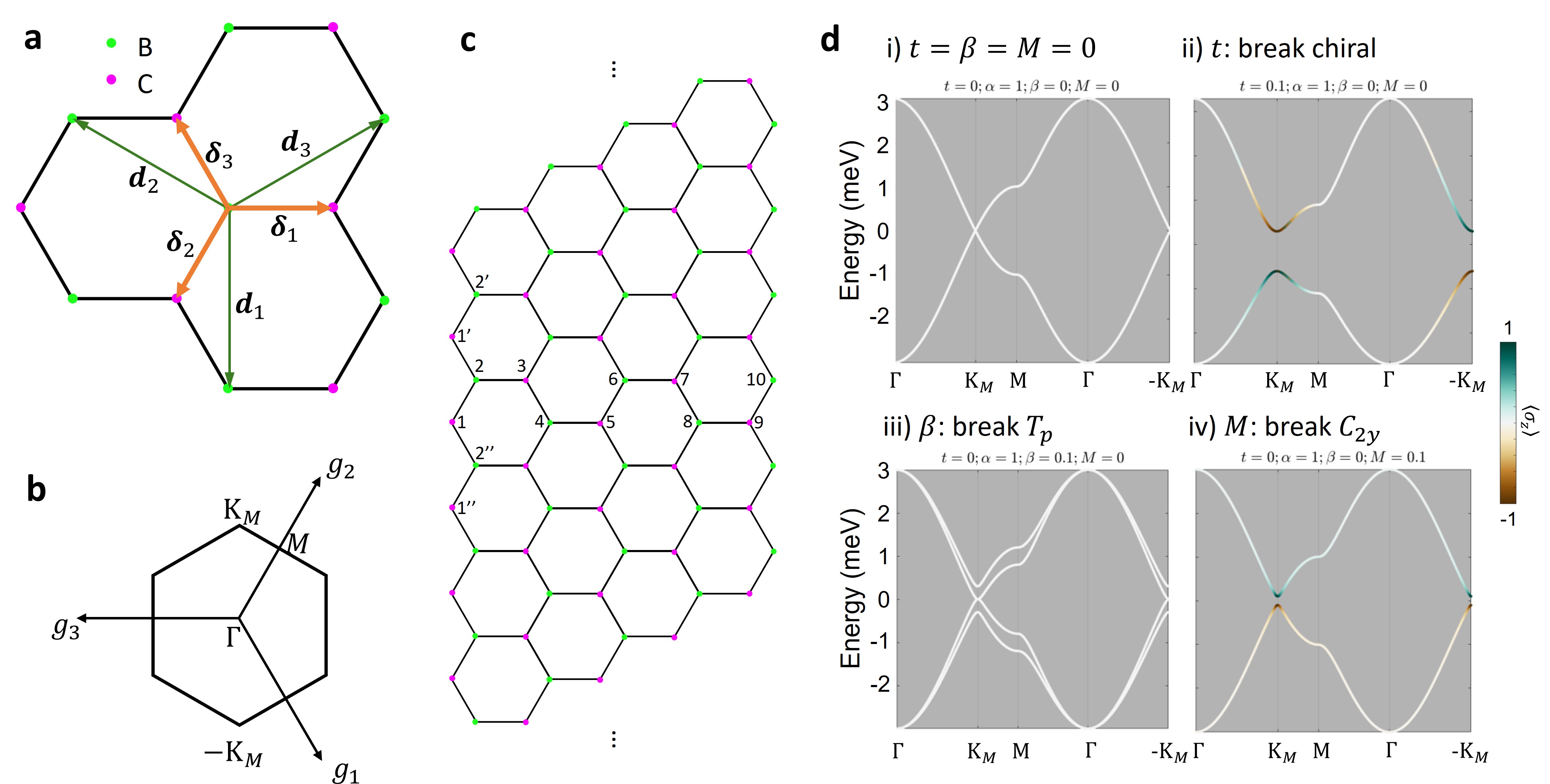}%
\caption{$\mathbf{a}$. Real space configuration of hybrid excitons.
$\mathbf{b}$. Moir\'e Brillouin zone of hybrid exciton lattice.
$\mathbf{c}$. Hybrid exciton lattice with open boundary condition along the armchair direction.
$\mathbf{d}$. Some exemplary bulk dispersion of hybrid moir\'e exciton in t-MoTe$_2$, color coded with the sublattice polarization $\braket{\sigma_z}$.}
\label{figS_lattice}
\end{figure}

In this section, we construct the tight binding Hamiltonian of the hybrid
moire excitons in t-MoTe$_2$. We first focus on F\"orster valley-orbit-coupling
(VOC) of two nearest neighboring sites $i$ and $i'$, and then expand to a
lattice model. Since only the hole hopping is considered, B (C) site has the
intralayer component on top (bottom) layer. The twist angle are inversed
$\theta_{\text{top}} = - \theta_{\text{bottom}} = \theta / 2$. There are four
coupling terms involved: i) A valley-conserving on-site coupling is present
for all states, which can be neglected by shifting the energy; ii) The
valley-filp on-site coupling vanishes according to Fig. 2c in the main text;
iii) The valley-conserving coupling between B and C: $(t_v \mathcal{C} /
\delta E)^2 e^{- i \theta} J_{1, 1}^{1, 1} = \alpha e^{- i \theta} =
\tilde{\alpha}$ ($J_{- 1, - 1}^{- 1, - 1}$ for time reversal counterpart),
with the amplitude being denoted by $\alpha$; iv) The valley-flip coupling
between B and C: $- (t_v \mathcal{C} / \delta E)^2 e^{4 i \varphi} J_{1, 1}^{-
1, - 1} = - \beta e^{- 4 i \varphi}$ ($e^{- 4 i \varphi} J_{- 1, - 1}^{1, 1}$
for time reversal counterpart), with the amplitude being denoted by $\beta$.
$\varphi$ denotes the azimuthal angle of $\mathbf{R}_\text{C} - \mathbf{R}_\text{B}$.
Thus, in the basis of $\{ | \Psi_{B, + K} \rangle, | \Psi_{B, - K} \rangle, |
\Psi_{C, + K} \rangle, | \Psi_{C, - K} \rangle \}$, the Hamiltonian reads
\begin{eqnarray}
  H^{v o c}_{\text{two} - \text{sites}} & = & \left( \begin{array}{cccc}
    0 & 0 & \tilde{\alpha} & - \beta e^{- 4 i \varphi}\\
    0 & 0 & - \beta e^{4 i \varphi} & \tilde{\alpha}^{\ast}\\
    \tilde{\alpha}^{\ast} & - \beta e^{- 4 i \varphi} & 0 & 0\\
    - \beta e^{+ 4 i \varphi} & \tilde{\alpha} & 0 & 0
  \end{array} \right). 
\end{eqnarray}
Using the basis $\{ e^{- i \theta / 2} | \Psi_{B, + K} \rangle, e^{i \theta /
2} | \Psi_{B, - K} \rangle, e^{i \theta / 2} | \Psi_{C, + K} \rangle, e^{- i
\theta / 2} | \Psi_{C, - K} \rangle \}$, the $\theta$-phase can be eliminated:
\begin{eqnarray}
  \tilde{H}^{v o c}_{\text{two} - \text{sites}} & = & \left(
  \begin{array}{cccc}
    0 & 0 & \alpha & - \beta e^{- 4 i \varphi}\\
    0 & 0 & - \beta e^{4 i \varphi} & \alpha\\
    \alpha & - \beta e^{- 4 i \varphi} & 0 & 0\\
    - \beta e^{+ 4 i \varphi} & \alpha & 0 & 0
  \end{array} \right). \label{eq-twosites} 
\end{eqnarray}
We then construct lattice model. The zigzag direction is chosen along $y$ axis
(Fig.~\ref{figS_lattice}a). Centered at B site, three nearest-neighbor (NN) bonds
$\mathbf{\delta}_j  (j = 1, 2, 3)$ and three next-nearest-neighbor (NNN)
bonds $\mathbf{d}_j  (j = 1, 2, 3)$ can be defined, where $- \mathbf{d}_2$
and $\mathbf{d}_3$ are primitive lattice vectors $\mathbf{a}_{M, 1}$ and
$\mathbf{a}_{M, 2}$, respectively (c.f. Fig. 3a in the main text). The
corresponding moire Brillouin zone (BZ) is shown in Fig.~\ref{figS_lattice}b. $\mathbf{g}_{1,2}$ are moire reciprocal vectors and $\mathbf{g}_3 = -(\mathbf{g}_1 + \mathbf{g}_2)$. High symmetry
points B and C are extended in a honeycomb lattice. The hybrid excitons are
written in
\begin{eqnarray}
  \ket{\Psi^{\text{hy}}_{\text{B}, + K}} & = & e^{- i \theta / 2} (| \psi^{\text{inter}}_{+ K, 0}
  (\mathbf{R}_\text{B}) \rangle + \mathcal{W}_{+ 1} | \psi^{\text{intra}}_{+ K, t, + 1}
  (\mathbf{R}_\text{B}) \rangle), \nonumber\\
  \ket{\Psi^{\text{hy}}_{\text{B}, - K}} & = & e^{i \theta / 2} (| \psi^{\text{inter}}_{- K, 0}
  (\mathbf{R}_\text{B}) \rangle + \mathcal{W}_{- 1} | \psi^{\text{intra}}_{- K, b, - 1}
  (\mathbf{R}_\text{B}) \rangle), \nonumber\\
  \ket{\Psi^{\text{hy}}_{\text{C}, + K}} & = & e^{i \theta / 2} (| \psi^{\text{inter}}_{+ K, 0}
  (\mathbf{R}_\text{C}) \rangle + \mathcal{W}_{+ 1} | \psi^{\text{intra}}_{+ K, b, + 1}
  (\mathbf{R}_\text{C}) \rangle), \nonumber\\
  \ket{\Psi^{\text{hy}}_{\text{C}, - K}} & = & e^{- i \theta / 2} (| \psi^{\text{inter}}_{- K, 0}
  (\mathbf{R}_\text{C}) \rangle + \mathcal{W}_{- 1} | \psi^{\text{intra}}_{- K, b, - 1}
  (\mathbf{R}_\text{C}) \rangle), 
\end{eqnarray}
where the sublattice and valley pseudospins are described by Pauli matrices
$\sigma_{x, y, z}$ and $\tau_{x, y, z}$, respectively. Besides, we use bosonic
operator $b_i = (b_{i, K}, b_{i, - K})$ to represent annihilation of hybrid
excitons from $\pm K$ valleys at site $\mathbf{R}_i$.
\begin{eqnarray}
  H & = & \sum_i \varepsilon_i b_i^{\dag} b_i + \sum_{\langle \langle i' i
  \rangle \rangle} b_{i'}^{\dag} \exp \left( i \frac{2 \pi}{3} \nu_{i' i}
  \tau_z \right) b_i + \sum_{\langle i' i \rangle} b_{i'}^{\dag} [\alpha -
  \beta (\tau_x \cos 4 \varphi_{i' i} + \tau_y \sin 4 \varphi_{i' i})] b_i. 
\end{eqnarray}
The first term is a staggered potential on site B ($\varepsilon_i = M$) and C
($\varepsilon_i = - M$), which can be introduced by an interlayer bias. The
second term is the valley-conserving kinetic propagation in the moire
potential landscape, which is absent between sublattices, and we keep the NNN
one as the leading order term within each sublattice. Its strength $t$ can be fitted from the ground state of continuum model. The hopping carries a complex phase, which originates from
the valley mismatch between the electron and hole separated in two layers.
$\nu_{i' i} = - \nu_{i i'} = \pm 1$, depending on the orientation of two NN
bonds the exciton traverses in going from site $i$ to $i'$, i.e. $\nu_{i' i} =
+ 1 (- 1)$ if the exciton makes a counterclockwise (clockwise) hopping in a
honeycomb to get to the second bond (see Fig. 3c in the main text).

The NN hopping term can be rewitten from Eq. (\ref{eq-twosites})
\begin{eqnarray}
  \mathcal{H}^{v o c}_{i' i} & = & b_{i'}^{\dag} [\alpha - \beta (\tau_x \cos
  4 \varphi_{i' i} + \tau_y \sin 4 \varphi_{i' i})] b_i. 
  \label{eqS_HamiltonianVOC}
\end{eqnarray}
The tight binding Hamiltonian can be written in the momentum space by making
the Fourier transformation
\begin{eqnarray}
  b_{\mathbf{k}} = \frac{1}{\sqrt{N}} \sum_i \left( \begin{array}{c}
    B_{i, + K}\\
    B_{i, - K}\\
    C_{i, + K}\\
    C_{i, - K}
  \end{array} \right) e^{i \mathbf{k} \cdot \mathbf{R}_i} & , &
  b_{\mathbf{k}}^{\dag} = \frac{1}{\sqrt{N}} \sum_i \left( \begin{array}{c}
    B_{i, + K}^{\dag}\\
    B_{i, - K}^{\dag}\\
    C^{\dag}_{i, + K}\\
    C^{\dag}_{i, - K}
  \end{array} \right)^T e^{- i \mathbf{k} \cdot \mathbf{R}_i}, 
\end{eqnarray}
where $N$ denotes the number of sites in total, and $B_{i, \pm K} (C_{i, \pm
K})$ denotes $b_{i, \pm K}$ operator at site B (C).

The Hamiltonian in momentum space is thus written as
\begin{eqnarray}
  H_{\mathbf{k}} & = & - t \sum_j \left( \begin{array}{cccc}
    -2t\cos \left( \mathbf{k} \cdot \mathbf{d}_j + \frac{2 \pi}{3} \right) & 0 & \alpha e^{i \mathbf{k} \cdot \mathbf{\delta}_j} &
    -\beta e^{i (\mathbf{k} \cdot \mathbf{\delta}_j - 4 \varphi_j)}\\
    0 & -2t\cos \left( \mathbf{k} \cdot \mathbf{d}_j - \frac{2 \pi}{3} \right) & -\beta e^{i (\mathbf{k} \cdot \mathbf{\delta}_j + 4 \varphi_j)} &
    \alpha e^{i \mathbf{k} \cdot \mathbf{\delta}_j}\\
    \alpha e^{-i \mathbf{k} \cdot \mathbf{\delta}_j} & -\beta e^{- i (\mathbf{k} \cdot \mathbf{\delta}_j + 4 \varphi_j)} & -2t\cos \left( \mathbf{k} \cdot \mathbf{d}_j - \frac{2 \pi}{3} \right) &
    0\\
    -\beta e^{- i (\mathbf{k} \cdot \mathbf{\delta}_j - 4 \varphi_j)} & \alpha e^{-i \mathbf{k} \cdot \mathbf{\delta}_j} & 0 & -2t\cos \left( \mathbf{k} \cdot \mathbf{d}_j + \frac{2 \pi}{3} \right)
  \end{array} \right) \nonumber\\
  &  & + M \left( \begin{array}{cccc}
    1 & 0 & 0 & 0\\
    0 & 1 & 0 & 0\\
    0 & 0 & - 1 & 0\\
    0 & 0 & 0 & - 1
  \end{array} \right),
  \label{eqS_TBHamiltonian}
\end{eqnarray}
where $\varphi_j$ is the azimuthal angle of $\mathbf{\delta}_j$.

\subsection{Symmetry analysis}

1. Under time reversal (TR) transformation, the valley pseudospin of the
hybrid exciton will flip while the sublattice pseudospin is kept invariant, so
the matrix representation $\hat{\mathcal{T}} = (\sigma_0 \otimes \tau_x)
\mathcal{K}$, where $\mathcal{K}$ is complex conjugate.
\begin{eqnarray}
  \hat{\mathcal{T}} H_{\mathbf{k}} \hat{\mathcal{T}}^{- 1} & = &
  (\sigma_0 \otimes \tau_x) H^{\ast}_{\mathbf{k}} (\sigma_0 \otimes
  \tau_x)^{- 1} = H_{- \mathbf{k}},
\end{eqnarray}
The Hamiltonian is TR invariant.

2. Under chiral (sublattice) transformation, $\hat{\mathcal{C}} = \sigma_z
\otimes \tau_0$, the NNN hopping and staggered potential term have even
chirality, the remaining terms have odd chirality $\hat{\mathcal{C}}
H_{\mathbf{k}} \hat{\mathcal{C}}^{- 1} = - H_{\mathbf{k}}$ (see Fig.~\ref{figS_lattice}d).

3. Under the pseudo-TR (pTR) operator $\hat{\mathcal{T}}_p = i (\sigma_0
\otimes \tau_y) \mathcal{K}$, the Hamiltonian has $\hat{\mathcal{T}}_p
H_{\mathbf{k}} \hat{\mathcal{T}}_p^{- 1} = H_{- \mathbf{k}}$ and
$\hat{\mathcal{T}_p}^2 = - 1$. The intervalley F\"orster coupling term will
break pTR, so the Kramer's degeneracy at M point of moire BZ will be
lifted.

4. The hybrid exciton lattice inherits the $C_{2y}$ symmetry of the R-stacking t-MoTe$_2$, which can be broken by staggered potential $M$.

\subsection{Minimal model ($k \cdot p$ expansion at moire BZ corners)}

Threefold rotation symmetry dictates that the Dirac points are located at moire BZ corners $\pm \text{K}_M$ where the band inversion occurs (Fig.~\ref{figS_lattice}d). Here, we make a $k \cdot p$ expansion around moire BZ corners to obtain the minimal model. We
define $\mathbf{k} = \mathbf{K}_M + \mathbf{\kappa}$, and make the
substitution in $H_{\mathbf{k}}$ (to the second order):
\begin{eqnarray}
  \sin (\mathbf{k} \cdot \mathbf{d}_j) & \rightarrow & \sin (\mathbf{K}_M
  \cdot \mathbf{d}_j) \left[ 1 - \frac{1}{2} (\mathbf{\kappa} \cdot
  \mathbf{d}_j)^2 \right] + \cos (\mathbf{K}_M \cdot \mathbf{d}_j)
  (\mathbf{\kappa} \cdot \mathbf{d}_j), \\
  \cos (\mathbf{k} \cdot \mathbf{d}_j) & \rightarrow & \cos (\mathbf{K}_M
  \cdot \mathbf{d}_j) \left[ 1 - \frac{1}{2} (\mathbf{\kappa} \cdot
  \mathbf{d}_j)^2 \right] - \sin (\mathbf{K}_M \cdot \mathbf{d}_j)
  (\mathbf{\kappa} \cdot \mathbf{d}_j).
\end{eqnarray}
The minimal model around $\mathbf{K}_M$ point (to the first order) is:
\begin{eqnarray}
  H^{+\text{K}_M}_{\kappa} & = & \left( \begin{array}{cccc}
    3 t + M & 0 & 0 & - 3 \beta\\
    0 & - 6 t + M & 0 & 0\\
    0 & 0 & - 6 t - M & 0\\
    - 3 \beta & 0 & 0 & 3 t - M
  \end{array} \right) \notag \\
  & + & \frac{\sqrt{3}}{2} \left( \begin{array}{cccc}
    0 & 0 & i \beta \kappa_+ a_M & 0\\
    0 & i \beta \kappa_+ a_M & 0 & i \beta \kappa_+ a_M\\
    - i \beta \kappa_- a_M & 0 & - i \beta \kappa_- a_M & 0\\
    0 & - i \beta \kappa_- a_M & 0 & 0
  \end{array} \right),
\end{eqnarray}
where $\kappa_{\pm} \equiv \kappa_x \pm i \kappa_y$. Similar operations can be done around $-\text{K}_M$ point.
At Dirac points ($\kappa = 0$), four eigenenergy can be obtained $E_{1, 2} = -
6 t \mp M$, $E_{3, 4} = 3 t \mp \sqrt{9 \beta^2 + M^2}$.

\subsection{F\"orster coupling of next nearest neighbor}

\begin{figure}[t]
\includegraphics[width=0.7\textwidth]{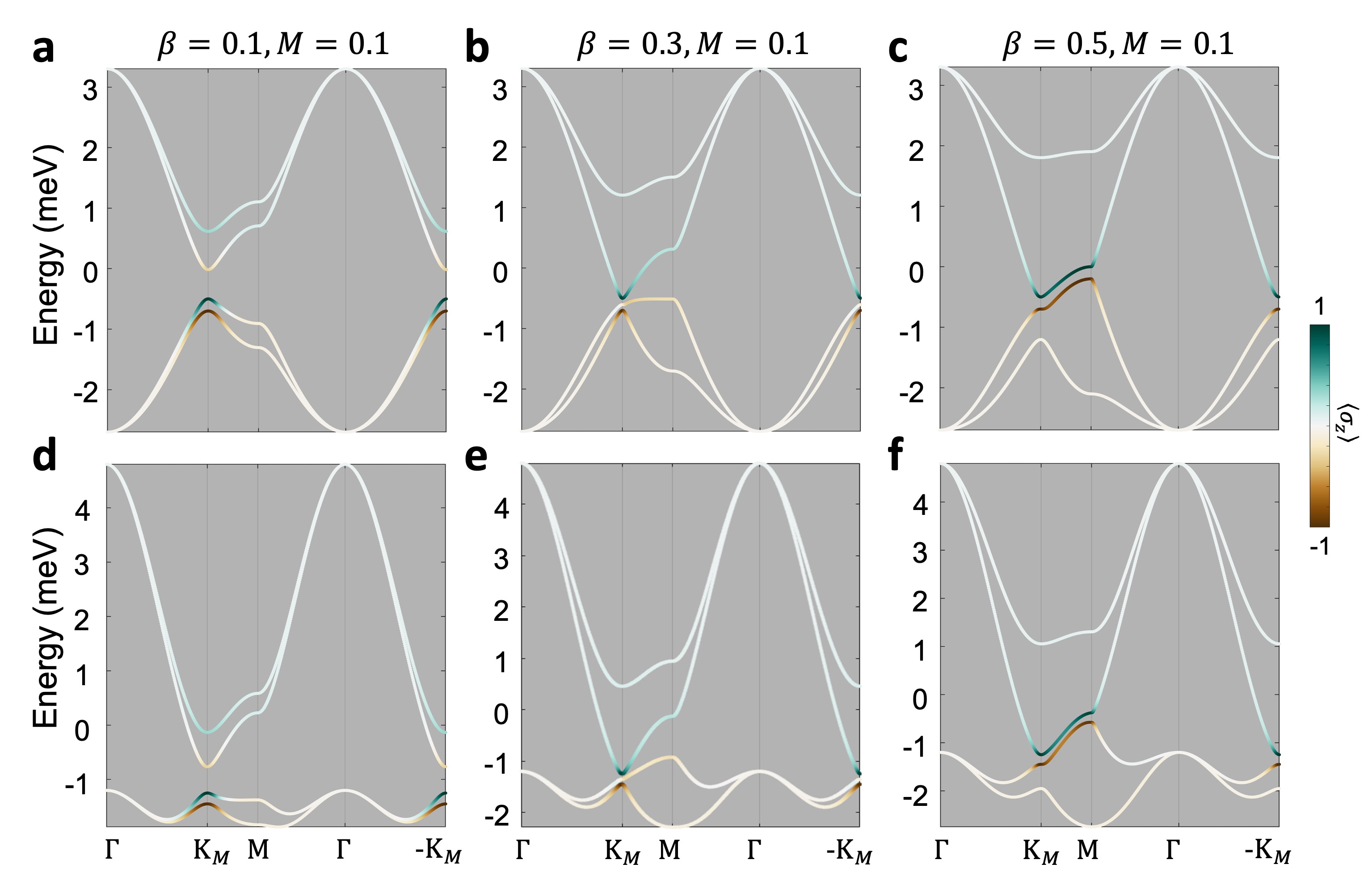}%
\caption{Dispersion with and without NNN F\"orster coupling.
$\mathbf{a}$-$\mathbf{c}$. Dispersion without NNN F\"orster coupling at different intervalley F\"orster coupling strength $\beta$. Parameters are taken as the blue points in the phase diagram Fig. 4c in the main text. $t = 0.1$ and $\alpha = 1$ for all cases. 
$\mathbf{d}$-$\mathbf{f}$. Dispersion with NNN F\"orster coupling, compared with ($\mathbf{a}$) to ($\mathbf{c}$), respectively. The NNN F\"orster coupling strengths are approximated as $\alpha' = \alpha/4, \beta' = \beta/4$. Color represents the sublattice polarization $\braket{\sigma_z}$.}
\label{figS_NNNForster}
\end{figure}

We consider the NNN F\"orster coupling between the intralayer components, where
the valley-conserving (valley-flip) strength is denoted as $\alpha' (\beta')$.
\begin{eqnarray}
  H^{N N N}_{\mathbf{k}} & = & \left( \begin{array}{cccc}
    \alpha' \sum_j e^{i \mathbf{k} \cdot \mathbf{d}_j} + c.c. & \beta'
    \sum_j e^{i (\mathbf{k} \cdot \mathbf{d}_j - 4 \phi_j)} & 0 & 0\\
    \beta' \sum_j e^{- i (\mathbf{k} \cdot \mathbf{d}_j - 4 \phi_j)} &
    \alpha' \sum_j e^{i \mathbf{k} \cdot \mathbf{d}_j} + c.c. & 0 & 0\\
    0 & 0 & \alpha' \sum_j e^{- i \mathbf{k} \cdot \mathbf{d}_j} + c.c. &
    \beta' \sum_j e^{- i (\mathbf{k} \cdot \mathbf{d}_j + 4 \phi_j)}\\
    0 & 0 & \beta' \sum_j e^{i (\mathbf{k} \cdot \mathbf{d}_j + 4 \phi_j)} &
    a' \sum_j e^{- i \mathbf{k} \cdot \mathbf{d}_j} + c.c.
  \end{array} \right),
\end{eqnarray}
where $\phi_j (j = 1, 2, 3)$ is the the hopping phase traversing intercell
$\mathbf{d}_j$. The Hamiltonian in real space is analogous to Eq.(\ref{eqS_HamiltonianVOC}) while coupling the same sublattices. We add this term into Eq.(\ref{eqS_TBHamiltonian}) by estimating $\alpha' = \alpha/4$ and $\beta' = \beta/4$, the dispersions are shown in Fig.~\ref{figS_NNNForster}d-f. Their comparison with Fig.~\ref{figS_NNNForster}a-c manifests that the involvement of NNN F\"orster coupling does not qualitatively change the eigenstates at Dirac cones.

\section{Supplementary Note 6: Zigzag edge states of hybrid excitons}

\begin{figure}
\includegraphics[width=0.8\textwidth]{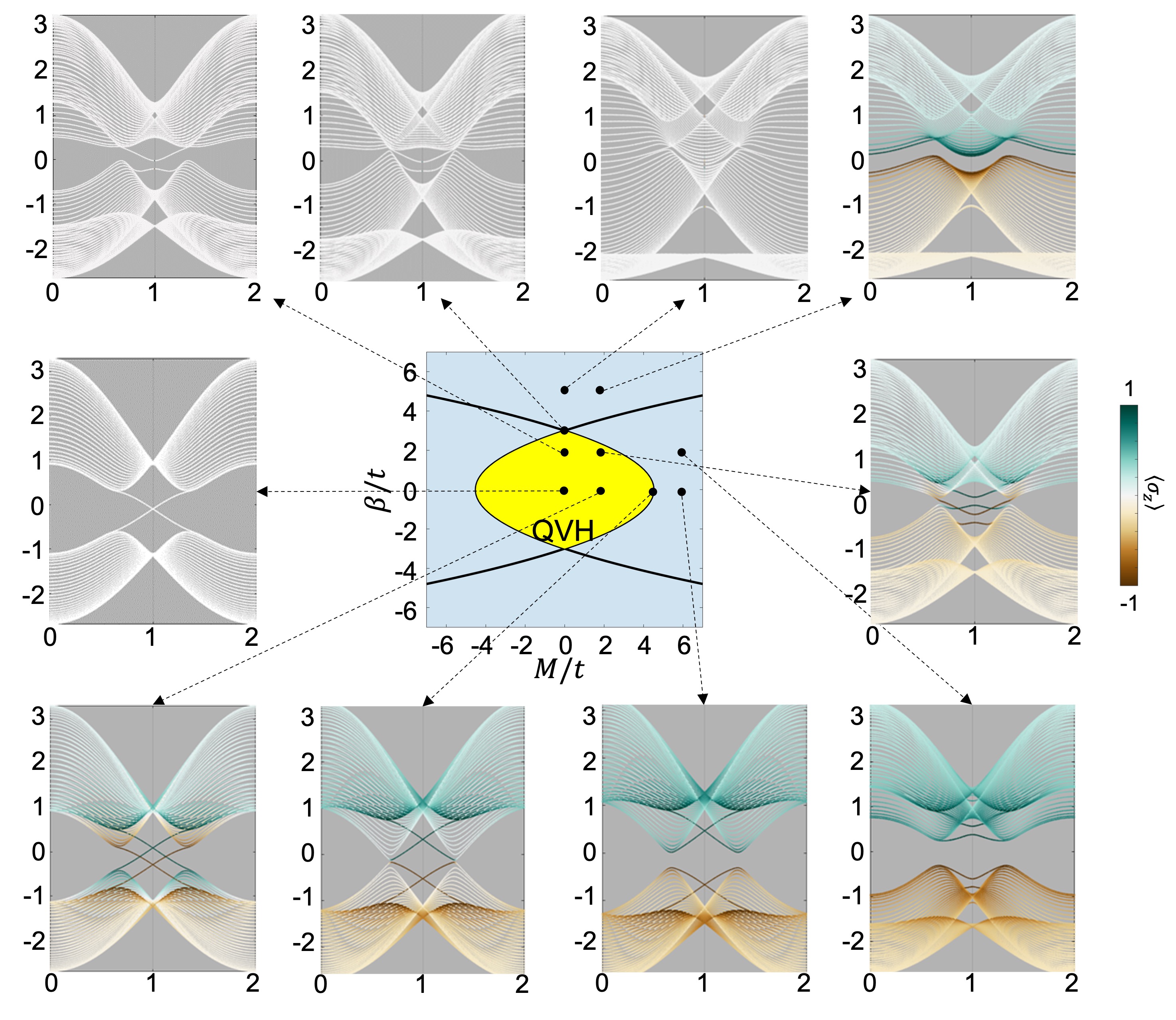}%
\caption{Zigzag modes of exciton dispersion, with parameters taken in the phase diagram. For each panel, the quantity shown on vertical axes is energy in meV, and shown on horizontal axes is $k_y$ in $\pi/a_M$. Color represents the sublattice polarization $\braket{\sigma_z}$.}
\label{figS_edgestates}
\end{figure}

The zigzag edge states of hybrid exciton can be obtained by opening the
armchair boundary condition of the tight binding lattice. Each zigzag mode is
along $y$ direction with good quantum number $k_y$. Along $x$ diraction, sites
are labeled by ascending numbers (Fig.~\ref{figS_lattice}c). The tighting binding Hamiltonian can be
written
\begin{eqnarray}
  H_{11} & = & \left( \begin{array}{cc}
    t \left( e^{i \frac{2 \pi}{3}} e^{i k_y a_M} + c.c. \right) & 0\\
    0 & t \left( e^{- i \frac{2 \pi}{3}} e^{i k_y a_M} + c.c. \right)
  \end{array} \right) - M, 
\end{eqnarray}
\begin{eqnarray}
  H_{12} & = & \left( \begin{array}{cc}
    \alpha \left( e^{i \frac{1}{2} k_y a_M} + c.c. \right) & \beta \left( e^{i
    \frac{1}{2} k_y a_M} e^{- 4 i \varphi_3} + e^{- i \frac{1}{2} k_y a_M}
    e^{- 4 i \varphi_2} \right)\\
    \beta \left( e^{i \frac{1}{2} k_y a_M} e^{4 i \varphi_3} + e^{- i
    \frac{1}{2} k_y a_M} e^{4 i \varphi_2} \right) & \alpha \left( e^{i
    \frac{1}{2} k_y a_M} + c.c. \right)
  \end{array} \right), 
\end{eqnarray}
\begin{eqnarray}
  H_{13} & = & \left( \begin{array}{cc}
    t \left( e^{i \frac{2 \pi}{3}} e^{- i \frac{1}{2} k_y a_M} + c.c. \right)
    & 0\\
    0 & t \left( e^{- i \frac{2 \pi}{3}} e^{- i \frac{1}{2} k_y a_M} + c.c.
    \right)
  \end{array} \right), 
\end{eqnarray}
\begin{eqnarray}
  H_{22} & = & \left( \begin{array}{cc}
    t \left( e^{i \frac{2 \pi}{3}} e^{- i k_y a_M} + c.c. \right) & 0\\
    0 & t \left( e^{- i \frac{2 \pi}{3}} e^{- i k_y a_M} + c.c. \right)
  \end{array} \right) + M, 
\end{eqnarray}
\begin{eqnarray}
  H_{23} & = & \left( \begin{array}{cc}
    \alpha & \beta e^{4 i \varphi_3}\\
    \beta e^{- 4 i \varphi_3} & \alpha
  \end{array} \right), 
\end{eqnarray}
\begin{eqnarray}
  H_{24} & = & \left( \begin{array}{cc}
    t \left( e^{- i \frac{2 \pi}{3}} e^{- i \frac{1}{2} k_y a_M} + c.c.
    \right) & 0\\
    0 & t \left( e^{i \frac{2 \pi}{3}} e^{i \frac{1}{2} k_y a_M} + c.c.
    \right)
  \end{array} \right), 
\end{eqnarray}
\begin{eqnarray}
  H_{33} & = & H_{11}, \\
  & \ldots &  \nonumber
\end{eqnarray}
Therefore, the Hamiltonian can be concluded
\begin{eqnarray}
  H_{\text{zigzag}} & = & \left( \begin{array}{ccccccc}
    H_{11} & H_{12} & H_{13} & 0 & 0 & 0 & \\
    H_{12}^{\dag} & H_{22} & H_{23} & H_{24} & 0 & 0 & \\
    H_{13}^{\dag} & H_{23}^{\dag} & H_{33} & H_{34} & H_{35} & 0 & \\
    0 & H_{24}^{\dag} & H_{34}^{\dag} & H_{44} & H_{45} & H_{46} & \\
    0 & 0 & H_{35}^{\dag} & H_{45}^{\dag} & H_{55} & H_{56} & \\
    0 & 0 & 0 & H_{46}^{\dag} & H_{56}^{\dag} & H_{66} & \\
    &  &  &  &  &  & \ddots
  \end{array} \right) = \left( \begin{array}{ccccccc}
    H_{11} & H_{12} & H_{13} & 0 & 0 & 0 & \\
    H_{12}^{\dag} & H_{22} & H_{23} & H_{24} & 0 & 0 & \\
    H_{13}^{\dag} & H_{23}^{\dag} & H_{11} & H_{12} & H_{13} & 0 & \\
    0 & H_{24}^{\dag} & H_{12}^{\dag} & H_{22} & H_{23} & H_{24} & \\
    0 & 0 & H_{13}^{\dag} & H_{23}^{\dag} & H_{11} & H_{12} & \\
    0 & 0 & 0 & H_{24}^{\dag} & H_{12}^{\dag} & H_{22} & \\
    &  &  &  &  &  & \ddots
  \end{array} \right). 
\end{eqnarray}
The dispersions are shown in Fig.~\ref{figS_edgestates}, with parameters taken in the phase diagram.

\clearpage

\section{Supplementary Note 7: The twist angle dependence of intra- and intervalley F\"orster coupling strength}

\begin{figure}[htb]
\includegraphics[width=0.5\textwidth]{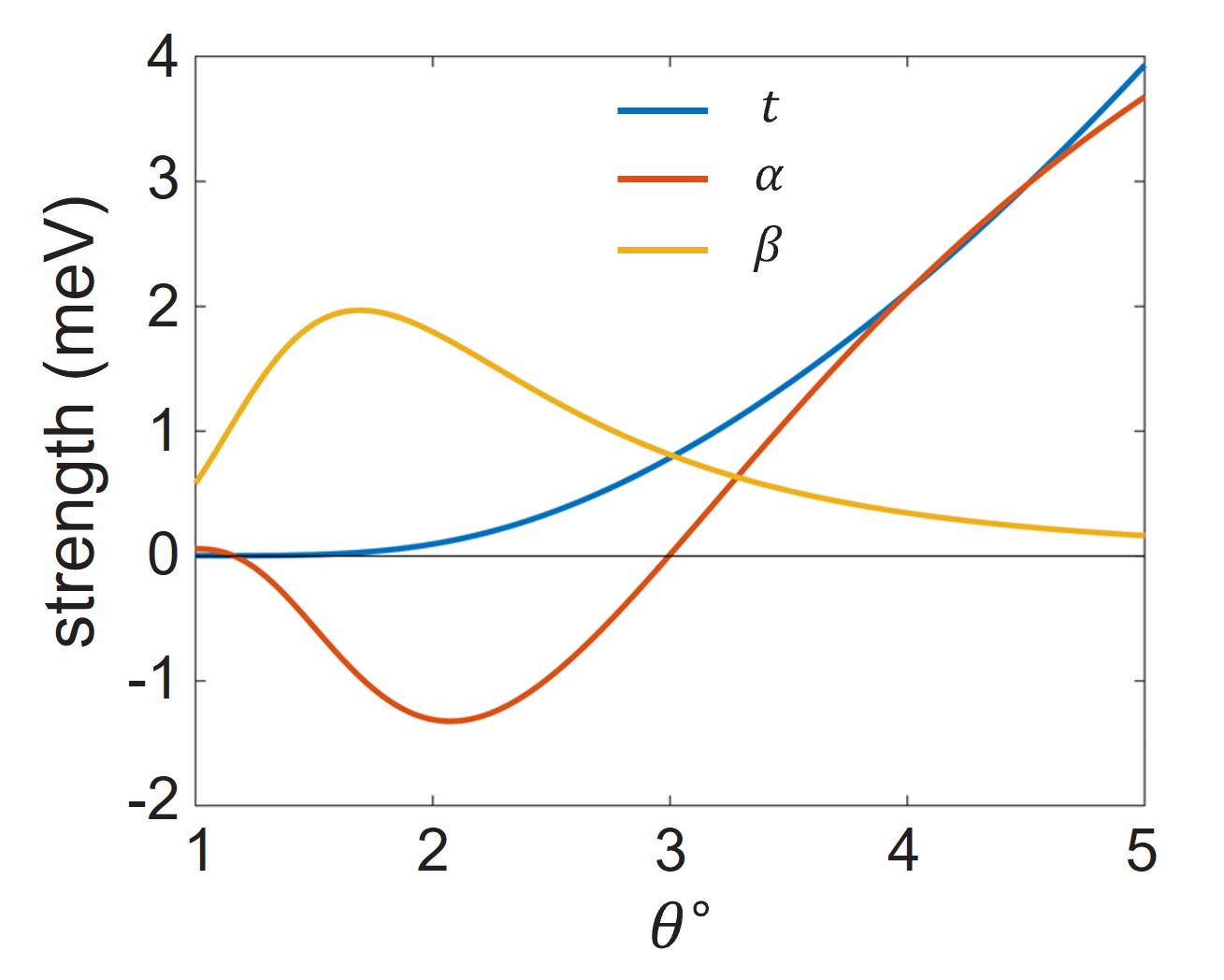}%
\caption{Kinetic propagation strength $t$ and intravalley (intervalley) F\"orster coupling strength $\alpha (\beta)$ as a function of twist angle $\theta$. The hybridization weight $\eta = 0.25$, and wavepacket width $w = 2$ nm. These parameter functions determine the phase diagram of Fig. 4e in the main text. }
\end{figure}

\bibliography{hybridX_ref}